\documentclass[aps,twocolumn,notitlepage,prl,superscriptaddress,nofootinbib]{revtex4-2}
\usepackage{amsmath,amssymb,bm}
\usepackage{graphicx}
\usepackage{epstopdf}
\usepackage{latexsym}
\usepackage[normalem]{ulem} 
\usepackage[usenames,dvipsnames]{color}
\usepackage{hyperref}
\usepackage{natbib}
\usepackage{braket}
\usepackage{mathtools}
\graphicspath{{Figures/}}
\begin{document}

\newcommand{\Rev }[1]{{\color{blue}{#1}\normalcolor}} 
\newcommand{\Com}[1]{{\color{red}{#1}\normalcolor}} 
\newcommand{\abs}[1]{\lvert #1 \rvert}

\title{Quantum-enhanced sensing of displacements and electric fields with large trapped-ion crystals}

\date{\today}

\author{K. A. Gilmore}
\altaffiliation{Current address: Honeywell Quantum Solutions, 303 S. Technology Ct., Broomfield, Colorado 80021, USA}
\email[]{kevin.gilmore@colorado.edu}
\affiliation{National Institute of Standards and Technology, Boulder, Colorado 80305, USA}
\affiliation{Department of Physics, University of Colorado, Boulder, Colorado, 80309, USA}

\author{M. Affolter}
\affiliation{National Institute of Standards and Technology, Boulder, Colorado 80305, USA}

\author{R.~J. Lewis-Swan}
\affiliation{Homer L. Dodge Department of Physics and Astronomy, The University of Oklahoma, Norman, Oklahoma 73019, USA}
\affiliation{Center for Quantum Research and Technology, The University of Oklahoma, Norman, Oklahoma 73019, USA}

\author{D. Barberena}
\affiliation{JILA, NIST and Department of Physics, University of Colorado, 440 UCB, Boulder, Colorado 80309, USA}
\affiliation{Center for Theory of Quantum Matter, University of Colorado, Boulder, CO 80309, USA}

\author{E. Jordan}
\altaffiliation{Current address: Physikalisch-Technische Bundesanstalt, Bundesallee 100, 38116 Braunschweig, Germany}
\affiliation{National Institute of Standards and Technology, Boulder, Colorado 80305, USA}

\author{A. M. Rey}
\email[]{arey@jila.colorado.edu}
\affiliation{JILA, NIST and Department of Physics, University of Colorado, 440 UCB, Boulder, Colorado 80309, USA}
\affiliation{Center for Theory of Quantum Matter, University of Colorado, Boulder, CO 80309, USA}

\author{J. J. Bollinger}
\email[]{john.bollinger@nist.gov}
\affiliation{National Institute of Standards and Technology, Boulder, Colorado 80305, USA}

\begin{abstract}
Developing the isolation and control of ultracold atomic systems to the level of single quanta has led to significant advances in quantum sensing, yet demonstrating a quantum advantage in real world   applications by harnessing entanglement remains a core task. Here, we realize a many-body quantum-enhanced sensor to detect weak displacements and electric fields using a large crystal of $\sim 150$ trapped ions. The center of mass vibrational mode of the crystal serves as high-Q mechanical oscillator and the collective electronic spin as the measurement device. 
By entangling  the oscillator and the collective spin before the displacement is applied and by controlling  the coherent dynamics via a many-body echo we are able to utilize  the delicate  spin-motion entanglement to map the displacement into a spin rotation such that we avoid quantum  back-action and cancel detrimental thermal noise. 
We report quantum  enhanced  sensitivity to displacements  of $8.8 \pm 0.4$~dB below the standard quantum limit and a sensitivity for measuring electric  fields of $240\pm10~\mathrm{nV}\mathrm{m}^{-1}$ in $1$ second ($240~\mathrm{nV}\mathrm{m}^{-1}/\sqrt{\mathrm{Hz}}$).
\end{abstract}

\maketitle

\noindent{\bf Introduction:} The development of protocols and quantum platforms that harness entanglement or correlations to sense or measure a physical quantity with an advantage relative to classical alternatives is of both practical and fundamental interest \cite{Dowling2003, Cappellaro_2017}. While practical quantum-enhanced sensors may take many years to come to fruition, they have the potential to enable the measurement of previously undetectable signals that could improve our understanding of the universe. Two prominent examples are gravitational wave detection, where non-classical states of light are now being adopted to achieve unparalleled sensitivity \cite{ligo2019,VIRGO2019}, and searches for dark matter, where a quantum advantage could enable the sensing of the weak, non-gravitational interaction of dark matter with normal matter \cite{ADMX,HAYSTAC,Malnou2019,Backes2021}.

In parallel, mechanical oscillators have become established as exquisite quantum tools to measure small displacements due to weak forces and electric fields \cite{Didi_2018,Wolf2019,Schreppler_2014,Kolkowitz_2012,Delaney2019,Gilmore2017,Wang2019,Affolter2020,Polzik2021}, such as those generated by axion-like dark matter and hidden photons \cite{Turner1990}. At the simplest level, a  weak, resonant force interacting with the mechanical oscillator generates a small coherent displacement $\beta$ of the oscillator amplitude over time. Typically, the displacement is then inferred by making a measurement on a complementary internal or coupled degree of freedom of the system.

The precision $\Delta\beta$ to which this displacement $\beta$ can be determined using \emph{classical} resources that use uncorrelated states, such as a vacuum or a coherent state, is fundamentally bounded by the so  called standard quantum limit (SQL) that limits the attainable sensitivity to $\Delta\beta \geq 1/2$. However, by introducing entanglement between the oscillator and the measurement degree of freedom before the oscillator is excited, one can attain sub-SQL precision \cite{Hempel_2013, Toscano_2006,Penasa_2016,Lewis-Swan2020}. Nevertheless, this requires subtle control over both the oscillator and the measurement system to minimize undesirable classical noise and to evade quantum back-action \cite{Kampel_2017}.

\begin{figure*}
    \centering
    \includegraphics[width = 1.6\columnwidth]{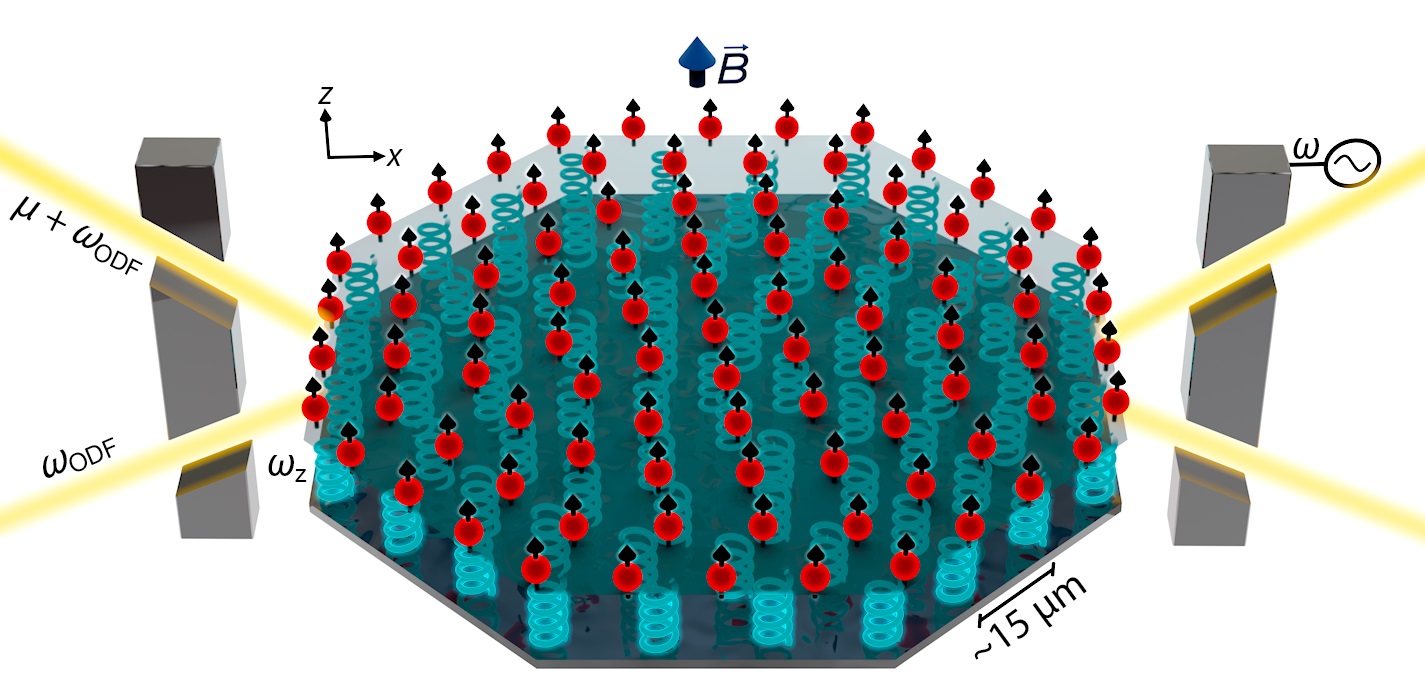}
    \caption[]{Trapped-ion crystal quantum sensor. An ensemble of beryllium ions (red dots) confined within a Penning trap self-arrange into a $2$D triangular lattice. The Penning trap is characterized by axial magnetic field $B = 4.45$~T and axial trap frequency $\omega_z = 2\pi \times 1.59$ MHz. Cylindrical electrodes (cross-section shown, in gray) generate a harmonic confining potential along the $\hat{Z}$-axis, while radial confinement is provided by the Lorentz force from $\vec{E} \times \vec{B}$-induced rotation in the axial magnetic field. The valence electron spin (black arrows) of the ions realize a collective spin that serves as the measurement device. The center-of-mass (COM) motional mode of the ion crystal realizes a high-Q mechanical oscillator (blue springs) with frequency $\omega_{z}$. Crossed optical beams (yellow lines) with a beat-note frequency $\mu \simeq \omega_z$ generate a spin-dependent optical-dipole force that couples the spins and COM oscillator. To generate a coherent axial oscillation with calibrated amplitude $Z_c$ and frequency $\omega = \omega_{z}$ an AC voltage source is applied to the trap endcap electrode.}
    \label{fig:trap}
\end{figure*}

In this work, we demonstrate a quantum advantage for both displacement and electric field sensing in a two-dimensional, 
$200~\mu$m diameter trapped-ion crystal of $\sim 150$ ions. The center-of-mass (COM) normal mode of the crystal realizes the mechanical oscillator and the internal electronic spin of the ions serves as the accessible measurement probe. A spin dependent, optical-dipole force (ODF) that resonantly couples the spins and the COM mode is used to generate metrologically useful entanglement which enables precise sensing of displacements of the oscillator \cite{Toscano_2006}. 
By implementing a many-body echo  protocol \cite{Lewis-Swan2020} to subsequently disentangle the spins and the oscillator, we are able to estimate the displacement - ideally free from thermal noise - through simple global measurements of the spins. With this technique, we achieve a sensitivity of $8.8 \pm 0.4$~dB  below the SQL for displacements  and $4.0 \pm 0.5$~dB below the SQL for electric fields. In practical terms we demonstrate an ultimate electric field measurement sensitivity of $240 \pm 10$ {\rm nV} $\mathrm{m}^{-1}/\sqrt{\mathrm{Hz}}$, an improvement by a factor of 300 over prior off-resonant classical protocols in trapped ions \cite{Gilmore2017,Affolter2020} and more than an order of magnitude compared to  state-of-the-art electrometers based on Rydberg atoms \cite{Jing2020}. Although similar protocols have been previously performed using a single Rydberg atom \cite{Penasa_2016} or a single trapped ion\cite{Hempel_2013,Burd_2019} to encode a spin-$1/2$, our protocol is the first to demonstrate an enhanced sensitivity resulting from quantum entanglement in a mesoscopic ion crystal, while also benefiting from the intrinsic reduction of zero-point motion in the collective mechanical oscillator due to the large ion number.

\noindent{\bf Entanglement-enhanced sensor:} Our quantum sensor consists of a single-plane Coulomb crystal of $N\sim 150$ $^{9}$Be$^{+}$ ions confined in a Penning trap  \cite{Bollinger2013,Sawyer2014,Bohnet2016,Gilmore2017,Affolter2020}, as shown in Fig.~\ref{fig:trap}. The $\prescript{2}{}{S}_{1/2}$ ground-state valence electron spin $\ket{\uparrow} (\ket{\downarrow}) \equiv \ket{m_{J}=+1/2} (\ket{m_{J}=-1/2})$ encodes a spin-$1/2$ degree of freedom in each ion, which can be coherently controlled by external microwaves  resonant with the $124$~GHz frequency splitting of the electronic spin states in the presence of a $B=4.5$ T magnetic field. The motion of the ion crystal can be decomposed into $2N$ in-plane modes and $N$ axial  modes, with the latter coupled to the spin degree of freedom by a spin-dependent ODF produced by a pair of off-resonant laser beams detuned from the nearest optical transitions by $\sim$20 GHz. The ODF beams generate a one-dimensional (1D) traveling-wave potential at a frequency $\mu$, which, in contrast to prior  settings for sensing in our experiment \cite{Gilmore2017,Affolter2020}, is now  chosen  to be near-resonant with the COM mode at a frequency $\omega_{z}/(2\pi)= 1.59$~MHz, such that the COM is the dominant motional contribution to the observed dynamics. In this limit, and assuming the ions have an axial extent that is small compared to the wavelength of the traveling wave optical potential, the system can be well approximated by the  Hamiltonian \cite{Bollinger_2018,SOM},
\begin{equation}
    \hat{H}_{\mathrm{ODF}} = \frac{\hbar g}{\sqrt{N}} \left( \hat{a}+\hat{a}^{\dagger} \right) \hat{J}_{z} + \hbar\delta \hat{a}^{\dagger} \hat{a} . \label{eqn:Hodf_res}
\end{equation}
Here, we have introduced collective spin operators $\hat{J}_{\alpha} = 1/2\sum_{j=1}^N \hat{\sigma}^{(j)}_{\alpha}$ where $\hat{\sigma}^{(j)}_{\alpha}$ are Pauli operators for the $j$th spin, $\hat{a}$ and $\hat{a}^{\dagger}$ are the COM phonon creation and annihilation operators that couple uniformly to all spins with strength $g$ and $\delta = \mu - \omega_{z}$ is  the detuning from the COM mode, ideally tuned to be on resonance in our protocol, $\delta=0$.

\begin{figure*}
    \centering
    \includegraphics[width = 1.5\columnwidth]{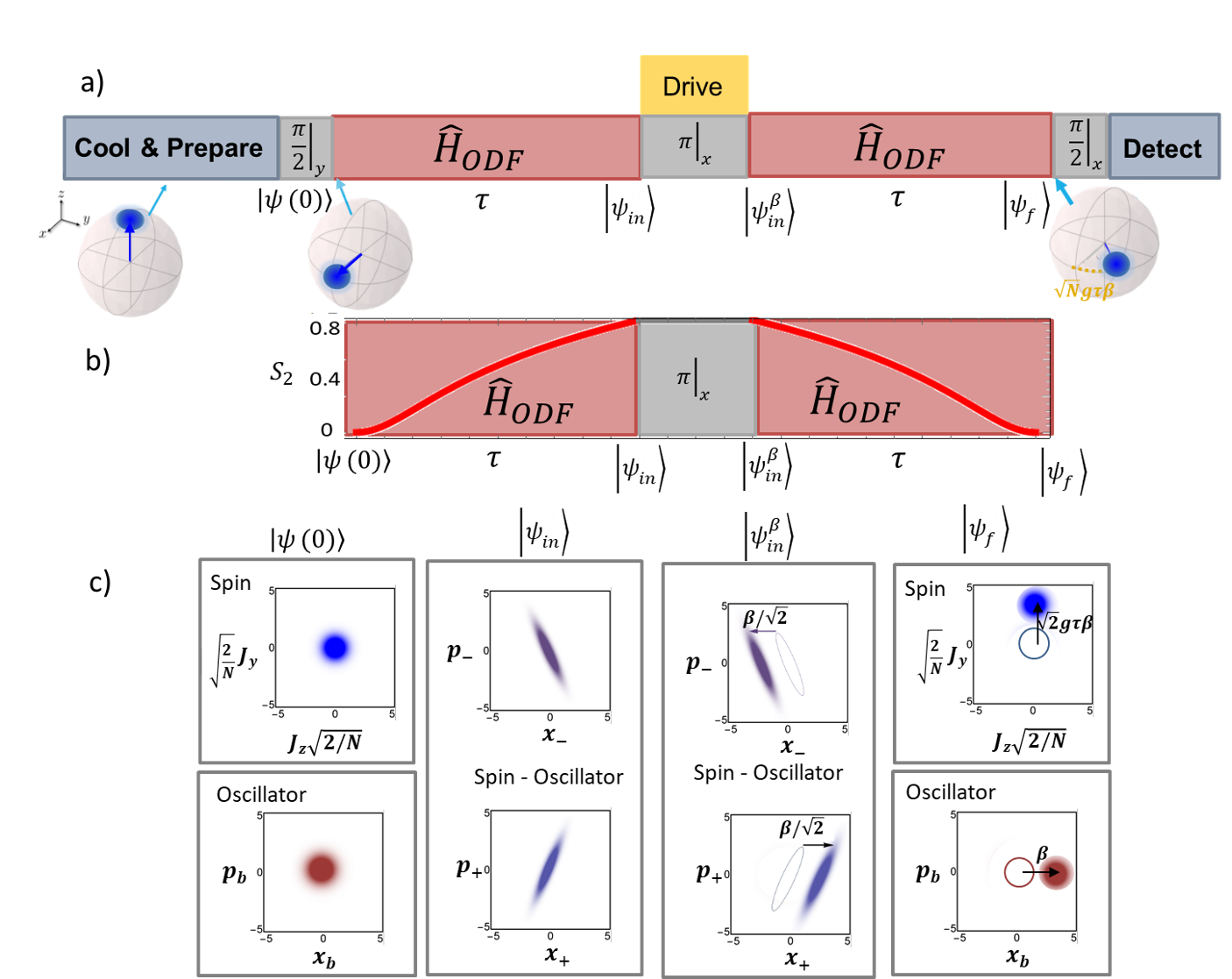}
    \caption[]{Displacement sensing protocol. a) The ions are Doppler cooled and optically pumped into the state $\vert \uparrow \rangle$, before a  microwave $\pi/2$-pulse rotates the spin ensemble to align along the $\hat{x}$-axis to prepare the initial state $\vert\psi(0)\rangle$. A phase-space illustration of the initial state is shown in c). It features the characteristic   Gaussian  and  isotropic   quantum noise distribution of a coherent state  in both  the spin and the oscillator degrees of freedom. A resonant ODF drive, described by $\hat{H}_{\mathrm{ODF}}$ [Eq.~(\ref{eqn:Hodf_res})], is then applied for a duration $\tau$ to yield the state $\vert\psi_{\mathrm{in}}\rangle$ in which the spin and the COM oscillator are entangled. This entanglement is illustrated by the growth of the R\'{e}nyi entropy $S_2$ \cite{SOM}, shown in b), which reaches a maximum after the first ODF drive. In panel c) we illustrate that this entanglement can be interpreted as squeezing of the composite spin-oscillator quantum noise in a coupled basis. To generate a small spin-independent displacement $\beta$ of the oscillator we apply a weak drive to an endcap electrode resonant with the COM frequency, yielding the perturbed state $\vert\psi_{\mathrm{in}}^{\beta}\rangle$. Simultaneously, a $\pi$-pulse is applied to flip $\vert\uparrow\rangle\to \vert\downarrow\rangle$  and thus effectively reverse the sign of $\hat{H}_{\mathrm{ODF}}$ when a subsequent resonant ODF drive is applied. This second application of the ODF realizes a many-body echo of the initial dynamics and disentangles the spin and the oscillator [demonstrated in b) by the vanishing R\'{e}nyi entropy after the second ODF drive], while also acting to map the small spin-independent displacement $\beta$ into a spin rotation of $\phi= g\tau\beta/\sqrt{N}$ about $\hat{z}$. The final decoupled state $\vert\psi_{f}\rangle$ is shown in c). A final $\pi/2$-pulse transforms the accumulated spin precession into a measurable change in the fraction of spins in $\vert \uparrow \rangle$ that is readout via a projective measurement. In b) and c) we use typical values of $g\tau = 2$ for illustration.}
    \label{fig:protocol}
\end{figure*}

The experimental sequence to sense small displacements of the COM oscillator is illustrated in Fig.~\ref{fig:protocol}(a). The ions are prepared in the state $\ket{\uparrow}_N$ by optical pumping before a microwave $\pi/2$-pulse is applied to rotate the spins to align along the $\hat{x}$-axis of the Bloch sphere. To entangle the spins and the phonons, we apply the ODF Hamiltonian  ($\hat{H}_{\mathrm{ODF}}$) for time $\tau$. Then, a calibrated AC voltage is applied to an endcap electrode for a time $t_{\mathrm{drive}} \ll \tau$ to drive up a small coherent displacement $\beta$ of the COM oscillator. Recent work \cite{Affolter2020} established a stable relative phase between the displacement $\beta$ and the ODF, which is experimentally optimized to ensure the largest signal. To optimally detect the displacement we implement a many-body echo to undo the entangling dynamics by applying a second ODF drive with opposite sign $-\hat{H}_{\mathrm{ODF}}$ for an identical time $\tau$. Experimentally this is done by  applying a microwave $\pi$-pulse about the $\hat{x}$-axis to map $\vert \uparrow \rangle \to \vert \downarrow \rangle$, which is equivalent to setting $g \to -g$. Finally, the displacement is estimated via a projective measurement on the spin degree of freedom. Specifically, to make an effective measurement of $\hat{J}_y$ we rotate the spins by a microwave $\pi/2$-pulse about the $\hat{x}$-axis and then measure $\hat{J}_z$ using state-dependent fluorescence imaging \cite{SOM}.

\begin{figure*}
    \centering
    \includegraphics[width=\textwidth]{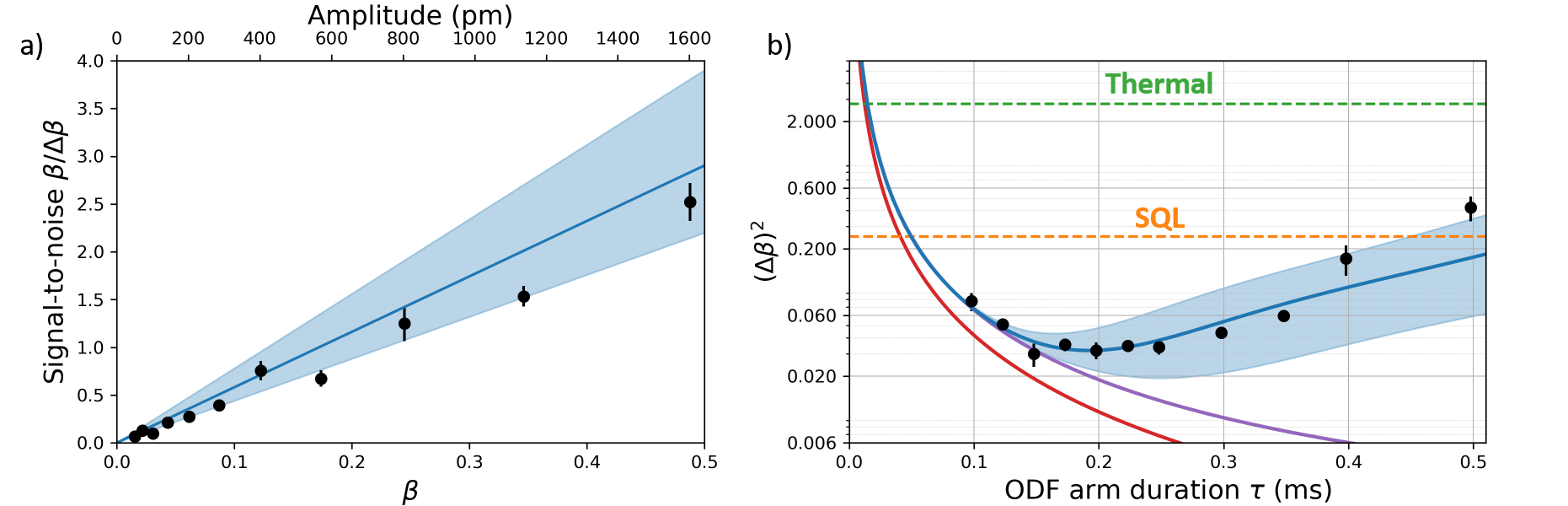}
    \caption[]{Performance of mechanical displacement sensor. a) Signal-to-noise ratio, $\beta/(\Delta\beta) \equiv Z_c/(\Delta Z_c)$, as a function of displacement $\beta$ (amplitude $Z_c \equiv 2z_0\beta/\sqrt{N}$ with $z_0 = \sqrt{\hbar/(2m\omega_z)}$ the spatial extent of the ground-state wavefunction). Each experiment data point (black markers, error bars indicate statistical and systematic uncertainty) corresponds to the SNR of a single measurement with fixed $\tau = 200~\mu$s. Good agreement is found with the theoretical model (blue line) using independently calibrated values of $g/(2\pi)=3.91$~kHz, COM frequency fluctuations of $\sigma/(2\pi) = 40$~Hz (shaded region indicates confidence region for $\sigma/(2\pi) \in [20,60]$~Hz), as well as spin depolarization characterized by $\Gamma = 500 ~\mathrm{s}^{-1}$. The theoretical calculation also includes an observed $18\%$ increase in background noise above the expected projection noise \cite{SOM}. b) Sensitivity $(\Delta\beta)^2$ to the dimensionless displacement $\beta$ as a function of ODF duration $\tau$. Experimental data (black markers, error bars indicate statistical and systematic uncertainty) is obtained from a small fixed physical displacement $Z_c = 775 \pm 28$~pm with a single measurement SNR of approximately $1$. We compare to the theoretical predictions of an idealized model (red line), a model that includes depolarization due to spin decoherence (purple line, $\Gamma = 610 ~\mathrm{s}^{-1}$), and a model which also includes COM fluctuations of $\sigma = 40$~Hz (solid blue line, shaded region indicates confidence region for $\sigma/(2\pi) \in [20,60]$~Hz). The SQL $(\Delta\beta)_{\mathrm{SQL}}^2 = 1/4$, corresponding the extent of the ground-state wavefunction, is indicated by a dashed orange line. For completeness, we also indicate the classically attainable sensitivity when accounting for excess thermal fluctuations, $\bar{n} = 5$ at the Doppler cooling limit, $(\Delta\beta)^2_{\mathrm{th}} = (2\bar{n}+1)(\Delta\beta)_{\mathrm{SQL}}^2$. The optimal experimental sensitivity of $8.8 \pm 0.4$ dB below the SQL ($\sim 19$ dB below the thermal noise limit) corresponds to the average of the five points centered around $\tau = 200~\mu$s \cite{SOM}.}
    \label{fig:amp_sens}
\end{figure*}

To simplify our analysis of the experimental protocol we assume the ions are initialized in the motional ground-state, such that the state of the spins and COM mode after the first microwave pulse is $\ket{ \psi \left( 0 \right)} = \ket{0}_{b} \otimes \ket{ \left( N/2 \right)_x}$. Here, we describe the spins using Dicke states $\hat{J}_{\alpha} \vert m_{\alpha} \rangle = m_{\alpha} \vert m_{\alpha} \rangle$  with $\alpha=x,y,z$ and $\ket{0}_{b}$ is the bosonic vacuum. The first ODF generates a \emph{spin-dependent} coherent displacement of the oscillator,
\begin{eqnarray}
    \vert \psi_{\mathrm{in}}\rangle &=&
   e^{-i \frac{\hbar g \tau}{\sqrt{N}} \left( \hat{a}+\hat{a}^{\dagger} \right) \hat{J}_{z}}\vert \psi(0)\rangle, \\&=& \sum_{m_z=-N/2}^{N/2} c_{m_z} \vert \alpha_{m_z} \rangle_{b} \otimes \vert m_z \rangle , \label{eqn:NcatState}
\end{eqnarray}
resulting in strong spin-motion entanglement.
This can be directly quantified by the purity of the reduced density matrix of the oscillator degree of freedom after tracing over the spin (e.g, R\'{e}nyi entanglement entropy $S_2$ \cite{SOM}) during the first ODF sequence and shown in Fig.~\ref{fig:protocol}b. In Eq.~(\ref{eqn:NcatState}) we introduce $\vert \alpha_{m_z} \rangle_b$ as coherent states with complex amplitude $\alpha_{m_z} = -i m_z g\tau/\sqrt{N}$, and $c_{m_z}$ are defined as the expansion coefficients $\ket{ \left( N/2 \right)_x} \equiv \sum_{m_z} c_{m_z} \vert m_z \rangle$.

As $N \gg 1$ in our $2$D trapped ion crystal we are able to formulate a particularly simple understanding of the metrological utility of the spin-boson entanglement created in the state $\vert \psi_{\mathrm{in}} \rangle$. First, we notice that the ODF Hamiltonian can be approximately re-cast as a \emph{squeezing} Hamiltonian $\hat{H}_{\mathrm{ODF}} \equiv g(\hat{x}_+^2 - \hat{x}_-^2)/2$ where $\hat{x}_{\pm} = (\hat{a}+\hat{a}^{\dagger} \pm \sqrt{4/N}\hat{J}_z)/2$ and $\hat{p}_{\pm} = i(\hat{a}^{\dagger}-\hat{a} \pm i\sqrt{4/N}\hat{J}_y)/2$. Here, $\hat{x}_{\pm}$ and $\hat{p}_{\pm}$ are quadrature operators of a pair of effective oscillators resulting from a hybridization of the spin and boson degrees of freedom. Note that the normalization of the spin operators in the definition of $\hat{x}_{\pm}$ and $\hat{p}_{\pm}$ means that the $\sim\sqrt{N}$ projection noise of the original coherent spin state contributes $\sim\mathcal{O}(1)$ noise to the hybrid quadratures, similar to the vacuum noise of the mechanical oscillator. In this language, the action of $\hat{H}_{\mathrm{ODF}}$ is to generate squeezing along specific quadratures in the independent $(\hat{x}_+,\hat{p}_+)$ and $(\hat{x}_-,\hat{p}_-)$ phase-spaces. This is illustrated in panel c of Fig.~\ref{fig:protocol}. In the original basis of the spin and boson degrees of freedom, this squeezing manifests as strong correlations and entanglement between the spins and bosons.

The subsequently applied displacement of the COM oscillator, assumed to be orthogonal to the spin-dependent displacements, is described by the unitary transformation $\hat{U}_{\beta} = e^{\beta(\hat{a}-\hat{a}^{\dagger})}=e^{i \sqrt{2}\beta \hat{p}_b}$, where $\sqrt{2} i \hat{p}_b=\hat{a}-\hat{a}^{\dagger}$ ($\sqrt{2}\hat{x}_b=\hat{a}+\hat{a}^{\dagger}$) are the oscillator quadratures. This leads to the displaced state $\vert \psi^{\beta}_{\rm in}\rangle =\hat{U}_{\beta}\vert \psi_{\rm in}\rangle$, which
can be equivalently framed as a shift in the spin-boson quadrature $\hat{x}_+ \to \hat{x}_+ + \beta/\sqrt{2}$ ($\hat{x}_- \to \hat{x}_- - \beta/\sqrt{2}$). The squeezing generated by the initial ODF drive means that, in principle, a small displacement quickly leads to a state that is distinguishable to $\vert \psi_{\mathrm{in}} \rangle$. This displacement, parametrized by $\beta$, can be estimated with a sensitivity (see Fig.~\ref{fig:protocol}) limited by the quantum Cramer-Rao bound for displacements \cite{Braunstein1994}, $(\Delta\beta)^2 = 1/(4 + 4g^2\tau^2)$ \cite{SOM}. This bound can be recast in the more familiar form of the Heisenberg limit for displacements, $(\Delta\beta)^2_{\mathrm{HL}} \approx 1/n$ \cite{Toscano_2006}, where $n \sim g^2 \tau^2$ is the occupancy of the effective oscillator before the displacement. This is to be contrasted with the SQL for displacements $(\Delta\beta)_{\mathrm{SQL}}^2 = 1/4$, which is the sensitivity attainable with a coherent state.

Nevertheless, fully attaining the sensitivity promised by the state $\vert \psi_{\mathrm{in}}\rangle$ is not straightforward. In particular, the high degree of spin-boson entanglement in $\vert \psi_{\mathrm{in}}\rangle$ means that a simple measurement of either the spins or phonons independently is insufficient to precisely infer $\beta$ \cite{SOM}. Instead, a more sophisticated measurement that accounts for correlations between the degrees of freedom is required. We address this challenge and remove undesirable quantum back-action effects by the application of a time-reversal step that perfectly disentangles the spin and the oscillator (see Fig.~\ref{fig:protocol}) and maps the displacement to accessible spin observables. Specifically, the final state, obtained after applying the  many-body echo to $\vert \psi_{\mathrm{in}}^{\beta} \rangle$ [Eq.~(\ref{eqn:NcatState})], is equivalent to \cite{SOM}
\begin{equation}
  \ket{\psi_f} = e^{i \frac{\hbar g\tau}{\sqrt{N}} \left( \hat{a}+\hat{a}^{\dagger} \right) \hat{J}_{z}}  \vert \psi^\beta_\mathrm{in} \rangle 
  \equiv e^{\frac{2ig\tau\beta}{\sqrt{N}}\hat{J}_z}\hat{U}_{\beta}\ket{\psi(0)} .
\end{equation}
Writing the final state in the latter form demonstrates the power of the many-body echo: The displacement of the COM mode is encoded into a collective spin rotation of angle $\varphi = 2g\tau\beta/\sqrt{N}$, which can be read out by simple collective spin measurements \cite{Lewis-Swan2020,Hosten_2016,Davis_2016,Nolan_2017}. Accounting for quantum projection noise of the collective spin, which limits angular resolution of small rotations to $\Delta\varphi \geq 1/\sqrt{N}$, the displacement can thus be estimated with a sensitivity $(\Delta\beta)^2 = 1/(4g^2\tau^2)$ that approaches the Cramer-Rao bound for $g\tau \gg 1$. Moreover, we note that the form of the collective spin rotation is \emph{independent} of the initial state of the phonons and hence is insensitive to any thermal occupation of the COM mode. 

\noindent{\bf Displacement sensing:} In Fig.~\ref{fig:amp_sens} we present the actual performance of the implemented  displacement sensing protocol. The achievable sensitivity of the current experiment is primarily, but not fundamentally, limited by small shot-to-shot fluctuations of the COM mode frequency away from resonance due to impurity ions in the crystal and a thermal occupation of the in-plane modes \cite{Shankar2020}. These fluctuations limit our ability to perfectly reverse the entangling dynamics. Including COM frequency fluctuations with an rms spread $\sigma$, as well as single-particle decoherence of the spins at rate $\Gamma$ due to light scattering generated by the applied ODF beams and an initial thermal occupation $\bar{n} \approx 5$ of the COM mode, the  SNR of a single measurement of $\beta$ is reduced to \cite{SOM},
\begin{equation}
    \frac{\beta}{\Delta\beta} = \frac{2g\tau \beta e^{-\Gamma\tau}}{\sqrt{1 + e^{-2\Gamma \tau}\big[(2\bar{n}+1)g^2\sigma^2\tau^4 + \frac{4}{9}g^4\sigma^2\tau^6\big]}} . \label{eqn:SignalNoise}
\end{equation}
Physically, the prefactor $e^{-\Gamma\tau}$ comes from depolarization of the collective spin due to decoherence, while the $\tau^4$ and $\tau^6$ terms of the denominator describe residual spin-phonon entanglement and excess projection noise introduced by the imperfect time-reversal. In Fig.~\ref{fig:amp_sens}(a) we find good agreement between the experimentally determined SNR and the theoretical prediction Eq.~(\ref{eqn:SignalNoise}) with independently calibrated values of $g, \sigma,$ and $\Gamma$ \cite{SOM}, justifying our understanding of the various noise processes. In particular, we observe $\beta/\Delta\beta \to 0$ as $\beta \to 0$, indicating there are no additional systematic errors from effects beyond the applied displacement. 

To assess the performance of the displacement sensing protocol  we determine the variance $(\Delta\beta)^2$ of a single measurement as a function of the ODF interaction time $\tau$ for a small fixed displacement $\beta = 0.24$. Again, we find good agreement with the theoretical estimates of $(\Delta\beta)^2$ presented in Fig.~\ref{fig:amp_sens}(b). Optimizing over $\tau$ we demonstrate a sensitivity of $(\Delta\beta)^2 = 3.3\times 10^{-2}$ or $8.8 \pm 0.4$~dB below the SQL. We emphasize that no technical noise has been subtracted in obtaining this result. Our result can be recast in terms of the absolute physical displacement $Z_c$ of the COM mode using the relation $Z_c=2z_0\beta/\sqrt{N}$ where $z_0= \sqrt{\hbar/(2m\omega_z)}$ is the size of the ground-state motional wavefunction of a single ion and the enhancement $\propto \sqrt{N}$ arises due to the increased mass of the COM mode when more ions are present. 
Accounting for the $8$~ms duration (e.g., accounting for preparation and readout) of a single measurement, we equivalently determine the optimal sensitivity is $\Delta Z_c = 36 \pm 1.5~\mathrm{pm}/\sqrt{\mathrm{Hz}}$. Note that in the absence of our time-reversal protocol, excess projection noise introduced by the thermal occupation of the phonon mode, $\bar{n}\approx 5$, would have limited the achievable sensitivity to $(\delta\beta)^2_{\mathrm{th}} \geq (2\bar{n}+1)(\delta\beta)^2_{\mathrm{SQL}}$ \cite{SOM}, or $\sim 19$~dB worse than our result.

\noindent{\bf Electric field sensing:} Our ability to resonantly drive a small displacement of the COM mode over a long period of time, while maintaining a stable phase lock to a resonant ODF drive
that generates spin-phonon entanglement, places our experiment in an excellent position for quantum-enhanced  measurements of weak AC electric fields at the frequency of the COM mode.

\begin{figure*}
    \includegraphics[width=\textwidth]{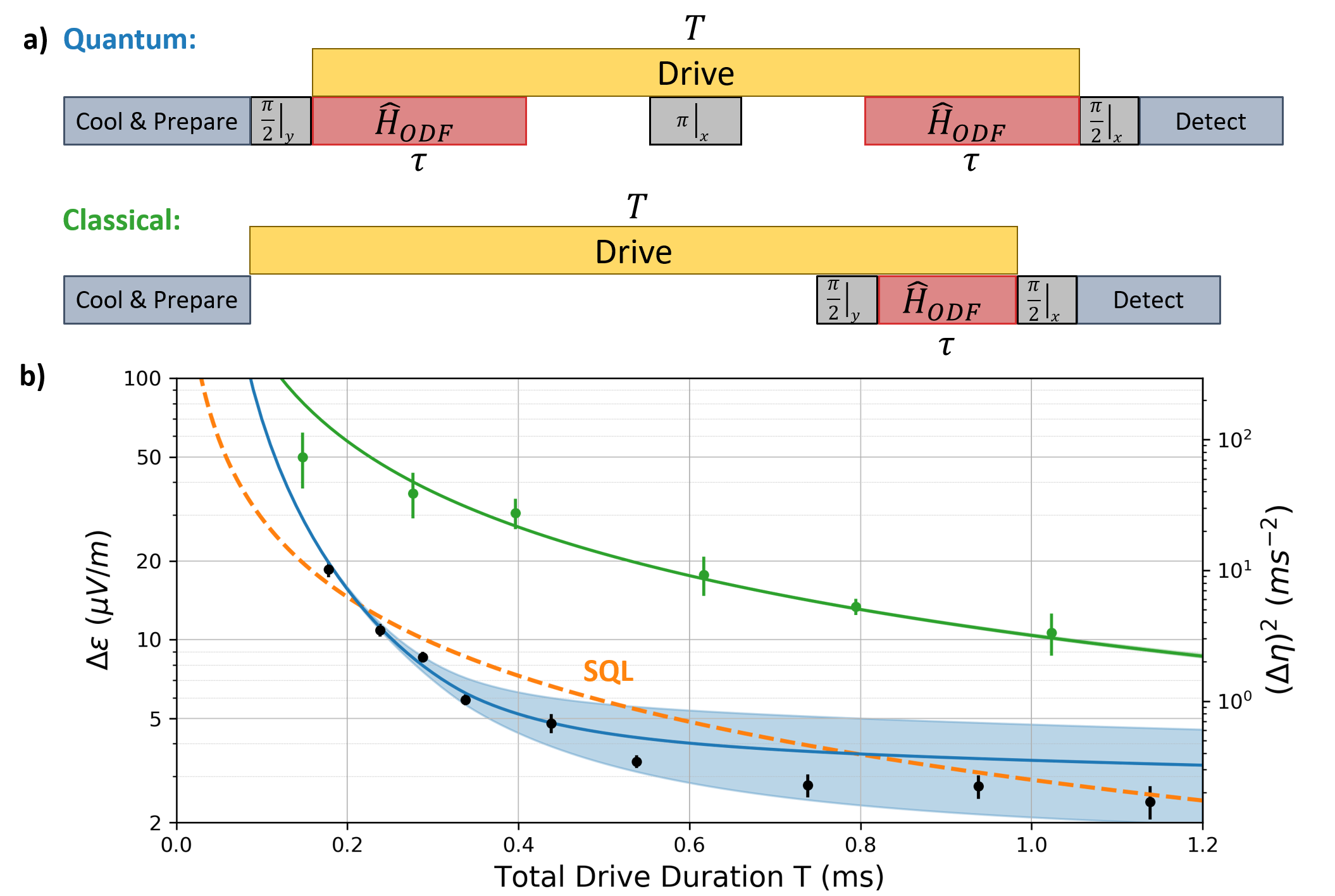}
\caption[]{Performance of the electric field sensor. a) Schematic timeline of the quantum and classical protocols used to sense weak electric fields. The quantum protocol involves sequentially applying a pair of ODF drives with equal duration $\tau$ to first entangle and then disentangle the spin and motional degrees of freedom for readout via the spins, while a weak spin-independent drive is applied concurrently for a duration $T \geq 2\tau$. The classical protocol identically applies a spin-independent drive for duration $T$, but instead the ODF is only turned on for $\tau \leq T$ at the end of the sequence to enable spin readout of the overall displacement. b) Electric field sensitivity $\Delta \epsilon$ (and $(\Delta\eta)^2$, right y-axis) as a function of the total spin-independent drive duration $T$. We plot experimental data from both classical and quantum protocols (green and black markers respectively, error bars indicate statistical and systematic uncertainty) with $\tau$ chosen to ideally optimise $(\Delta\eta)^2$ \cite{SOM}. Good agreement is found with theoretical models of both protocols (green and blue solid lines) using independently calibrated values of $g/(2\pi)=3.88$~kHz and COM mode frequency fluctuations of $\sigma/(2\pi) \approx 40$~Hz (shaded region indicates confidence region of $\sigma/(2\pi)\in[20,60]$~Hz, which is not visible on this scale for the classical protocol), as well as $\Gamma = 520\text{ s}^{-1}$. As reference, we contrast with the SQL $(\Delta\eta)^2 = (2T)^{-2}$ (orange dashed line). Experimental data indicates that the quantum protocol attains a sensitivity that is optimally $4$~dB below the SQL \cite{SOM}, or up to $14$~dB better than the sensitivity attainable with the classical protocol. Taking the average of the 3 points with longest drive duration $T$ results in an electric field sensitivity of $\Delta E$ = $240 \pm 10$~nV m$^{-1}/\sqrt{\mathrm{Hz}}$.}
  \label{fig:quan_class_seq}
\end{figure*}

The protocol to sense an electric field is illustrated in Fig.~\ref{fig:quan_class_seq}. A fixed AC voltage is applied to an endcap electrode to drive a continuous displacement of the COM mode for a total time $T$. Simultaneously, an ODF drive is initially turned on for a time $\tau \leq T/2$ to generate spin-boson entanglement. At some later time, a second ODF pulse with inverted coupling $g \to -g$ (realized again by a $\pi$-pulse about $\hat{x}$ of the spins after the first ODF drive) is applied for an identical duration $\tau$ and timed to finish with the conclusion of the COM displacement. Collectively, the dynamics of the system is described by the modified Hamiltonian \cite{SOM} 
\begin{equation}
\hat{H}_{\mathrm{sens}} = \hat{H}_{\mathrm{ODF}}(t) + i\eta(\hat{a}-\hat{a}^{\dagger}) ,
\end{equation} 
where $\eta$ is a parameter related to the electric field to be measured. Note the time-dependence of $\hat{H}_{\mathrm{ODF}}(t)$ indicates the modulation $g \to g(t)$.

For comparison purposes and to emphasize the quantum advantage, we compare the above protocol with a purely `classical' scheme where the continuous displacement of the COM mode is similarly applied for a total time $T$ but we only employ a single ODF drive for readout. Such a protocol ideally gives the sensitivity $(\Delta\eta)^2_{\mathrm{c}} = (1+g^2\tau^2)/[g^2\tau^2(2T-\tau)^2]$, which is always worse than the SQL defined as $(\Delta\eta)_{\mathrm{SQL}}^2 = 1/(4T^2)$. Note that the total displacement of the oscillator is ideally $\beta \equiv \eta T$. 

In contrast, our quantum-enhanced sensing scheme leads to $(\Delta\eta)^2 = 1/[4g^2\tau^2(T-\tau)^2]$, which yields sub-SQL scaling $(\Delta\eta)^2 \to 4/(g^2T^4)$ for $g\tau = gT/2 \gg 1$.

Figure \ref{fig:quan_class_seq} shows the experimentally obtained single-measurement electric field sensitivity of our protocol, in comparison to the SQL and the classical scheme. The ODF duration $\tau$ is chosen by numerically optimising the theoretically predicted sensitivity including technical noise due to COM fluctuations $\sigma$ and the thermal phonon occupation $\bar{n}$ \cite{SOM}. Under current experimental conditions, for intermediate driving time $T = 538~\mu$s the quantum protocol allows us to attain a sensitivity $\sim 4.0 \pm 0.5$~dB below the SQL. Moreover, we find that  in contrast to the quantum  protocol, which is ideally insensitive to the initial phonon state, the classical scheme is strongly limited by thermal noise. Quantitatively,  the quantum protocol  provides a $\sim 14$~dB improvement in sensitivity compared to the comparable classical scheme, demonstrating that entanglement can provide not only a fundamental but also practical advantage when all relevant noise sources are taken into account.

For long drive times $T \gtrsim 1$~ms we find the sensitivity is ultimately bounded by COM frequency noise, $(\Delta\eta)^2 \gtrsim \sigma^2(2\bar{n}+1)/4$ \cite{SOM}. Experimentally we determine a single-measurement sensitivity of $\Delta \epsilon = 2.3 \pm 0.1 ~\mu\mathrm{V}\mathrm{m}^{-1}$ at $T = 1.14$~ms, in good agreement with theory (see Fig. \ref{fig:quan_class_seq}). Accounting for the total duration of the experimental trial, $T_{\mathrm{shot}} = 8.73~$ms, we thus obtain a best electric field sensitivity of $\Delta E = \Delta \epsilon ~\sqrt{T_{\mathrm{shot}}}$ =$220 \pm 10$~nV m$^{-1}/\sqrt{\mathrm{Hz}}$. Averaging the three experimental results with the longest drive time $T$ gives a sensitivity of $\Delta E$ = $240 \pm 10$~nV m$^{-1}/\sqrt{\mathrm{Hz}}$ at $\sim 1.6$~MHz. This compares favorably with electric field sensors using Rydberg atoms~\cite{Facon2016}
that can reach sensitivities of $5.5~\mathrm{\mu V m}^{-1}/\sqrt{\mathrm{Hz}}$~\cite{Jing2020} at frequencies $\sim 10$~GHz, and also approaches the sensitivity of meter length classical antennas \cite{Meyer_2020} but in a micrometer scale device.
 
\noindent{\bf Conclusions and Outlook:} We have demonstrated a quantum-enhanced sensor of mechanical displacements and weak electric fields in a crystal composed of $\sim 150$ trapped ions. Currently, the primary limitations to the sensor's performance are COM frequency fluctuations and a thermal phonon population. Nevertheless, these issues are not fundamental in nature and could be resolved in next-generation experiments. In fact, feasible improvements to the stability of the COM mode frequency to the level of $\sigma/(2\pi)\approx 1$~Hz combined with a reduction of the phonon temperature via EIT cooling, which has already been demonstrated in the same platform \cite{Jordan_2019}, should allow us to reach an electric field sensitivity of $\sim 10$~nV~m$^{-1}/\sqrt{\mathrm{Hz}}$. Such performance would lead to a reduction of two orders of magnitude in the required averaging time to sense electric fields as weak as $1$~nV~m$^{-1}$. This could enable trapped ion crystals as detectors of hidden photons and axions with a frequency range (effective mass) compatible with current COM mode frequencies (between 10 kHz to  10 MHz), although shielding due to the interaction of the dark matter with the surrounding apparatus must be carefully considered \cite{Chaudhuri2015}. In the case of axions, a coupling to photons in the presence of a strong magnetic field, which is already inherently present in our experimental platform, is predicted to generate a classical field that will excite the COM mode \cite{Turner1990}. Further, although the achieved electric field sensitivity already takes advantage of the mesoscopic system size of our $2$D crystal, increasing the ion number to $N \gtrsim 10^5$ by the use of $3$D crystals \cite{Itano1998,Mortensen2006} should allow us to enhance our sensing capability by at least two orders of magnitude.

\begin{acknowledgments}
\noindent{\textit{Acknowledgements:}} We acknowledge helpful discussions with James Thompson and Athreya Shankar, and thank John Teufel and Kevin Cox for careful review of our manuscript.  
This work is supported by U.S. Department of Energy (DOE), Office of Science, National Quantum Information Science Research Centers, Quantum Systems Accelerator (QSA), a DOE Office of Science HEP QuantISED award, AFOSR grants FA9550-18-1-0319 and FA9550-20-1-0019, by the DARPA and ARO grant W911NF-16-1-0576, the ARO single investigator award W911NF-19-1-0210, the NSF PHY1820885, NSF JILA-PFC PHY-1734006 and NSF QLCI-2016244 grants, and by NIST. K.A.G., M. A., E.J., and J.J.B collected and analysed the experimental data. R.J.L.-S., D.B., and A.M.R. developed the theoretical model. All authors discussed the results and contributed to the preparation of the manuscript. The authors declare no competing interests. All data is available in the manuscript or supplementary materials.

\end{acknowledgments}

\bibliography{library}

\begin{thebibliography}{41}%
\makeatletter
\providecommand \@ifxundefined [1]{%
 \@ifx{#1\undefined}
}%
\providecommand \@ifnum [1]{%
 \ifnum #1\expandafter \@firstoftwo
 \else \expandafter \@secondoftwo
 \fi
}%
\providecommand \@ifx [1]{%
 \ifx #1\expandafter \@firstoftwo
 \else \expandafter \@secondoftwo
 \fi
}%
\providecommand \natexlab [1]{#1}%
\providecommand \enquote  [1]{``#1''}%
\providecommand \bibnamefont  [1]{#1}%
\providecommand \bibfnamefont [1]{#1}%
\providecommand \citenamefont [1]{#1}%
\providecommand \href@noop [0]{\@secondoftwo}%
\providecommand \href [0]{\begingroup \@sanitize@url \@href}%
\providecommand \@href[1]{\@@startlink{#1}\@@href}%
\providecommand \@@href[1]{\endgroup#1\@@endlink}%
\providecommand \@sanitize@url [0]{\catcode `\\12\catcode `\$12\catcode
  `\&12\catcode `\#12\catcode `\^12\catcode `\_12\catcode `\%12\relax}%
\providecommand \@@startlink[1]{}%
\providecommand \@@endlink[0]{}%
\providecommand \url  [0]{\begingroup\@sanitize@url \@url }%
\providecommand \@url [1]{\endgroup\@href {#1}{\urlprefix }}%
\providecommand \urlprefix  [0]{URL }%
\providecommand \Eprint [0]{\href }%
\providecommand \doibase [0]{https://doi.org/}%
\providecommand \selectlanguage [0]{\@gobble}%
\providecommand \bibinfo  [0]{\@secondoftwo}%
\providecommand \bibfield  [0]{\@secondoftwo}%
\providecommand \translation [1]{[#1]}%
\providecommand \BibitemOpen [0]{}%
\providecommand \bibitemStop [0]{}%
\providecommand \bibitemNoStop [0]{.\EOS\space}%
\providecommand \EOS [0]{\spacefactor3000\relax}%
\providecommand \BibitemShut  [1]{\csname bibitem#1\endcsname}%
\let\auto@bib@innerbib\@empty
\bibitem [{\citenamefont {Dowling}\ and\ \citenamefont
  {Milburn}(2003)}]{Dowling2003}%
  \BibitemOpen
  \bibfield  {author} {\bibinfo {author} {\bibfnamefont {J.~P.}\ \bibnamefont
  {Dowling}}\ and\ \bibinfo {author} {\bibfnamefont {G.~J.}\ \bibnamefont
  {Milburn}},\ }\bibfield  {title} {\bibinfo {title} {Quantum technology: the
  second quantum revolution},\ }\href {https://doi.org/10.1098/rsta.2003.1227}
  {\bibfield  {journal} {\bibinfo  {journal} {Phil. Trans. R. Soc. A.}\
  }\textbf {\bibinfo {volume} {361}},\ \bibinfo {pages} {1655–1674} (\bibinfo
  {year} {2003})}\BibitemShut {NoStop}%
\bibitem [{\citenamefont {Degen}\ \emph {et~al.}(2017)\citenamefont {Degen},
  \citenamefont {Reinhard},\ and\ \citenamefont
  {Cappellaro}}]{Cappellaro_2017}%
  \BibitemOpen
  \bibfield  {author} {\bibinfo {author} {\bibfnamefont {C.~L.}\ \bibnamefont
  {Degen}}, \bibinfo {author} {\bibfnamefont {F.}~\bibnamefont {Reinhard}},\
  and\ \bibinfo {author} {\bibfnamefont {P.}~\bibnamefont {Cappellaro}},\
  }\bibfield  {title} {\bibinfo {title} {Quantum sensing},\ }\href
  {https://doi.org/10.1103/RevModPhys.89.035002} {\bibfield  {journal}
  {\bibinfo  {journal} {Rev. Mod. Phys.}\ }\textbf {\bibinfo {volume} {89}},\
  \bibinfo {pages} {035002} (\bibinfo {year} {2017})}\BibitemShut {NoStop}%
\bibitem [{\citenamefont {Tse}\ \emph {et~al.}(2019)\citenamefont {Tse} \emph
  {et~al.}}]{ligo2019}%
  \BibitemOpen
  \bibfield  {author} {\bibinfo {author} {\bibfnamefont {M.}~\bibnamefont
  {Tse}} \emph {et~al.},\ }\bibfield  {title} {\bibinfo {title}
  {Quantum-enhanced advanced ligo detectors in the era of gravitational-wave
  astronomy},\ }\href {https://doi.org/10.1103/PhysRevLett.123.231107}
  {\bibfield  {journal} {\bibinfo  {journal} {Phys. Rev. Lett.}\ }\textbf
  {\bibinfo {volume} {123}},\ \bibinfo {pages} {231107} (\bibinfo {year}
  {2019})}\BibitemShut {NoStop}%
\bibitem [{\citenamefont {Acernese}(2019)}]{VIRGO2019}%
  \BibitemOpen
  \bibfield  {author} {\bibinfo {author} {\bibfnamefont {F.~e.~a.}\
  \bibnamefont {Acernese}} (\bibinfo {collaboration} {Virgo Collaboration}),\
  }\bibfield  {title} {\bibinfo {title} {Increasing the astrophysical reach of
  the advanced virgo detector via the application of squeezed vacuum states of
  light},\ }\href {https://doi.org/10.1103/PhysRevLett.123.231108} {\bibfield
  {journal} {\bibinfo  {journal} {Phys. Rev. Lett.}\ }\textbf {\bibinfo
  {volume} {123}},\ \bibinfo {pages} {231108} (\bibinfo {year}
  {2019})}\BibitemShut {NoStop}%
\bibitem [{\citenamefont {Du}\ \emph {et~al.}(2018)\citenamefont {Du},
  \citenamefont {Force}, \citenamefont {Khatiwada}, \citenamefont {Lentz},
  \citenamefont {Ottens}, \citenamefont {Rosenberg}, \citenamefont {Rybka},
  \citenamefont {Carosi}, \citenamefont {Woollett}, \citenamefont {Bowring},
  \citenamefont {Chou}, \citenamefont {Sonnenschein}, \citenamefont {Wester},
  \citenamefont {Boutan}, \citenamefont {Oblath}, \citenamefont {Bradley},
  \citenamefont {Daw}, \citenamefont {Dixit}, \citenamefont {Clarke},
  \citenamefont {O'Kelley}, \citenamefont {Crisosto}, \citenamefont {Gleason},
  \citenamefont {Jois}, \citenamefont {Sikivie}, \citenamefont {Stern},
  \citenamefont {Sullivan}, \citenamefont {Tanner},\ and\ \citenamefont
  {Hilton}}]{ADMX}%
  \BibitemOpen
  \bibfield  {author} {\bibinfo {author} {\bibfnamefont {N.}~\bibnamefont
  {Du}}, \bibinfo {author} {\bibfnamefont {N.}~\bibnamefont {Force}}, \bibinfo
  {author} {\bibfnamefont {R.}~\bibnamefont {Khatiwada}}, \bibinfo {author}
  {\bibfnamefont {E.}~\bibnamefont {Lentz}}, \bibinfo {author} {\bibfnamefont
  {R.}~\bibnamefont {Ottens}}, \bibinfo {author} {\bibfnamefont {L.~J.}\
  \bibnamefont {Rosenberg}}, \bibinfo {author} {\bibfnamefont {G.}~\bibnamefont
  {Rybka}}, \bibinfo {author} {\bibfnamefont {G.}~\bibnamefont {Carosi}},
  \bibinfo {author} {\bibfnamefont {N.}~\bibnamefont {Woollett}}, \bibinfo
  {author} {\bibfnamefont {D.}~\bibnamefont {Bowring}}, \bibinfo {author}
  {\bibfnamefont {A.~S.}\ \bibnamefont {Chou}}, \bibinfo {author}
  {\bibfnamefont {A.}~\bibnamefont {Sonnenschein}}, \bibinfo {author}
  {\bibfnamefont {W.}~\bibnamefont {Wester}}, \bibinfo {author} {\bibfnamefont
  {C.}~\bibnamefont {Boutan}}, \bibinfo {author} {\bibfnamefont {N.~S.}\
  \bibnamefont {Oblath}}, \bibinfo {author} {\bibfnamefont {R.}~\bibnamefont
  {Bradley}}, \bibinfo {author} {\bibfnamefont {E.~J.}\ \bibnamefont {Daw}},
  \bibinfo {author} {\bibfnamefont {A.~V.}\ \bibnamefont {Dixit}}, \bibinfo
  {author} {\bibfnamefont {J.}~\bibnamefont {Clarke}}, \bibinfo {author}
  {\bibfnamefont {S.~R.}\ \bibnamefont {O'Kelley}}, \bibinfo {author}
  {\bibfnamefont {N.}~\bibnamefont {Crisosto}}, \bibinfo {author}
  {\bibfnamefont {J.~R.}\ \bibnamefont {Gleason}}, \bibinfo {author}
  {\bibfnamefont {S.}~\bibnamefont {Jois}}, \bibinfo {author} {\bibfnamefont
  {P.}~\bibnamefont {Sikivie}}, \bibinfo {author} {\bibfnamefont
  {I.}~\bibnamefont {Stern}}, \bibinfo {author} {\bibfnamefont {N.~S.}\
  \bibnamefont {Sullivan}}, \bibinfo {author} {\bibfnamefont {D.~B.}\
  \bibnamefont {Tanner}},\ and\ \bibinfo {author} {\bibfnamefont {G.~C.}\
  \bibnamefont {Hilton}} (\bibinfo {collaboration} {ADMX Collaboration}),\
  }\bibfield  {title} {\bibinfo {title} {Search for invisible axion dark matter
  with the axion dark matter experiment},\ }\href
  {https://doi.org/10.1103/PhysRevLett.120.151301} {\bibfield  {journal}
  {\bibinfo  {journal} {Phys. Rev. Lett.}\ }\textbf {\bibinfo {volume} {120}},\
  \bibinfo {pages} {151301} (\bibinfo {year} {2018})}\BibitemShut {NoStop}%
\bibitem [{\citenamefont {Zhong}\ \emph {et~al.}(2018)\citenamefont {Zhong},
  \citenamefont {Al~Kenany}, \citenamefont {Backes}, \citenamefont {Brubaker},
  \citenamefont {Cahn}, \citenamefont {Carosi}, \citenamefont {Gurevich},
  \citenamefont {Kindel}, \citenamefont {Lamoreaux}, \citenamefont {Lehnert},
  \citenamefont {Lewis}, \citenamefont {Malnou}, \citenamefont {Maruyama},
  \citenamefont {Palken}, \citenamefont {Rapidis}, \citenamefont {Root},
  \citenamefont {Simanovskaia}, \citenamefont {Shokair}, \citenamefont
  {Speller}, \citenamefont {Urdinaran},\ and\ \citenamefont {van
  Bibber}}]{HAYSTAC}%
  \BibitemOpen
  \bibfield  {author} {\bibinfo {author} {\bibfnamefont {L.}~\bibnamefont
  {Zhong}}, \bibinfo {author} {\bibfnamefont {S.}~\bibnamefont {Al~Kenany}},
  \bibinfo {author} {\bibfnamefont {K.~M.}\ \bibnamefont {Backes}}, \bibinfo
  {author} {\bibfnamefont {B.~M.}\ \bibnamefont {Brubaker}}, \bibinfo {author}
  {\bibfnamefont {S.~B.}\ \bibnamefont {Cahn}}, \bibinfo {author}
  {\bibfnamefont {G.}~\bibnamefont {Carosi}}, \bibinfo {author} {\bibfnamefont
  {Y.~V.}\ \bibnamefont {Gurevich}}, \bibinfo {author} {\bibfnamefont {W.~F.}\
  \bibnamefont {Kindel}}, \bibinfo {author} {\bibfnamefont {S.~K.}\
  \bibnamefont {Lamoreaux}}, \bibinfo {author} {\bibfnamefont {K.~W.}\
  \bibnamefont {Lehnert}}, \bibinfo {author} {\bibfnamefont {S.~M.}\
  \bibnamefont {Lewis}}, \bibinfo {author} {\bibfnamefont {M.}~\bibnamefont
  {Malnou}}, \bibinfo {author} {\bibfnamefont {R.~H.}\ \bibnamefont
  {Maruyama}}, \bibinfo {author} {\bibfnamefont {D.~A.}\ \bibnamefont
  {Palken}}, \bibinfo {author} {\bibfnamefont {N.~M.}\ \bibnamefont {Rapidis}},
  \bibinfo {author} {\bibfnamefont {J.~R.}\ \bibnamefont {Root}}, \bibinfo
  {author} {\bibfnamefont {M.}~\bibnamefont {Simanovskaia}}, \bibinfo {author}
  {\bibfnamefont {T.~M.}\ \bibnamefont {Shokair}}, \bibinfo {author}
  {\bibfnamefont {D.~H.}\ \bibnamefont {Speller}}, \bibinfo {author}
  {\bibfnamefont {I.}~\bibnamefont {Urdinaran}},\ and\ \bibinfo {author}
  {\bibfnamefont {K.~A.}\ \bibnamefont {van Bibber}},\ }\bibfield  {title}
  {\bibinfo {title} {Results from phase 1 of the haystac microwave cavity axion
  experiment},\ }\href {https://doi.org/10.1103/PhysRevD.97.092001} {\bibfield
  {journal} {\bibinfo  {journal} {Phys. Rev. D}\ }\textbf {\bibinfo {volume}
  {97}},\ \bibinfo {pages} {092001} (\bibinfo {year} {2018})}\BibitemShut
  {NoStop}%
\bibitem [{\citenamefont {Malnou}\ \emph {et~al.}(2019)\citenamefont {Malnou},
  \citenamefont {Palken}, \citenamefont {Brubaker}, \citenamefont {Vale},
  \citenamefont {Hilton},\ and\ \citenamefont {Lehnert}}]{Malnou2019}%
  \BibitemOpen
  \bibfield  {author} {\bibinfo {author} {\bibfnamefont {M.}~\bibnamefont
  {Malnou}}, \bibinfo {author} {\bibfnamefont {D.~A.}\ \bibnamefont {Palken}},
  \bibinfo {author} {\bibfnamefont {B.~M.}\ \bibnamefont {Brubaker}}, \bibinfo
  {author} {\bibfnamefont {L.~R.}\ \bibnamefont {Vale}}, \bibinfo {author}
  {\bibfnamefont {G.~C.}\ \bibnamefont {Hilton}},\ and\ \bibinfo {author}
  {\bibfnamefont {K.~W.}\ \bibnamefont {Lehnert}},\ }\bibfield  {title}
  {\bibinfo {title} {Squeezed vacuum used to accelerate the search for a weak
  classical signal},\ }\href {https://doi.org/10.1103/PhysRevX.9.021023}
  {\bibfield  {journal} {\bibinfo  {journal} {Phys. Rev. X}\ }\textbf {\bibinfo
  {volume} {9}},\ \bibinfo {pages} {021023} (\bibinfo {year}
  {2019})}\BibitemShut {NoStop}%
\bibitem [{\citenamefont {Backes}\ \emph {et~al.}(2021)\citenamefont {Backes},
  \citenamefont {Palken}, \citenamefont {Kenany}, \citenamefont {Brubaker},
  \citenamefont {Cahn}, \citenamefont {Droster}, \citenamefont {Hilton},
  \citenamefont {Ghosh}, \citenamefont {Jackson}, \citenamefont {Lamoreaux},
  \citenamefont {Leder}, \citenamefont {Lehnert}, \citenamefont {Lewis},
  \citenamefont {Malnou}, \citenamefont {Maruyama}, \citenamefont {Rapidis},
  \citenamefont {Simanovskaia}, \citenamefont {Singh}, \citenamefont {Speller},
  \citenamefont {Urdinaran}, \citenamefont {Vale}, \citenamefont {van
  Assendelft}, \citenamefont {van Bibber},\ and\ \citenamefont
  {Wang}}]{Backes2021}%
  \BibitemOpen
  \bibfield  {author} {\bibinfo {author} {\bibfnamefont {K.~M.}\ \bibnamefont
  {Backes}}, \bibinfo {author} {\bibfnamefont {D.~A.}\ \bibnamefont {Palken}},
  \bibinfo {author} {\bibfnamefont {S.~A.}\ \bibnamefont {Kenany}}, \bibinfo
  {author} {\bibfnamefont {B.~M.}\ \bibnamefont {Brubaker}}, \bibinfo {author}
  {\bibfnamefont {S.~B.}\ \bibnamefont {Cahn}}, \bibinfo {author}
  {\bibfnamefont {A.}~\bibnamefont {Droster}}, \bibinfo {author} {\bibfnamefont
  {G.~C.}\ \bibnamefont {Hilton}}, \bibinfo {author} {\bibfnamefont
  {S.}~\bibnamefont {Ghosh}}, \bibinfo {author} {\bibfnamefont
  {H.}~\bibnamefont {Jackson}}, \bibinfo {author} {\bibfnamefont {S.~K.}\
  \bibnamefont {Lamoreaux}}, \bibinfo {author} {\bibfnamefont {A.~F.}\
  \bibnamefont {Leder}}, \bibinfo {author} {\bibfnamefont {K.~W.}\ \bibnamefont
  {Lehnert}}, \bibinfo {author} {\bibfnamefont {S.~M.}\ \bibnamefont {Lewis}},
  \bibinfo {author} {\bibfnamefont {M.}~\bibnamefont {Malnou}}, \bibinfo
  {author} {\bibfnamefont {R.~H.}\ \bibnamefont {Maruyama}}, \bibinfo {author}
  {\bibfnamefont {N.~M.}\ \bibnamefont {Rapidis}}, \bibinfo {author}
  {\bibfnamefont {M.}~\bibnamefont {Simanovskaia}}, \bibinfo {author}
  {\bibfnamefont {S.}~\bibnamefont {Singh}}, \bibinfo {author} {\bibfnamefont
  {D.~H.}\ \bibnamefont {Speller}}, \bibinfo {author} {\bibfnamefont
  {I.}~\bibnamefont {Urdinaran}}, \bibinfo {author} {\bibfnamefont {L.~R.}\
  \bibnamefont {Vale}}, \bibinfo {author} {\bibfnamefont {E.~C.}\ \bibnamefont
  {van Assendelft}}, \bibinfo {author} {\bibfnamefont {K.}~\bibnamefont {van
  Bibber}},\ and\ \bibinfo {author} {\bibfnamefont {H.}~\bibnamefont {Wang}},\
  }\bibfield  {title} {\bibinfo {title} {A quantum enhanced search for dark
  matter axions},\ }\href {https://doi.org/10.1038/s41586-021-03226-7}
  {\bibfield  {journal} {\bibinfo  {journal} {Nature}\ }\textbf {\bibinfo
  {volume} {590}},\ \bibinfo {pages} {238} (\bibinfo {year}
  {2021})}\BibitemShut {NoStop}%
\bibitem [{\citenamefont {McCormick}\ \emph {et~al.}(2019)\citenamefont
  {McCormick}, \citenamefont {Keller}, \citenamefont {Burd}, \citenamefont
  {Wineland}, \citenamefont {Wilson},\ and\ \citenamefont
  {Leibfried}}]{Didi_2018}%
  \BibitemOpen
  \bibfield  {author} {\bibinfo {author} {\bibfnamefont {K.~C.}\ \bibnamefont
  {McCormick}}, \bibinfo {author} {\bibfnamefont {J.}~\bibnamefont {Keller}},
  \bibinfo {author} {\bibfnamefont {S.~C.}\ \bibnamefont {Burd}}, \bibinfo
  {author} {\bibfnamefont {D.~J.}\ \bibnamefont {Wineland}}, \bibinfo {author}
  {\bibfnamefont {A.~C.}\ \bibnamefont {Wilson}},\ and\ \bibinfo {author}
  {\bibfnamefont {D.}~\bibnamefont {Leibfried}},\ }\bibfield  {title} {\bibinfo
  {title} {Quantum-enhanced sensing of a single-ion mechanical oscillator},\
  }\href {https://doi.org/10.1038/s41586-019-1421-y} {\bibfield  {journal}
  {\bibinfo  {journal} {Nature}\ }\textbf {\bibinfo {volume} {572}},\ \bibinfo
  {pages} {86} (\bibinfo {year} {2019})}\BibitemShut {NoStop}%
\bibitem [{\citenamefont {Wolf}\ \emph {et~al.}(2019)\citenamefont {Wolf},
  \citenamefont {Shi}, \citenamefont {Heip}, \citenamefont {Gessner},
  \citenamefont {Pezz{\`e}}, \citenamefont {Smerzi}, \citenamefont {Schulte},
  \citenamefont {Hammerer},\ and\ \citenamefont {Schmidt}}]{Wolf2019}%
  \BibitemOpen
  \bibfield  {author} {\bibinfo {author} {\bibfnamefont {F.}~\bibnamefont
  {Wolf}}, \bibinfo {author} {\bibfnamefont {C.}~\bibnamefont {Shi}}, \bibinfo
  {author} {\bibfnamefont {J.~C.}\ \bibnamefont {Heip}}, \bibinfo {author}
  {\bibfnamefont {M.}~\bibnamefont {Gessner}}, \bibinfo {author} {\bibfnamefont
  {L.}~\bibnamefont {Pezz{\`e}}}, \bibinfo {author} {\bibfnamefont
  {A.}~\bibnamefont {Smerzi}}, \bibinfo {author} {\bibfnamefont
  {M.}~\bibnamefont {Schulte}}, \bibinfo {author} {\bibfnamefont
  {K.}~\bibnamefont {Hammerer}},\ and\ \bibinfo {author} {\bibfnamefont
  {P.~O.}\ \bibnamefont {Schmidt}},\ }\bibfield  {title} {\bibinfo {title}
  {Motional fock states for quantum-enhanced amplitude and phase measurements
  with trapped ions},\ }\href {https://doi.org/10.1038/s41467-019-10576-4}
  {\bibfield  {journal} {\bibinfo  {journal} {Nature Communications}\ }\textbf
  {\bibinfo {volume} {10}},\ \bibinfo {pages} {2929} (\bibinfo {year}
  {2019})}\BibitemShut {NoStop}%
\bibitem [{\citenamefont {Schreppler}\ \emph {et~al.}(2014)\citenamefont
  {Schreppler}, \citenamefont {Spethmann}, \citenamefont {Brahms},
  \citenamefont {Botter}, \citenamefont {Barrios},\ and\ \citenamefont
  {Stamper-Kurn}}]{Schreppler_2014}%
  \BibitemOpen
  \bibfield  {author} {\bibinfo {author} {\bibfnamefont {S.}~\bibnamefont
  {Schreppler}}, \bibinfo {author} {\bibfnamefont {N.}~\bibnamefont
  {Spethmann}}, \bibinfo {author} {\bibfnamefont {N.}~\bibnamefont {Brahms}},
  \bibinfo {author} {\bibfnamefont {T.}~\bibnamefont {Botter}}, \bibinfo
  {author} {\bibfnamefont {M.}~\bibnamefont {Barrios}},\ and\ \bibinfo {author}
  {\bibfnamefont {D.~M.}\ \bibnamefont {Stamper-Kurn}},\ }\bibfield  {title}
  {\bibinfo {title} {Optically measuring force near the standard quantum
  limit},\ }\href {https://doi.org/10.1126/science.1249850} {\bibfield
  {journal} {\bibinfo  {journal} {Science}\ }\textbf {\bibinfo {volume}
  {344}},\ \bibinfo {pages} {1486} (\bibinfo {year} {2014})}\BibitemShut
  {NoStop}%
\bibitem [{\citenamefont {Kolkowitz}\ \emph {et~al.}(2012)\citenamefont
  {Kolkowitz}, \citenamefont {Bleszynski~Jayich}, \citenamefont
  {Unterreithmeier}, \citenamefont {Bennett}, \citenamefont {Rabl},
  \citenamefont {Harris},\ and\ \citenamefont {Lukin}}]{Kolkowitz_2012}%
  \BibitemOpen
  \bibfield  {author} {\bibinfo {author} {\bibfnamefont {S.}~\bibnamefont
  {Kolkowitz}}, \bibinfo {author} {\bibfnamefont {A.~C.}\ \bibnamefont
  {Bleszynski~Jayich}}, \bibinfo {author} {\bibfnamefont {Q.~P.}\ \bibnamefont
  {Unterreithmeier}}, \bibinfo {author} {\bibfnamefont {S.~D.}\ \bibnamefont
  {Bennett}}, \bibinfo {author} {\bibfnamefont {P.}~\bibnamefont {Rabl}},
  \bibinfo {author} {\bibfnamefont {J.~G.~E.}\ \bibnamefont {Harris}},\ and\
  \bibinfo {author} {\bibfnamefont {M.~D.}\ \bibnamefont {Lukin}},\ }\bibfield
  {title} {\bibinfo {title} {Coherent sensing of a mechanical resonator with a
  single-spin qubit},\ }\href {https://doi.org/10.1126/science.1216821}
  {\bibfield  {journal} {\bibinfo  {journal} {Science}\ }\textbf {\bibinfo
  {volume} {335}},\ \bibinfo {pages} {1603} (\bibinfo {year}
  {2012})}\BibitemShut {NoStop}%
\bibitem [{\citenamefont {Delaney}\ \emph {et~al.}(2019)\citenamefont
  {Delaney}, \citenamefont {Reed}, \citenamefont {Andrews},\ and\ \citenamefont
  {Lehnert}}]{Delaney2019}%
  \BibitemOpen
  \bibfield  {author} {\bibinfo {author} {\bibfnamefont {R.~D.}\ \bibnamefont
  {Delaney}}, \bibinfo {author} {\bibfnamefont {A.~P.}\ \bibnamefont {Reed}},
  \bibinfo {author} {\bibfnamefont {R.~W.}\ \bibnamefont {Andrews}},\ and\
  \bibinfo {author} {\bibfnamefont {K.~W.}\ \bibnamefont {Lehnert}},\
  }\bibfield  {title} {\bibinfo {title} {Measurement of motion beyond the
  quantum limit by transient amplification},\ }\href
  {https://doi.org/10.1103/PhysRevLett.123.183603} {\bibfield  {journal}
  {\bibinfo  {journal} {Phys. Rev. Lett.}\ }\textbf {\bibinfo {volume} {123}},\
  \bibinfo {pages} {183603} (\bibinfo {year} {2019})}\BibitemShut {NoStop}%
\bibitem [{\citenamefont {Gilmore}\ \emph {et~al.}(2017)\citenamefont
  {Gilmore}, \citenamefont {Bohnet}, \citenamefont {Sawyer}, \citenamefont
  {Britton},\ and\ \citenamefont {Bollinger}}]{Gilmore2017}%
  \BibitemOpen
  \bibfield  {author} {\bibinfo {author} {\bibfnamefont {K.~A.}\ \bibnamefont
  {Gilmore}}, \bibinfo {author} {\bibfnamefont {J.~G.}\ \bibnamefont {Bohnet}},
  \bibinfo {author} {\bibfnamefont {B.~C.}\ \bibnamefont {Sawyer}}, \bibinfo
  {author} {\bibfnamefont {J.~W.}\ \bibnamefont {Britton}},\ and\ \bibinfo
  {author} {\bibfnamefont {J.~J.}\ \bibnamefont {Bollinger}},\ }\bibfield
  {title} {\bibinfo {title} {Amplitude sensing below the zero-point
  fluctuations with a two-dimensional trapped-ion mechanical oscillator},\
  }\href {https://doi.org/10.1103/PhysRevLett.118.263602} {\bibfield  {journal}
  {\bibinfo  {journal} {Phys. Rev. Lett.}\ }\textbf {\bibinfo {volume} {118}},\
  \bibinfo {pages} {263602} (\bibinfo {year} {2017})}\BibitemShut {NoStop}%
\bibitem [{\citenamefont {Wang}\ \emph {et~al.}(2019)\citenamefont {Wang},
  \citenamefont {Wu}, \citenamefont {Ma}, \citenamefont {Cai}, \citenamefont
  {Hu}, \citenamefont {Mu}, \citenamefont {Xu}, \citenamefont {Chen},
  \citenamefont {Wang}, \citenamefont {Song}, \citenamefont {Yuan},
  \citenamefont {Zou}, \citenamefont {Duan},\ and\ \citenamefont
  {Sun}}]{Wang2019}%
  \BibitemOpen
  \bibfield  {author} {\bibinfo {author} {\bibfnamefont {W.}~\bibnamefont
  {Wang}}, \bibinfo {author} {\bibfnamefont {Y.}~\bibnamefont {Wu}}, \bibinfo
  {author} {\bibfnamefont {Y.}~\bibnamefont {Ma}}, \bibinfo {author}
  {\bibfnamefont {W.}~\bibnamefont {Cai}}, \bibinfo {author} {\bibfnamefont
  {L.}~\bibnamefont {Hu}}, \bibinfo {author} {\bibfnamefont {X.}~\bibnamefont
  {Mu}}, \bibinfo {author} {\bibfnamefont {Y.}~\bibnamefont {Xu}}, \bibinfo
  {author} {\bibfnamefont {Z.-J.}\ \bibnamefont {Chen}}, \bibinfo {author}
  {\bibfnamefont {H.}~\bibnamefont {Wang}}, \bibinfo {author} {\bibfnamefont
  {Y.~P.}\ \bibnamefont {Song}}, \bibinfo {author} {\bibfnamefont
  {H.}~\bibnamefont {Yuan}}, \bibinfo {author} {\bibfnamefont {C.-L.}\
  \bibnamefont {Zou}}, \bibinfo {author} {\bibfnamefont {L.-M.}\ \bibnamefont
  {Duan}},\ and\ \bibinfo {author} {\bibfnamefont {L.}~\bibnamefont {Sun}},\
  }\bibfield  {title} {\bibinfo {title} {Heisenberg-limited single-mode quantum
  metrology in a superconducting circuit},\ }\href
  {https://doi.org/10.1038/s41467-019-12290-7} {\bibfield  {journal} {\bibinfo
  {journal} {Nature Communications}\ }\textbf {\bibinfo {volume} {10}},\
  \bibinfo {pages} {4382} (\bibinfo {year} {2019})}\BibitemShut {NoStop}%
\bibitem [{\citenamefont {Affolter}\ \emph {et~al.}(2020)\citenamefont
  {Affolter}, \citenamefont {Gilmore}, \citenamefont {Jordan},\ and\
  \citenamefont {Bollinger}}]{Affolter2020}%
  \BibitemOpen
  \bibfield  {author} {\bibinfo {author} {\bibfnamefont {M.}~\bibnamefont
  {Affolter}}, \bibinfo {author} {\bibfnamefont {K.~A.}\ \bibnamefont
  {Gilmore}}, \bibinfo {author} {\bibfnamefont {J.~E.}\ \bibnamefont
  {Jordan}},\ and\ \bibinfo {author} {\bibfnamefont {J.~J.}\ \bibnamefont
  {Bollinger}},\ }\bibfield  {title} {\bibinfo {title} {Phase-coherent sensing
  of the center-of-mass motion of trapped-ion crystals},\ }\href
  {https://doi.org/10.1103/PhysRevA.102.052609} {\bibfield  {journal} {\bibinfo
   {journal} {Phys. Rev. A}\ }\textbf {\bibinfo {volume} {102}},\ \bibinfo
  {pages} {052609} (\bibinfo {year} {2020})}\BibitemShut {NoStop}%
\bibitem [{\citenamefont {Thomas}\ \emph {et~al.}(2021)\citenamefont {Thomas},
  \citenamefont {Parniak}, \citenamefont {{\O}stfeldt}, \citenamefont
  {M{\o}ller}, \citenamefont {B{\ae}rentsen}, \citenamefont {Tsaturyan},
  \citenamefont {Schliesser}, \citenamefont {Appel}, \citenamefont {Zeuthen},\
  and\ \citenamefont {Polzik}}]{Polzik2021}%
  \BibitemOpen
  \bibfield  {author} {\bibinfo {author} {\bibfnamefont {R.~A.}\ \bibnamefont
  {Thomas}}, \bibinfo {author} {\bibfnamefont {M.}~\bibnamefont {Parniak}},
  \bibinfo {author} {\bibfnamefont {C.}~\bibnamefont {{\O}stfeldt}}, \bibinfo
  {author} {\bibfnamefont {C.~B.}\ \bibnamefont {M{\o}ller}}, \bibinfo {author}
  {\bibfnamefont {C.}~\bibnamefont {B{\ae}rentsen}}, \bibinfo {author}
  {\bibfnamefont {Y.}~\bibnamefont {Tsaturyan}}, \bibinfo {author}
  {\bibfnamefont {A.}~\bibnamefont {Schliesser}}, \bibinfo {author}
  {\bibfnamefont {J.}~\bibnamefont {Appel}}, \bibinfo {author} {\bibfnamefont
  {E.}~\bibnamefont {Zeuthen}},\ and\ \bibinfo {author} {\bibfnamefont {E.~S.}\
  \bibnamefont {Polzik}},\ }\bibfield  {title} {\bibinfo {title} {Entanglement
  between distant macroscopic mechanical and spin systems},\ }\href
  {https://doi.org/10.1038/s41567-020-1031-5} {\bibfield  {journal} {\bibinfo
  {journal} {Nature Physics}\ }\textbf {\bibinfo {volume} {17}},\ \bibinfo
  {pages} {228} (\bibinfo {year} {2021})}\BibitemShut {NoStop}%
\bibitem [{\citenamefont {Turner}(1990)}]{Turner1990}%
  \BibitemOpen
  \bibfield  {author} {\bibinfo {author} {\bibfnamefont {M.~S.}\ \bibnamefont
  {Turner}},\ }\bibfield  {title} {\bibinfo {title} {Windows on the axion},\
  }\href {https://doi.org/https://doi.org/10.1016/0370-1573(90)90172-X}
  {\bibfield  {journal} {\bibinfo  {journal} {Physics Reports}\ }\textbf
  {\bibinfo {volume} {197}},\ \bibinfo {pages} {67} (\bibinfo {year}
  {1990})}\BibitemShut {NoStop}%
\bibitem [{\citenamefont {Hempel}\ \emph {et~al.}(2013)\citenamefont {Hempel},
  \citenamefont {Lanyon}, \citenamefont {Jurcevic}, \citenamefont {Gerritsma},
  \citenamefont {Blatt},\ and\ \citenamefont {Roos}}]{Hempel_2013}%
  \BibitemOpen
  \bibfield  {author} {\bibinfo {author} {\bibfnamefont {C.}~\bibnamefont
  {Hempel}}, \bibinfo {author} {\bibfnamefont {B.~P.}\ \bibnamefont {Lanyon}},
  \bibinfo {author} {\bibfnamefont {P.}~\bibnamefont {Jurcevic}}, \bibinfo
  {author} {\bibfnamefont {R.}~\bibnamefont {Gerritsma}}, \bibinfo {author}
  {\bibfnamefont {R.}~\bibnamefont {Blatt}},\ and\ \bibinfo {author}
  {\bibfnamefont {C.~F.}\ \bibnamefont {Roos}},\ }\bibfield  {title} {\bibinfo
  {title} {Entanglement-enhanced detection of single-photon scattering
  events},\ }\href {https://doi.org/10.1038/nphoton.2013.172} {\bibfield
  {journal} {\bibinfo  {journal} {Nature Photonics}\ }\textbf {\bibinfo
  {volume} {7}},\ \bibinfo {pages} {630} (\bibinfo {year} {2013})}\BibitemShut
  {NoStop}%
\bibitem [{\citenamefont {Toscano}\ \emph {et~al.}(2006)\citenamefont
  {Toscano}, \citenamefont {Dalvit}, \citenamefont {Davidovich},\ and\
  \citenamefont {Zurek}}]{Toscano_2006}%
  \BibitemOpen
  \bibfield  {author} {\bibinfo {author} {\bibfnamefont {F.}~\bibnamefont
  {Toscano}}, \bibinfo {author} {\bibfnamefont {D.~A.~R.}\ \bibnamefont
  {Dalvit}}, \bibinfo {author} {\bibfnamefont {L.}~\bibnamefont {Davidovich}},\
  and\ \bibinfo {author} {\bibfnamefont {W.~H.}\ \bibnamefont {Zurek}},\
  }\bibfield  {title} {\bibinfo {title} {Sub-planck phase-space structures and
  heisenberg-limited measurements},\ }\href
  {https://doi.org/10.1103/PhysRevA.73.023803} {\bibfield  {journal} {\bibinfo
  {journal} {Phys. Rev. A}\ }\textbf {\bibinfo {volume} {73}},\ \bibinfo
  {pages} {023803} (\bibinfo {year} {2006})}\BibitemShut {NoStop}%
\bibitem [{\citenamefont {Penasa}\ \emph {et~al.}(2016)\citenamefont {Penasa},
  \citenamefont {Gerlich}, \citenamefont {Rybarczyk}, \citenamefont
  {M\'etillon}, \citenamefont {Brune}, \citenamefont {Raimond}, \citenamefont
  {Haroche}, \citenamefont {Davidovich},\ and\ \citenamefont
  {Dotsenko}}]{Penasa_2016}%
  \BibitemOpen
  \bibfield  {author} {\bibinfo {author} {\bibfnamefont {M.}~\bibnamefont
  {Penasa}}, \bibinfo {author} {\bibfnamefont {S.}~\bibnamefont {Gerlich}},
  \bibinfo {author} {\bibfnamefont {T.}~\bibnamefont {Rybarczyk}}, \bibinfo
  {author} {\bibfnamefont {V.}~\bibnamefont {M\'etillon}}, \bibinfo {author}
  {\bibfnamefont {M.}~\bibnamefont {Brune}}, \bibinfo {author} {\bibfnamefont
  {J.~M.}\ \bibnamefont {Raimond}}, \bibinfo {author} {\bibfnamefont
  {S.}~\bibnamefont {Haroche}}, \bibinfo {author} {\bibfnamefont
  {L.}~\bibnamefont {Davidovich}},\ and\ \bibinfo {author} {\bibfnamefont
  {I.}~\bibnamefont {Dotsenko}},\ }\bibfield  {title} {\bibinfo {title}
  {Measurement of a microwave field amplitude beyond the standard quantum
  limit},\ }\href {https://doi.org/10.1103/PhysRevA.94.022313} {\bibfield
  {journal} {\bibinfo  {journal} {Phys. Rev. A}\ }\textbf {\bibinfo {volume}
  {94}},\ \bibinfo {pages} {022313} (\bibinfo {year} {2016})}\BibitemShut
  {NoStop}%
\bibitem [{\citenamefont {Lewis-Swan}\ \emph {et~al.}(2020)\citenamefont
  {Lewis-Swan}, \citenamefont {Barberena}, \citenamefont {Muniz}, \citenamefont
  {Cline}, \citenamefont {Young}, \citenamefont {Thompson},\ and\ \citenamefont
  {Rey}}]{Lewis-Swan2020}%
  \BibitemOpen
  \bibfield  {author} {\bibinfo {author} {\bibfnamefont {R.~J.}\ \bibnamefont
  {Lewis-Swan}}, \bibinfo {author} {\bibfnamefont {D.}~\bibnamefont
  {Barberena}}, \bibinfo {author} {\bibfnamefont {J.~A.}\ \bibnamefont
  {Muniz}}, \bibinfo {author} {\bibfnamefont {J.~R.~K.}\ \bibnamefont {Cline}},
  \bibinfo {author} {\bibfnamefont {D.}~\bibnamefont {Young}}, \bibinfo
  {author} {\bibfnamefont {J.~K.}\ \bibnamefont {Thompson}},\ and\ \bibinfo
  {author} {\bibfnamefont {A.~M.}\ \bibnamefont {Rey}},\ }\bibfield  {title}
  {\bibinfo {title} {Protocol for precise field sensing in the optical domain
  with cold atoms in a cavity},\ }\href
  {https://doi.org/10.1103/PhysRevLett.124.193602} {\bibfield  {journal}
  {\bibinfo  {journal} {Phys. Rev. Lett.}\ }\textbf {\bibinfo {volume} {124}},\
  \bibinfo {pages} {193602} (\bibinfo {year} {2020})}\BibitemShut {NoStop}%
\bibitem [{\citenamefont {Kampel}\ \emph {et~al.}(2017)\citenamefont {Kampel},
  \citenamefont {Peterson}, \citenamefont {Fischer}, \citenamefont {Yu},
  \citenamefont {Cicak}, \citenamefont {Simmonds}, \citenamefont {Lehnert},\
  and\ \citenamefont {Regal}}]{Kampel_2017}%
  \BibitemOpen
  \bibfield  {author} {\bibinfo {author} {\bibfnamefont {N.~S.}\ \bibnamefont
  {Kampel}}, \bibinfo {author} {\bibfnamefont {R.~W.}\ \bibnamefont
  {Peterson}}, \bibinfo {author} {\bibfnamefont {R.}~\bibnamefont {Fischer}},
  \bibinfo {author} {\bibfnamefont {P.-L.}\ \bibnamefont {Yu}}, \bibinfo
  {author} {\bibfnamefont {K.}~\bibnamefont {Cicak}}, \bibinfo {author}
  {\bibfnamefont {R.~W.}\ \bibnamefont {Simmonds}}, \bibinfo {author}
  {\bibfnamefont {K.~W.}\ \bibnamefont {Lehnert}},\ and\ \bibinfo {author}
  {\bibfnamefont {C.~A.}\ \bibnamefont {Regal}},\ }\bibfield  {title} {\bibinfo
  {title} {Improving broadband displacement detection with quantum
  correlations},\ }\href {https://doi.org/10.1103/PhysRevX.7.021008} {\bibfield
   {journal} {\bibinfo  {journal} {Phys. Rev. X}\ }\textbf {\bibinfo {volume}
  {7}},\ \bibinfo {pages} {021008} (\bibinfo {year} {2017})}\BibitemShut
  {NoStop}%
\bibitem [{\citenamefont {Jing}\ \emph {et~al.}(2020)\citenamefont {Jing},
  \citenamefont {Hu}, \citenamefont {Ma}, \citenamefont {Zhang}, \citenamefont
  {Zhang}, \citenamefont {Xiao},\ and\ \citenamefont {Jia}}]{Jing2020}%
  \BibitemOpen
  \bibfield  {author} {\bibinfo {author} {\bibfnamefont {M.}~\bibnamefont
  {Jing}}, \bibinfo {author} {\bibfnamefont {Y.}~\bibnamefont {Hu}}, \bibinfo
  {author} {\bibfnamefont {J.}~\bibnamefont {Ma}}, \bibinfo {author}
  {\bibfnamefont {H.}~\bibnamefont {Zhang}}, \bibinfo {author} {\bibfnamefont
  {L.}~\bibnamefont {Zhang}}, \bibinfo {author} {\bibfnamefont
  {L.}~\bibnamefont {Xiao}},\ and\ \bibinfo {author} {\bibfnamefont
  {S.}~\bibnamefont {Jia}},\ }\bibfield  {title} {\bibinfo {title} {Atomic
  superheterodyne receiver based on microwave-dressed rydberg spectroscopy},\
  }\href {https://doi.org/10.1038/s41567-020-0918-5} {\bibfield  {journal}
  {\bibinfo  {journal} {Nature Physics}\ }\textbf {\bibinfo {volume} {16}},\
  \bibinfo {pages} {911} (\bibinfo {year} {2020})}\BibitemShut {NoStop}%
\bibitem [{\citenamefont {Burd}\ \emph {et~al.}(2019)\citenamefont {Burd},
  \citenamefont {Srinivas}, \citenamefont {Bollinger}, \citenamefont {Wilson},
  \citenamefont {Wineland}, \citenamefont {Leibfried}, \citenamefont
  {Slichter},\ and\ \citenamefont {Allcock}}]{Burd_2019}%
  \BibitemOpen
  \bibfield  {author} {\bibinfo {author} {\bibfnamefont {S.~C.}\ \bibnamefont
  {Burd}}, \bibinfo {author} {\bibfnamefont {R.}~\bibnamefont {Srinivas}},
  \bibinfo {author} {\bibfnamefont {J.~J.}\ \bibnamefont {Bollinger}}, \bibinfo
  {author} {\bibfnamefont {A.~C.}\ \bibnamefont {Wilson}}, \bibinfo {author}
  {\bibfnamefont {D.~J.}\ \bibnamefont {Wineland}}, \bibinfo {author}
  {\bibfnamefont {D.}~\bibnamefont {Leibfried}}, \bibinfo {author}
  {\bibfnamefont {D.~H.}\ \bibnamefont {Slichter}},\ and\ \bibinfo {author}
  {\bibfnamefont {D.~T.~C.}\ \bibnamefont {Allcock}},\ }\bibfield  {title}
  {\bibinfo {title} {Quantum amplification of mechanical oscillator motion},\
  }\href {https://doi.org/10.1126/science.aaw2884} {\bibfield  {journal}
  {\bibinfo  {journal} {Science}\ }\textbf {\bibinfo {volume} {364}},\ \bibinfo
  {pages} {1163} (\bibinfo {year} {2019})},\ \Eprint
  {https://arxiv.org/abs/https://science.sciencemag.org/content/364/6446/1163.full.pdf}
  {https://science.sciencemag.org/content/364/6446/1163.full.pdf} \BibitemShut
  {NoStop}%
\bibitem [{\citenamefont {Bollinger}\ \emph {et~al.}(2013)\citenamefont
  {Bollinger}, \citenamefont {Britton},\ and\ \citenamefont
  {Sawyer}}]{Bollinger2013}%
  \BibitemOpen
  \bibfield  {author} {\bibinfo {author} {\bibfnamefont {J.~J.}\ \bibnamefont
  {Bollinger}}, \bibinfo {author} {\bibfnamefont {J.~W.}\ \bibnamefont
  {Britton}},\ and\ \bibinfo {author} {\bibfnamefont {B.~C.}\ \bibnamefont
  {Sawyer}},\ }\bibfield  {title} {\bibinfo {title} {{Simulating quantum
  magnetism with correlated non-neutral ion plasmas}}\ }(\bibinfo {year}
  {2013})\ pp.\ \bibinfo {pages} {200--209}\BibitemShut {NoStop}%
\bibitem [{\citenamefont {Sawyer}\ \emph {et~al.}(2014)\citenamefont {Sawyer},
  \citenamefont {Britton},\ and\ \citenamefont {Bollinger}}]{Sawyer2014}%
  \BibitemOpen
  \bibfield  {author} {\bibinfo {author} {\bibfnamefont {B.~C.}\ \bibnamefont
  {Sawyer}}, \bibinfo {author} {\bibfnamefont {J.~W.}\ \bibnamefont
  {Britton}},\ and\ \bibinfo {author} {\bibfnamefont {J.~J.}\ \bibnamefont
  {Bollinger}},\ }\bibfield  {title} {\bibinfo {title} {{Spin dephasing as a
  probe of mode temperature, motional state distributions, and heating rates in
  a two-dimensional ion crystal}},\ }\href
  {https://doi.org/10.1103/PhysRevA.89.033408} {\bibfield  {journal} {\bibinfo
  {journal} {Phys. Rev. A}\ }\textbf {\bibinfo {volume} {89}},\ \bibinfo
  {pages} {033408} (\bibinfo {year} {2014})}\BibitemShut {NoStop}%
\bibitem [{\citenamefont {Bohnet}\ \emph {et~al.}(2016)\citenamefont {Bohnet},
  \citenamefont {Sawyer}, \citenamefont {Britton}, \citenamefont {Wall},
  \citenamefont {Rey}, \citenamefont {Foss-Feig},\ and\ \citenamefont
  {Bollinger}}]{Bohnet2016}%
  \BibitemOpen
  \bibfield  {author} {\bibinfo {author} {\bibfnamefont {J.~G.}\ \bibnamefont
  {Bohnet}}, \bibinfo {author} {\bibfnamefont {B.~C.}\ \bibnamefont {Sawyer}},
  \bibinfo {author} {\bibfnamefont {J.~W.}\ \bibnamefont {Britton}}, \bibinfo
  {author} {\bibfnamefont {M.~L.}\ \bibnamefont {Wall}}, \bibinfo {author}
  {\bibfnamefont {A.~M.}\ \bibnamefont {Rey}}, \bibinfo {author} {\bibfnamefont
  {M.}~\bibnamefont {Foss-Feig}},\ and\ \bibinfo {author} {\bibfnamefont
  {J.~J.}\ \bibnamefont {Bollinger}},\ }\bibfield  {title} {\bibinfo {title}
  {Quantum spin dynamics and entanglement generation with hundreds of trapped
  ions},\ }\href {https://doi.org/10.1126/science.aad9958} {\bibfield
  {journal} {\bibinfo  {journal} {Science}\ }\textbf {\bibinfo {volume}
  {352}},\ \bibinfo {pages} {1297} (\bibinfo {year} {2016})}\BibitemShut
  {NoStop}%
\bibitem [{\citenamefont {Safavi-Naini}\ \emph {et~al.}(2018)\citenamefont
  {Safavi-Naini}, \citenamefont {Lewis-Swan}, \citenamefont {Bohnet},
  \citenamefont {G\"arttner}, \citenamefont {Gilmore}, \citenamefont {Jordan},
  \citenamefont {Cohn}, \citenamefont {Freericks}, \citenamefont {Rey},\ and\
  \citenamefont {Bollinger}}]{Bollinger_2018}%
  \BibitemOpen
  \bibfield  {author} {\bibinfo {author} {\bibfnamefont {A.}~\bibnamefont
  {Safavi-Naini}}, \bibinfo {author} {\bibfnamefont {R.~J.}\ \bibnamefont
  {Lewis-Swan}}, \bibinfo {author} {\bibfnamefont {J.~G.}\ \bibnamefont
  {Bohnet}}, \bibinfo {author} {\bibfnamefont {M.}~\bibnamefont {G\"arttner}},
  \bibinfo {author} {\bibfnamefont {K.~A.}\ \bibnamefont {Gilmore}}, \bibinfo
  {author} {\bibfnamefont {J.~E.}\ \bibnamefont {Jordan}}, \bibinfo {author}
  {\bibfnamefont {J.}~\bibnamefont {Cohn}}, \bibinfo {author} {\bibfnamefont
  {J.~K.}\ \bibnamefont {Freericks}}, \bibinfo {author} {\bibfnamefont {A.~M.}\
  \bibnamefont {Rey}},\ and\ \bibinfo {author} {\bibfnamefont {J.~J.}\
  \bibnamefont {Bollinger}},\ }\bibfield  {title} {\bibinfo {title}
  {Verification of a many-ion simulator of the dicke model through slow
  quenches across a phase transition},\ }\href
  {https://doi.org/10.1103/PhysRevLett.121.040503} {\bibfield  {journal}
  {\bibinfo  {journal} {Phys. Rev. Lett.}\ }\textbf {\bibinfo {volume} {121}},\
  \bibinfo {pages} {040503} (\bibinfo {year} {2018})}\BibitemShut {NoStop}%
\bibitem [{SOM()}]{SOM}%
  \BibitemOpen
  \href@noop {} {\bibinfo {title} {{See Supplemental Material at [URL will be
  inserted by publisher].}}}\BibitemShut {Stop}%
\bibitem [{\citenamefont {Braunstein}\ and\ \citenamefont
  {Caves}(1994)}]{Braunstein1994}%
  \BibitemOpen
  \bibfield  {author} {\bibinfo {author} {\bibfnamefont {S.~L.}\ \bibnamefont
  {Braunstein}}\ and\ \bibinfo {author} {\bibfnamefont {C.~M.}\ \bibnamefont
  {Caves}},\ }\bibfield  {title} {\bibinfo {title} {Statistical distance and
  the geometry of quantum states},\ }\href
  {https://doi.org/10.1103/PhysRevLett.72.3439} {\bibfield  {journal} {\bibinfo
   {journal} {Phys. Rev. Lett.}\ }\textbf {\bibinfo {volume} {72}},\ \bibinfo
  {pages} {3439} (\bibinfo {year} {1994})}\BibitemShut {NoStop}%
\bibitem [{\citenamefont {Hosten}\ \emph {et~al.}(2016)\citenamefont {Hosten},
  \citenamefont {Krishnakumar}, \citenamefont {Engelsen},\ and\ \citenamefont
  {Kasevich}}]{Hosten_2016}%
  \BibitemOpen
  \bibfield  {author} {\bibinfo {author} {\bibfnamefont {O.}~\bibnamefont
  {Hosten}}, \bibinfo {author} {\bibfnamefont {R.}~\bibnamefont
  {Krishnakumar}}, \bibinfo {author} {\bibfnamefont {N.~J.}\ \bibnamefont
  {Engelsen}},\ and\ \bibinfo {author} {\bibfnamefont {M.~A.}\ \bibnamefont
  {Kasevich}},\ }\bibfield  {title} {\bibinfo {title} {Quantum phase
  magnification},\ }\href {https://doi.org/10.1126/science.aaf3397} {\bibfield
  {journal} {\bibinfo  {journal} {Science}\ }\textbf {\bibinfo {volume}
  {352}},\ \bibinfo {pages} {1552} (\bibinfo {year} {2016})}\BibitemShut
  {NoStop}%
\bibitem [{\citenamefont {Davis}\ \emph {et~al.}(2016)\citenamefont {Davis},
  \citenamefont {Bentsen},\ and\ \citenamefont {Schleier-Smith}}]{Davis_2016}%
  \BibitemOpen
  \bibfield  {author} {\bibinfo {author} {\bibfnamefont {E.}~\bibnamefont
  {Davis}}, \bibinfo {author} {\bibfnamefont {G.}~\bibnamefont {Bentsen}},\
  and\ \bibinfo {author} {\bibfnamefont {M.}~\bibnamefont {Schleier-Smith}},\
  }\bibfield  {title} {\bibinfo {title} {Approaching the heisenberg limit
  without single-particle detection},\ }\href
  {https://doi.org/10.1103/PhysRevLett.116.053601} {\bibfield  {journal}
  {\bibinfo  {journal} {Phys. Rev. Lett.}\ }\textbf {\bibinfo {volume} {116}},\
  \bibinfo {pages} {053601} (\bibinfo {year} {2016})}\BibitemShut {NoStop}%
\bibitem [{\citenamefont {Nolan}\ \emph {et~al.}(2017)\citenamefont {Nolan},
  \citenamefont {Szigeti},\ and\ \citenamefont {Haine}}]{Nolan_2017}%
  \BibitemOpen
  \bibfield  {author} {\bibinfo {author} {\bibfnamefont {S.~P.}\ \bibnamefont
  {Nolan}}, \bibinfo {author} {\bibfnamefont {S.~S.}\ \bibnamefont {Szigeti}},\
  and\ \bibinfo {author} {\bibfnamefont {S.~A.}\ \bibnamefont {Haine}},\
  }\bibfield  {title} {\bibinfo {title} {Optimal and robust quantum metrology
  using interaction-based readouts},\ }\href
  {https://doi.org/10.1103/PhysRevLett.119.193601} {\bibfield  {journal}
  {\bibinfo  {journal} {Phys. Rev. Lett.}\ }\textbf {\bibinfo {volume} {119}},\
  \bibinfo {pages} {193601} (\bibinfo {year} {2017})}\BibitemShut {NoStop}%
\bibitem [{\citenamefont {Shankar}\ \emph {et~al.}(2020)\citenamefont
  {Shankar}, \citenamefont {Tang}, \citenamefont {Affolter}, \citenamefont
  {Gilmore}, \citenamefont {Dubin}, \citenamefont {Parker}, \citenamefont
  {Holland},\ and\ \citenamefont {Bollinger}}]{Shankar2020}%
  \BibitemOpen
  \bibfield  {author} {\bibinfo {author} {\bibfnamefont {A.}~\bibnamefont
  {Shankar}}, \bibinfo {author} {\bibfnamefont {C.}~\bibnamefont {Tang}},
  \bibinfo {author} {\bibfnamefont {M.}~\bibnamefont {Affolter}}, \bibinfo
  {author} {\bibfnamefont {K.}~\bibnamefont {Gilmore}}, \bibinfo {author}
  {\bibfnamefont {D.~H.~E.}\ \bibnamefont {Dubin}}, \bibinfo {author}
  {\bibfnamefont {S.}~\bibnamefont {Parker}}, \bibinfo {author} {\bibfnamefont
  {M.~J.}\ \bibnamefont {Holland}},\ and\ \bibinfo {author} {\bibfnamefont
  {J.~J.}\ \bibnamefont {Bollinger}},\ }\bibfield  {title} {\bibinfo {title}
  {Broadening of the drumhead-mode spectrum due to in-plane thermal
  fluctuations of two-dimensional trapped ion crystals in a penning trap},\
  }\href {https://doi.org/10.1103/PhysRevA.102.053106} {\bibfield  {journal}
  {\bibinfo  {journal} {Phys. Rev. A}\ }\textbf {\bibinfo {volume} {102}},\
  \bibinfo {pages} {053106} (\bibinfo {year} {2020})}\BibitemShut {NoStop}%
\bibitem [{\citenamefont {Facon}\ \emph {et~al.}(2016)\citenamefont {Facon},
  \citenamefont {Dietsche}, \citenamefont {Grosso}, \citenamefont {Haroche},
  \citenamefont {Raimond}, \citenamefont {Brune},\ and\ \citenamefont
  {Gleyzes}}]{Facon2016}%
  \BibitemOpen
  \bibfield  {author} {\bibinfo {author} {\bibfnamefont {A.}~\bibnamefont
  {Facon}}, \bibinfo {author} {\bibfnamefont {E.-K.}\ \bibnamefont {Dietsche}},
  \bibinfo {author} {\bibfnamefont {D.}~\bibnamefont {Grosso}}, \bibinfo
  {author} {\bibfnamefont {S.}~\bibnamefont {Haroche}}, \bibinfo {author}
  {\bibfnamefont {J.-M.}\ \bibnamefont {Raimond}}, \bibinfo {author}
  {\bibfnamefont {M.}~\bibnamefont {Brune}},\ and\ \bibinfo {author}
  {\bibfnamefont {S.}~\bibnamefont {Gleyzes}},\ }\bibfield  {title} {\bibinfo
  {title} {A sensitive electrometer based on a rydberg atom in a
  schr{\"o}dinger-cat state},\ }\href {https://doi.org/10.1038/nature18327}
  {\bibfield  {journal} {\bibinfo  {journal} {Nature}\ }\textbf {\bibinfo
  {volume} {535}},\ \bibinfo {pages} {262} (\bibinfo {year}
  {2016})}\BibitemShut {NoStop}%
\bibitem [{\citenamefont {Meyer}\ \emph {et~al.}(2020)\citenamefont {Meyer},
  \citenamefont {Castillo}, \citenamefont {Cox},\ and\ \citenamefont
  {Kunz}}]{Meyer_2020}%
  \BibitemOpen
  \bibfield  {author} {\bibinfo {author} {\bibfnamefont {D.~H.}\ \bibnamefont
  {Meyer}}, \bibinfo {author} {\bibfnamefont {Z.~A.}\ \bibnamefont {Castillo}},
  \bibinfo {author} {\bibfnamefont {K.~C.}\ \bibnamefont {Cox}},\ and\ \bibinfo
  {author} {\bibfnamefont {P.~D.}\ \bibnamefont {Kunz}},\ }\bibfield  {title}
  {\bibinfo {title} {Assessment of rydberg atoms for wideband electric field
  sensing},\ }\href {https://doi.org/10.1088/1361-6455/ab6051} {\bibfield
  {journal} {\bibinfo  {journal} {Journal of Physics B: Atomic, Molecular and
  Optical Physics}\ }\textbf {\bibinfo {volume} {53}},\ \bibinfo {pages}
  {034001} (\bibinfo {year} {2020})}\BibitemShut {NoStop}%
\bibitem [{\citenamefont {Jordan}\ \emph {et~al.}(2019)\citenamefont {Jordan},
  \citenamefont {Gilmore}, \citenamefont {Shankar}, \citenamefont
  {Safavi-Naini}, \citenamefont {Bohnet}, \citenamefont {Holland},\ and\
  \citenamefont {Bollinger}}]{Jordan_2019}%
  \BibitemOpen
  \bibfield  {author} {\bibinfo {author} {\bibfnamefont {E.}~\bibnamefont
  {Jordan}}, \bibinfo {author} {\bibfnamefont {K.~A.}\ \bibnamefont {Gilmore}},
  \bibinfo {author} {\bibfnamefont {A.}~\bibnamefont {Shankar}}, \bibinfo
  {author} {\bibfnamefont {A.}~\bibnamefont {Safavi-Naini}}, \bibinfo {author}
  {\bibfnamefont {J.~G.}\ \bibnamefont {Bohnet}}, \bibinfo {author}
  {\bibfnamefont {M.~J.}\ \bibnamefont {Holland}},\ and\ \bibinfo {author}
  {\bibfnamefont {J.~J.}\ \bibnamefont {Bollinger}},\ }\bibfield  {title}
  {\bibinfo {title} {Near ground-state cooling of two-dimensional trapped-ion
  crystals with more than 100 ions},\ }\href
  {https://doi.org/10.1103/PhysRevLett.122.053603} {\bibfield  {journal}
  {\bibinfo  {journal} {Phys. Rev. Lett.}\ }\textbf {\bibinfo {volume} {122}},\
  \bibinfo {pages} {053603} (\bibinfo {year} {2019})}\BibitemShut {NoStop}%
\bibitem [{\citenamefont {Chaudhuri}\ \emph {et~al.}(2015)\citenamefont
  {Chaudhuri}, \citenamefont {Graham}, \citenamefont {Irwin}, \citenamefont
  {Mardon}, \citenamefont {Rajendran},\ and\ \citenamefont
  {Zhao}}]{Chaudhuri2015}%
  \BibitemOpen
  \bibfield  {author} {\bibinfo {author} {\bibfnamefont {S.}~\bibnamefont
  {Chaudhuri}}, \bibinfo {author} {\bibfnamefont {P.~W.}\ \bibnamefont
  {Graham}}, \bibinfo {author} {\bibfnamefont {K.}~\bibnamefont {Irwin}},
  \bibinfo {author} {\bibfnamefont {J.}~\bibnamefont {Mardon}}, \bibinfo
  {author} {\bibfnamefont {S.}~\bibnamefont {Rajendran}},\ and\ \bibinfo
  {author} {\bibfnamefont {Y.}~\bibnamefont {Zhao}},\ }\bibfield  {title}
  {\bibinfo {title} {Radio for hidden-photon dark matter detection},\ }\href
  {https://doi.org/10.1103/PhysRevD.92.075012} {\bibfield  {journal} {\bibinfo
  {journal} {Phys. Rev. D}\ }\textbf {\bibinfo {volume} {92}},\ \bibinfo
  {pages} {075012} (\bibinfo {year} {2015})}\BibitemShut {NoStop}%
\bibitem [{\citenamefont {Itano}\ \emph {et~al.}(1998)\citenamefont {Itano},
  \citenamefont {Bollinger}, \citenamefont {Tan}, \citenamefont
  {Jelenkovi{\'c}}, \citenamefont {Huang},\ and\ \citenamefont
  {Wineland}}]{Itano1998}%
  \BibitemOpen
  \bibfield  {author} {\bibinfo {author} {\bibfnamefont {W.~M.}\ \bibnamefont
  {Itano}}, \bibinfo {author} {\bibfnamefont {J.~J.}\ \bibnamefont
  {Bollinger}}, \bibinfo {author} {\bibfnamefont {J.~N.}\ \bibnamefont {Tan}},
  \bibinfo {author} {\bibfnamefont {B.}~\bibnamefont {Jelenkovi{\'c}}},
  \bibinfo {author} {\bibfnamefont {X.-P.}\ \bibnamefont {Huang}},\ and\
  \bibinfo {author} {\bibfnamefont {D.~J.}\ \bibnamefont {Wineland}},\
  }\bibfield  {title} {\bibinfo {title} {Bragg diffraction from crystallized
  ion plasmas},\ }\href {https://doi.org/10.1126/science.279.5351.686}
  {\bibfield  {journal} {\bibinfo  {journal} {Science}\ }\textbf {\bibinfo
  {volume} {279}},\ \bibinfo {pages} {686} (\bibinfo {year}
  {1998})}\BibitemShut {NoStop}%
\bibitem [{\citenamefont {Mortensen}\ \emph {et~al.}(2006)\citenamefont
  {Mortensen}, \citenamefont {Nielsen}, \citenamefont {Matthey},\ and\
  \citenamefont {Drewsen}}]{Mortensen2006}%
  \BibitemOpen
  \bibfield  {author} {\bibinfo {author} {\bibfnamefont {A.}~\bibnamefont
  {Mortensen}}, \bibinfo {author} {\bibfnamefont {E.}~\bibnamefont {Nielsen}},
  \bibinfo {author} {\bibfnamefont {T.}~\bibnamefont {Matthey}},\ and\ \bibinfo
  {author} {\bibfnamefont {M.}~\bibnamefont {Drewsen}},\ }\bibfield  {title}
  {\bibinfo {title} {Observation of three-dimensional long-range order in small
  ion coulomb crystals in an rf trap},\ }\href
  {https://doi.org/10.1103/PhysRevLett.96.103001} {\bibfield  {journal}
  {\bibinfo  {journal} {Phys. Rev. Lett.}\ }\textbf {\bibinfo {volume} {96}},\
  \bibinfo {pages} {103001} (\bibinfo {year} {2006})}\BibitemShut {NoStop}%
\end{thebibliography}%


\begin{thebibliography}{14}%
\makeatletter
\providecommand \@ifxundefined [1]{%
 \@ifx{#1\undefined}
}%
\providecommand \@ifnum [1]{%
 \ifnum #1\expandafter \@firstoftwo
 \else \expandafter \@secondoftwo
 \fi
}%
\providecommand \@ifx [1]{%
 \ifx #1\expandafter \@firstoftwo
 \else \expandafter \@secondoftwo
 \fi
}%
\providecommand \natexlab [1]{#1}%
\providecommand \enquote  [1]{``#1''}%
\providecommand \bibnamefont  [1]{#1}%
\providecommand \bibfnamefont [1]{#1}%
\providecommand \citenamefont [1]{#1}%
\providecommand \href@noop [0]{\@secondoftwo}%
\providecommand \href [0]{\begingroup \@sanitize@url \@href}%
\providecommand \@href[1]{\@@startlink{#1}\@@href}%
\providecommand \@@href[1]{\endgroup#1\@@endlink}%
\providecommand \@sanitize@url [0]{\catcode `\\12\catcode `\$12\catcode
  `\&12\catcode `\#12\catcode `\^12\catcode `\_12\catcode `\%12\relax}%
\providecommand \@@startlink[1]{}%
\providecommand \@@endlink[0]{}%
\providecommand \url  [0]{\begingroup\@sanitize@url \@url }%
\providecommand \@url [1]{\endgroup\@href {#1}{\urlprefix }}%
\providecommand \urlprefix  [0]{URL }%
\providecommand \Eprint [0]{\href }%
\providecommand \doibase [0]{https://doi.org/}%
\providecommand \selectlanguage [0]{\@gobble}%
\providecommand \bibinfo  [0]{\@secondoftwo}%
\providecommand \bibfield  [0]{\@secondoftwo}%
\providecommand \translation [1]{[#1]}%
\providecommand \BibitemOpen [0]{}%
\providecommand \bibitemStop [0]{}%
\providecommand \bibitemNoStop [0]{.\EOS\space}%
\providecommand \EOS [0]{\spacefactor3000\relax}%
\providecommand \BibitemShut  [1]{\csname bibitem#1\endcsname}%
\let\auto@bib@innerbib\@empty
\bibitem [{\citenamefont {Lewis-Swan}\ \emph {et~al.}(2020)\citenamefont
  {Lewis-Swan}, \citenamefont {Barberena}, \citenamefont {Muniz}, \citenamefont
  {Cline}, \citenamefont {Young}, \citenamefont {Thompson},\ and\ \citenamefont
  {Rey}}]{Lewis-Swan2020}%
  \BibitemOpen
  \bibfield  {author} {\bibinfo {author} {\bibfnamefont {R.~J.}\ \bibnamefont
  {Lewis-Swan}}, \bibinfo {author} {\bibfnamefont {D.}~\bibnamefont
  {Barberena}}, \bibinfo {author} {\bibfnamefont {J.~A.}\ \bibnamefont
  {Muniz}}, \bibinfo {author} {\bibfnamefont {J.~R.~K.}\ \bibnamefont {Cline}},
  \bibinfo {author} {\bibfnamefont {D.}~\bibnamefont {Young}}, \bibinfo
  {author} {\bibfnamefont {J.~K.}\ \bibnamefont {Thompson}},\ and\ \bibinfo
  {author} {\bibfnamefont {A.~M.}\ \bibnamefont {Rey}},\ }\bibfield  {title}
  {\bibinfo {title} {Protocol for precise field sensing in the optical domain
  with cold atoms in a cavity},\ }\href
  {https://doi.org/10.1103/PhysRevLett.124.193602} {\bibfield  {journal}
  {\bibinfo  {journal} {Phys. Rev. Lett.}\ }\textbf {\bibinfo {volume} {124}},\
  \bibinfo {pages} {193602} (\bibinfo {year} {2020})}\BibitemShut {NoStop}%
\bibitem [{\citenamefont {Zurek}(2001)}]{Zurek2001}%
  \BibitemOpen
  \bibfield  {author} {\bibinfo {author} {\bibfnamefont {W.~H.}\ \bibnamefont
  {Zurek}},\ }\bibfield  {title} {\bibinfo {title} {Sub-planck structure in
  phase space and its relevance for quantum decoherence},\ }\href
  {https://doi.org/10.1038/35089017} {\bibfield  {journal} {\bibinfo  {journal}
  {Nature}\ }\textbf {\bibinfo {volume} {412}},\ \bibinfo {pages} {712}
  (\bibinfo {year} {2001})}\BibitemShut {NoStop}%
\bibitem [{\citenamefont {Toscano}\ \emph {et~al.}(2006)\citenamefont
  {Toscano}, \citenamefont {Dalvit}, \citenamefont {Davidovich},\ and\
  \citenamefont {Zurek}}]{Toscano_2006}%
  \BibitemOpen
  \bibfield  {author} {\bibinfo {author} {\bibfnamefont {F.}~\bibnamefont
  {Toscano}}, \bibinfo {author} {\bibfnamefont {D.~A.~R.}\ \bibnamefont
  {Dalvit}}, \bibinfo {author} {\bibfnamefont {L.}~\bibnamefont {Davidovich}},\
  and\ \bibinfo {author} {\bibfnamefont {W.~H.}\ \bibnamefont {Zurek}},\
  }\bibfield  {title} {\bibinfo {title} {Sub-planck phase-space structures and
  heisenberg-limited measurements},\ }\href
  {https://doi.org/10.1103/PhysRevA.73.023803} {\bibfield  {journal} {\bibinfo
  {journal} {Phys. Rev. A}\ }\textbf {\bibinfo {volume} {73}},\ \bibinfo
  {pages} {023803} (\bibinfo {year} {2006})}\BibitemShut {NoStop}%
\bibitem [{\citenamefont {Walls}\ and\ \citenamefont
  {Milburn}(2008)}]{walls_quantum_2008}%
  \BibitemOpen
  \bibfield  {author} {\bibinfo {author} {\bibfnamefont {D.~F.}\ \bibnamefont
  {Walls}}\ and\ \bibinfo {author} {\bibfnamefont {G.~J.}\ \bibnamefont
  {Milburn}},\ }\href@noop {} {\emph {\bibinfo {title} {{Quantum Optics}}}},\
  \bibinfo {edition} {2nd}\ ed.\ (\bibinfo  {publisher} {Springer},\ \bibinfo
  {year} {2008})\BibitemShut {NoStop}%
\bibitem [{\citenamefont {Gilmore}\ \emph {et~al.}(2017)\citenamefont
  {Gilmore}, \citenamefont {Bohnet}, \citenamefont {Sawyer}, \citenamefont
  {Britton},\ and\ \citenamefont {Bollinger}}]{Gilmore2017}%
  \BibitemOpen
  \bibfield  {author} {\bibinfo {author} {\bibfnamefont {K.~A.}\ \bibnamefont
  {Gilmore}}, \bibinfo {author} {\bibfnamefont {J.~G.}\ \bibnamefont {Bohnet}},
  \bibinfo {author} {\bibfnamefont {B.~C.}\ \bibnamefont {Sawyer}}, \bibinfo
  {author} {\bibfnamefont {J.~W.}\ \bibnamefont {Britton}},\ and\ \bibinfo
  {author} {\bibfnamefont {J.~J.}\ \bibnamefont {Bollinger}},\ }\bibfield
  {title} {\bibinfo {title} {Amplitude sensing below the zero-point
  fluctuations with a two-dimensional trapped-ion mechanical oscillator},\
  }\href {https://doi.org/10.1103/PhysRevLett.118.263602} {\bibfield  {journal}
  {\bibinfo  {journal} {Phys. Rev. Lett.}\ }\textbf {\bibinfo {volume} {118}},\
  \bibinfo {pages} {263602} (\bibinfo {year} {2017})}\BibitemShut {NoStop}%
\bibitem [{\citenamefont {Affolter}\ \emph {et~al.}(2020)\citenamefont
  {Affolter}, \citenamefont {Gilmore}, \citenamefont {Jordan},\ and\
  \citenamefont {Bollinger}}]{Affolter2020}%
  \BibitemOpen
  \bibfield  {author} {\bibinfo {author} {\bibfnamefont {M.}~\bibnamefont
  {Affolter}}, \bibinfo {author} {\bibfnamefont {K.~A.}\ \bibnamefont
  {Gilmore}}, \bibinfo {author} {\bibfnamefont {J.~E.}\ \bibnamefont
  {Jordan}},\ and\ \bibinfo {author} {\bibfnamefont {J.~J.}\ \bibnamefont
  {Bollinger}},\ }\bibfield  {title} {\bibinfo {title} {Phase-coherent sensing
  of the center-of-mass motion of trapped-ion crystals},\ }\href
  {https://doi.org/10.1103/PhysRevA.102.052609} {\bibfield  {journal} {\bibinfo
   {journal} {Phys. Rev. A}\ }\textbf {\bibinfo {volume} {102}},\ \bibinfo
  {pages} {052609} (\bibinfo {year} {2020})}\BibitemShut {NoStop}%
\bibitem [{\citenamefont {Rey}\ \emph {et~al.}(2008)\citenamefont {Rey},
  \citenamefont {Jiang}, \citenamefont {Fleischhauer}, \citenamefont {Demler},\
  and\ \citenamefont {Lukin}}]{Rey_2008}%
  \BibitemOpen
  \bibfield  {author} {\bibinfo {author} {\bibfnamefont {A.~M.}\ \bibnamefont
  {Rey}}, \bibinfo {author} {\bibfnamefont {L.}~\bibnamefont {Jiang}}, \bibinfo
  {author} {\bibfnamefont {M.}~\bibnamefont {Fleischhauer}}, \bibinfo {author}
  {\bibfnamefont {E.}~\bibnamefont {Demler}},\ and\ \bibinfo {author}
  {\bibfnamefont {M.~D.}\ \bibnamefont {Lukin}},\ }\bibfield  {title} {\bibinfo
  {title} {Many-body protected entanglement generation in interacting spin
  systems},\ }\href {https://doi.org/10.1103/PhysRevA.77.052305} {\bibfield
  {journal} {\bibinfo  {journal} {Phys. Rev. A}\ }\textbf {\bibinfo {volume}
  {77}},\ \bibinfo {pages} {052305} (\bibinfo {year} {2008})}\BibitemShut
  {NoStop}%
\bibitem [{\citenamefont {Biercuk}\ \emph {et~al.}(2009)\citenamefont
  {Biercuk}, \citenamefont {Uys}, \citenamefont {Vandevender}, \citenamefont
  {Shiga}, \citenamefont {Itano},\ and\ \citenamefont
  {Bollinger}}]{Biercuk2009}%
  \BibitemOpen
  \bibfield  {author} {\bibinfo {author} {\bibfnamefont {M.~J.}\ \bibnamefont
  {Biercuk}}, \bibinfo {author} {\bibfnamefont {H.}~\bibnamefont {Uys}},
  \bibinfo {author} {\bibfnamefont {A.~P.}\ \bibnamefont {Vandevender}},
  \bibinfo {author} {\bibfnamefont {N.}~\bibnamefont {Shiga}}, \bibinfo
  {author} {\bibfnamefont {W.~M.}\ \bibnamefont {Itano}},\ and\ \bibinfo
  {author} {\bibfnamefont {J.~J.}\ \bibnamefont {Bollinger}},\ }\bibfield
  {title} {\bibinfo {title} {{High-fidelity Quantum Control Using Ion Crystals
  in a Penning Trap}},\ }\href
  {http://dl.acm.org/citation.cfm?id=2012098.2012100} {\bibfield  {journal}
  {\bibinfo  {journal} {Quantum Info. Comput.}\ }\textbf {\bibinfo {volume}
  {9}},\ \bibinfo {pages} {920} (\bibinfo {year} {2009})}\BibitemShut {NoStop}%
\bibitem [{\citenamefont {Huang}\ \emph {et~al.}(1998)\citenamefont {Huang},
  \citenamefont {Bollinger}, \citenamefont {Mitchell}, \citenamefont {Itano},\
  and\ \citenamefont {Dubin}}]{Huang1998b}%
  \BibitemOpen
  \bibfield  {author} {\bibinfo {author} {\bibfnamefont {X.-P.}\ \bibnamefont
  {Huang}}, \bibinfo {author} {\bibfnamefont {J.~J.}\ \bibnamefont
  {Bollinger}}, \bibinfo {author} {\bibfnamefont {T.~B.}\ \bibnamefont
  {Mitchell}}, \bibinfo {author} {\bibfnamefont {W.~M.}\ \bibnamefont
  {Itano}},\ and\ \bibinfo {author} {\bibfnamefont {D.~H.~E.}\ \bibnamefont
  {Dubin}},\ }\bibfield  {title} {\bibinfo {title} {{Precise control of the
  global rotation of strongly coupled ion plasmas in a Penning trap}},\ }\href
  {https://doi.org/10.1063/1.872834} {\bibfield  {journal} {\bibinfo  {journal}
  {Phys. Plasmas}\ }\textbf {\bibinfo {volume} {5}},\ \bibinfo {pages} {1656}
  (\bibinfo {year} {1998})}\BibitemShut {NoStop}%
\bibitem [{\citenamefont {Uys}\ \emph {et~al.}(2010)\citenamefont {Uys},
  \citenamefont {Biercuk}, \citenamefont {VanDevender}, \citenamefont
  {Ospelkaus}, \citenamefont {Meiser}, \citenamefont {Ozeri},\ and\
  \citenamefont {Bollinger}}]{Uys2010}%
  \BibitemOpen
  \bibfield  {author} {\bibinfo {author} {\bibfnamefont {H.}~\bibnamefont
  {Uys}}, \bibinfo {author} {\bibfnamefont {M.~J.}\ \bibnamefont {Biercuk}},
  \bibinfo {author} {\bibfnamefont {A.~P.}\ \bibnamefont {VanDevender}},
  \bibinfo {author} {\bibfnamefont {C.}~\bibnamefont {Ospelkaus}}, \bibinfo
  {author} {\bibfnamefont {D.}~\bibnamefont {Meiser}}, \bibinfo {author}
  {\bibfnamefont {R.}~\bibnamefont {Ozeri}},\ and\ \bibinfo {author}
  {\bibfnamefont {J.~J.}\ \bibnamefont {Bollinger}},\ }\bibfield  {title}
  {\bibinfo {title} {{Decoherence due to elastic Rayleigh scattering}},\ }\href
  {https://doi.org/10.1103/PhysRevLett.105.200401} {\bibfield  {journal}
  {\bibinfo  {journal} {Phys. Rev. Lett.}\ }\textbf {\bibinfo {volume} {105}},\
  \bibinfo {pages} {200401} (\bibinfo {year} {2010})}\BibitemShut {NoStop}%
\bibitem [{\citenamefont {Britton}\ \emph {et~al.}(2012)\citenamefont
  {Britton}, \citenamefont {Sawyer}, \citenamefont {Keith}, \citenamefont
  {Wang}, \citenamefont {Freericks}, \citenamefont {Uys}, \citenamefont
  {Biercuk},\ and\ \citenamefont {Bollinger}}]{Britton2012}%
  \BibitemOpen
  \bibfield  {author} {\bibinfo {author} {\bibfnamefont {J.~W.}\ \bibnamefont
  {Britton}}, \bibinfo {author} {\bibfnamefont {B.~C.}\ \bibnamefont {Sawyer}},
  \bibinfo {author} {\bibfnamefont {A.~C.}\ \bibnamefont {Keith}}, \bibinfo
  {author} {\bibfnamefont {C.-C.~J.}\ \bibnamefont {Wang}}, \bibinfo {author}
  {\bibfnamefont {J.~K.}\ \bibnamefont {Freericks}}, \bibinfo {author}
  {\bibfnamefont {H.}~\bibnamefont {Uys}}, \bibinfo {author} {\bibfnamefont
  {M.~J.}\ \bibnamefont {Biercuk}},\ and\ \bibinfo {author} {\bibfnamefont
  {J.~J.}\ \bibnamefont {Bollinger}},\ }\bibfield  {title} {\bibinfo {title}
  {{Engineered two-dimensional Ising interactions in a trapped-ion quantum
  simulator with hundreds of spins}},\ }\href
  {https://doi.org/10.1038/nature10981} {\bibfield  {journal} {\bibinfo
  {journal} {Nature}\ }\textbf {\bibinfo {volume} {484}},\ \bibinfo {pages}
  {489} (\bibinfo {year} {2012})}\BibitemShut {NoStop}%
\bibitem [{\citenamefont {Bohnet}\ \emph {et~al.}(2016)\citenamefont {Bohnet},
  \citenamefont {Sawyer}, \citenamefont {Britton}, \citenamefont {Wall},
  \citenamefont {Rey}, \citenamefont {Foss-Feig},\ and\ \citenamefont
  {Bollinger}}]{Bohnet2016}%
  \BibitemOpen
  \bibfield  {author} {\bibinfo {author} {\bibfnamefont {J.~G.}\ \bibnamefont
  {Bohnet}}, \bibinfo {author} {\bibfnamefont {B.~C.}\ \bibnamefont {Sawyer}},
  \bibinfo {author} {\bibfnamefont {J.~W.}\ \bibnamefont {Britton}}, \bibinfo
  {author} {\bibfnamefont {M.~L.}\ \bibnamefont {Wall}}, \bibinfo {author}
  {\bibfnamefont {A.~M.}\ \bibnamefont {Rey}}, \bibinfo {author} {\bibfnamefont
  {M.}~\bibnamefont {Foss-Feig}},\ and\ \bibinfo {author} {\bibfnamefont
  {J.~J.}\ \bibnamefont {Bollinger}},\ }\bibfield  {title} {\bibinfo {title}
  {Quantum spin dynamics and entanglement generation with hundreds of trapped
  ions},\ }\href {https://doi.org/10.1126/science.aad9958} {\bibfield
  {journal} {\bibinfo  {journal} {Science}\ }\textbf {\bibinfo {volume}
  {352}},\ \bibinfo {pages} {1297} (\bibinfo {year} {2016})}\BibitemShut
  {NoStop}%
\bibitem [{\citenamefont {Leibfried}\ \emph {et~al.}(2003)\citenamefont
  {Leibfried}, \citenamefont {DeMarco}, \citenamefont {Meyer}, \citenamefont
  {Lucas}, \citenamefont {Barrett}, \citenamefont {Britton}, \citenamefont
  {Itano}, \citenamefont {Jelenkovi{\'{c}}}, \citenamefont {Langer},
  \citenamefont {Rosenband},\ and\ \citenamefont {Wineland}}]{Leibfried2003}%
  \BibitemOpen
  \bibfield  {author} {\bibinfo {author} {\bibfnamefont {D.}~\bibnamefont
  {Leibfried}}, \bibinfo {author} {\bibfnamefont {B.}~\bibnamefont {DeMarco}},
  \bibinfo {author} {\bibfnamefont {V.}~\bibnamefont {Meyer}}, \bibinfo
  {author} {\bibfnamefont {D.}~\bibnamefont {Lucas}}, \bibinfo {author}
  {\bibfnamefont {M.}~\bibnamefont {Barrett}}, \bibinfo {author} {\bibfnamefont
  {J.}~\bibnamefont {Britton}}, \bibinfo {author} {\bibfnamefont {W.~M.}\
  \bibnamefont {Itano}}, \bibinfo {author} {\bibfnamefont {B.}~\bibnamefont
  {Jelenkovi{\'{c}}}}, \bibinfo {author} {\bibfnamefont {C.}~\bibnamefont
  {Langer}}, \bibinfo {author} {\bibfnamefont {T.}~\bibnamefont {Rosenband}},\
  and\ \bibinfo {author} {\bibfnamefont {D.~J.}\ \bibnamefont {Wineland}},\
  }\bibfield  {title} {\bibinfo {title} {{Experimental demonstration of a
  robust, high-fidelity geometric two ion-qubit phase gate.}},\ }\href
  {https://doi.org/10.1038/nature01492} {\bibfield  {journal} {\bibinfo
  {journal} {Nature}\ }\textbf {\bibinfo {volume} {422}},\ \bibinfo {pages}
  {412} (\bibinfo {year} {2003})}\BibitemShut {NoStop}%
\bibitem [{\citenamefont {Brownnutt}\ \emph {et~al.}(2015)\citenamefont
  {Brownnutt}, \citenamefont {Kumph}, \citenamefont {Rabl},\ and\ \citenamefont
  {Blatt}}]{Brownnutt2015}%
  \BibitemOpen
  \bibfield  {author} {\bibinfo {author} {\bibfnamefont {M.}~\bibnamefont
  {Brownnutt}}, \bibinfo {author} {\bibfnamefont {M.}~\bibnamefont {Kumph}},
  \bibinfo {author} {\bibfnamefont {P.}~\bibnamefont {Rabl}},\ and\ \bibinfo
  {author} {\bibfnamefont {R.}~\bibnamefont {Blatt}},\ }\bibfield  {title}
  {\bibinfo {title} {Ion-trap measurements of electric-field noise near
  surfaces},\ }\href {https://doi.org/10.1103/RevModPhys.87.1419} {\bibfield
  {journal} {\bibinfo  {journal} {Rev. Mod. Phys.}\ }\textbf {\bibinfo {volume}
  {87}},\ \bibinfo {pages} {1419} (\bibinfo {year} {2015})}\BibitemShut
  {NoStop}%
\end{thebibliography}%

\end{document}


\newcommand{\Rev }[1]{{\color{blue}{#1}\normalcolor}} 
\newcommand{\Com}[1]{{\color{red}{#1}\normalcolor}} 
\newcommand{\ketbra}[2]{|#1\rangle\langle #2|}
\newcommand{\abs}[1]{\lvert #1 \rvert}

\title{Supplementary Online Material: Quantum-enhanced sensing of displacements and electric fields with large trapped-ion crystals}
\date{\today}

\author{K. A. Gilmore}
\affiliation{National Institute of Standards and Technology, Boulder, Colorado 80305, USA}
\affiliation{Department of Physics, University of Colorado, Boulder, Colorado, 80309, USA}

\author{M. Affolter}
\affiliation{National Institute of Standards and Technology, Boulder, Colorado 80305, USA}

\author{R.~J. Lewis-Swan}
\affiliation{Homer L. Dodge Department of Physics and Astronomy, The University of Oklahoma, Norman, Oklahoma 73019, USA}
\affiliation{Center for Quantum Research and Technology, The University of Oklahoma, Norman, Oklahoma 73019, USA}

\author{D. Barberena}
\affiliation{JILA, NIST and Department of Physics, University of Colorado, 440 UCB, Boulder, Colorado 80309, USA}
\affiliation{Center for Theory of Quantum Matter, University of Colorado, Boulder, CO 80309, USA}

\author{E. Jordan}
\affiliation{National Institute of Standards and Technology, Boulder, Colorado 80305, USA}

\author{A. M. Rey}
\affiliation{JILA, NIST and Department of Physics, University of Colorado, 440 UCB, Boulder, Colorado 80309, USA}
\affiliation{Center for Theory of Quantum Matter, University of Colorado, Boulder, CO 80309, USA}

\author{J. J. Bollinger}
\affiliation{National Institute of Standards and Technology, Boulder, Colorado 80305, USA}

\maketitle 

\section{Ideal model of displacement sensor}
The ideal displacement sensing protocol is realized by the implementation of the ODF Hamiltonian,
\begin{equation}
    \hat{H}_{\text{ODF}}=\frac{g}{\sqrt{N}}(\hat{a}+\hat{a}^\dagger)\hat{J}_z.
\end{equation}
Here, $\hat{a}$ ($\hat{a}^\dagger$) is an annihilation (creation) operator for the center of mass (COM) bosonic mode that couples uniformly to all spins with strength $g$ and $\hat{J}_{x,y,z}=\sum_{i=1}^N\hat{\sigma}_{x,y,z}^j/2$ are collective spin operators, with $\hat{\sigma}_{x,y,z}^j$ Pauli operators acting on ion $j$. As described in the main text, the protocol involves three steps.

\begin{enumerate}
    \item The application of $\hat{H}_{\text{ODF}}$ for a time $\tau$ on an initially unentangled product state $\ket{\psi(0)}$ of the spins and phonons, which creates an entangled state with quantum-enhanced sensitivity:
    \begin{equation}
        \ket{\psi_\text{in}}=e^{-i\tau\hat{H}_{\text{ODF}}}\ket{\psi(0)}. \label{eqn:GenericPsiIn}
    \end{equation}
    This entangled state can be understood by considering the ideal initial condition of the main text, $\ket{\psi(0)}=\ket{\left( N/2 \right)_x}\otimes\ket{0}_{b}$ where $\ket{0}_{b}$ is the bosonic vacuum and coherent spin state, defined by $\hat{J}_\alpha\ket{m_\alpha}=m_\alpha\ket{m_\alpha}$. Then, from Eq.~(\ref{eqn:GenericPsiIn}) it is straightforward to obtain
    \begin{equation}
        \ket{\psi_\text{in}}=\sum_{m_z=-N/2}^{N/2}c_{m_z}\ket{\alpha_{m_z}}_{b}\otimes \ket{m_z},
    \end{equation}
    Here, $\ket{\alpha_{m_z}}_{b}$ is a coherent state with amplitude $\alpha_{m_z}=-im_z g\tau/\sqrt{N}$, while $c_{m_z}$ are expansion coefficients defined by $\ket{\left( N/2 \right)_x}=\sum_{m_z}c_{m_z}\ket{m_z}$. This is a type of generalized spin-phonon cat state, known to be useful for metrology \cite{Lewis-Swan2020,Zurek2001,Toscano_2006}. 
    \item The instantaneous application of a displacement $\beta\in\mathbb{R}$. This is equivalent to the application of the unitary transformation $\hat{U}_{\beta} = e^{\beta(\hat{a}-\hat{a}^{\dagger})}$ on $\ket{\psi_\text{in}}$, to give the intermediate state
    \begin{equation}
        \ket{\psi^\beta_{\text{in}}}=e^{\beta(\hat{a}^\dagger-\hat{a})}\ket{\psi_{\text{in}}}.
    \end{equation}
    \item Application of $-\hat{H}_{\text{ODF}}$ for a time $\tau$, a process that is  equivalent to time-reversal of the initial entangling operation, 
    \begin{align}\begin{split}
        \ket{\psi_f}&=e^{i\tau\hat{H}_{\text{ODF}}}\ket{\psi_\beta}=\sum_{m_z=-N/2}^{N/2}c_{m_z}e^{2i\beta\alpha_{m_z}}\ket{\beta}_{b}\otimes \ket{m_z}. \label{eqn:PsiFIdeal}
    \end{split}\end{align}
\end{enumerate}


The final state (\ref{eqn:PsiFIdeal}) can be written in an equivalent form that better elucidates how the time-reversal maps the displacement to an overall rotation of the spin degree of freedom. This result is actually general, independent of the initial spin-boson state. To show this, consider the final state at the end of the protocol 
\begin{equation}
    \ket{\psi_f}=e^{i\tau\hat{H}_{\text{ODF}}}e^{\beta(\hat{a}-\hat{a}^\dagger)}e^{-i\tau\hat{H}_{\text{ODF}}}\ket{\psi(0)}.
\end{equation}

Using that the ODF interaction is equivalent to an $J_z$-dependent displacement of the COM mode, $\exp(-i\tau\hat{H}_{\text{ODF}})=\mathcal{D}(-ig\tau\hat{J}_z)$ where $\mathcal{D}(\alpha) = e^{\alpha\hat{a}^{\dagger} - \alpha^*\hat{a}}$ is the usual bosonic displacement operator, combined with the identity $\mathcal{D}(\beta)^\dagger\hat{a}\mathcal{D}(\beta)=\hat{a}+\beta$ we can rewrite the last equation as
\begin{equation}\label{eqn:MethodsPsiFinalIdeal}
    \ket{\psi_f}=e^{\frac{2ig\tau\beta}{\sqrt{N}}\hat{J}_z}e^{\beta(\hat{a}^\dagger-\hat{a})}\ket{\psi(0)}.
\end{equation}
In this form it is clear that the spins and bosons will be completely disentangled at the conclusion of the protocol so long as the initial state $\ket{\psi(0)}$ is also unentangled. Moreover, the initial state of the bosons is irrelevant and the displacement manifests independently as: i) a displacement of the boson spin state and ii) a rotation of the initial spin state about $\hat{z}$ by an angle $\varphi=2g \tau\beta/\sqrt{N}$. 

By initializing the spins along the $\hat{x}$-axis, e.g., in the state $\ket{\left( N/2 \right)_x}$, we are thus able to observe this rotation through a measurement of simple spin observables. For example, a small displacement $\beta$ will lead to a small rotation that generates a nonzero value of $\braket{\hat{J}_y}_f\equiv\bra{\psi_f}\hat{J}_y\ket{\psi_f}$. 

The usefulness of this observable is determined by computing the expected signal $\braket{\hat{J}_y}_f$ at the end of the protocol and comparing it to its associated quantum variance $\text{var}(\hat{J}_y)_f$ computed also at the end of the protocol (note that both quantities are functions of $\beta$). This is quantified by the signal to noise ratio
\begin{equation}
    \text{SNR}=\frac{\braket{\hat{J}_y}_f}{\sqrt{\text{var}(\hat{J}_y)_f}}\approx \frac{\frac{\partial\braket{\hat{J}_y}_f}{\partial\beta}}{\sqrt{\text{var}(\hat{J}_y)_f}}\Bigg|_{\beta=0}\beta, 
\end{equation}
where $\beta=0$ is the working point of our protocol, having the property that $\braket{\hat{J}_y}_f|_{\beta=0}=0$. The inverse of the slope (with respect to $\beta$) of the SNR is the sensitivity $\Delta\beta$ with which $\beta$ can be estimated using $\hat{J}_y$ as a measurement observable, given by
\begin{equation}\label{eqn:MethodsIdealSensitivity}
    (\Delta\beta)^2\equiv\frac{\text{Var}(\hat{J}_y)_f}{\Big(\frac{\partial\braket{\hat{J}_y}_f}{\partial\beta}\Big)^2}\bigg|_{\beta=0}=\frac{1}{4g^2\tau^2}.
\end{equation}
This expression quantifies the ratio of: i) how fast the signal varies as a function of $\beta$ ($\partial_{\beta}\langle\hat{J}_y\rangle_f$) and ii) the fundamental noise present in any measurement of $\hat{J}_y$ due to quantum fluctuations [$\text{Var}(\hat{J}_y)_f$]. For interaction times $g\tau \gg 1$ the change in the signal, $\partial_{\beta}\langle\hat{J}_y\rangle_f = -g\tau\sqrt{N}$, is much larger than the intrinsic quantum noise $\sqrt{\text{Var}(\hat{J}_y)_f} \sim \sqrt{N}$ and therefore sensing displacements with $\beta<1/4$ (SQL) is possible. 
\section{Role of entanglement}
In this section we analyze the precise role that entanglement plays in the time reversal protocol. To keep the analysis simple, we work in the limit of large $N$, which we can do because of the many body nature of the system. Then, since the initial spin state is polarized along $+x$, we can approximate the spin degree of freedom by a canonically conjugate pair $(\hat{x}_s,\hat{p}_s)$ according to
\begin{equation}
    \hat{J}_x\approx N/2\hspace{2cm}\hat{J}_z\approx \sqrt{\frac{N}{2}}\,\hat{x}_s\hspace{2cm}\hat{J}_y\approx- \sqrt{\frac{N}{2}}\,\hat{p}_s,
\end{equation}
with the initial state $\ket{\left( N/2 \right)_x}$ corresponging to the bosonic vacuum $\ket{0}_s$. It is also convenient to analyze the COM using quadratures instead of creation and annihilation operators:
\begin{equation}
    \hat{x}_b=\frac{\hat{a}+\hat{a}^\dagger}{\sqrt{2}}\hspace{2cm}\hat{p}_b=\frac{\hat{a}-\hat{a}^\dagger}{i\sqrt{2}}.
\end{equation}
\subsection{Properties of the state $\ket{\psi_\text{in}}$}
\begin{figure}
    \centering
    \includegraphics[width=0.8\textwidth]{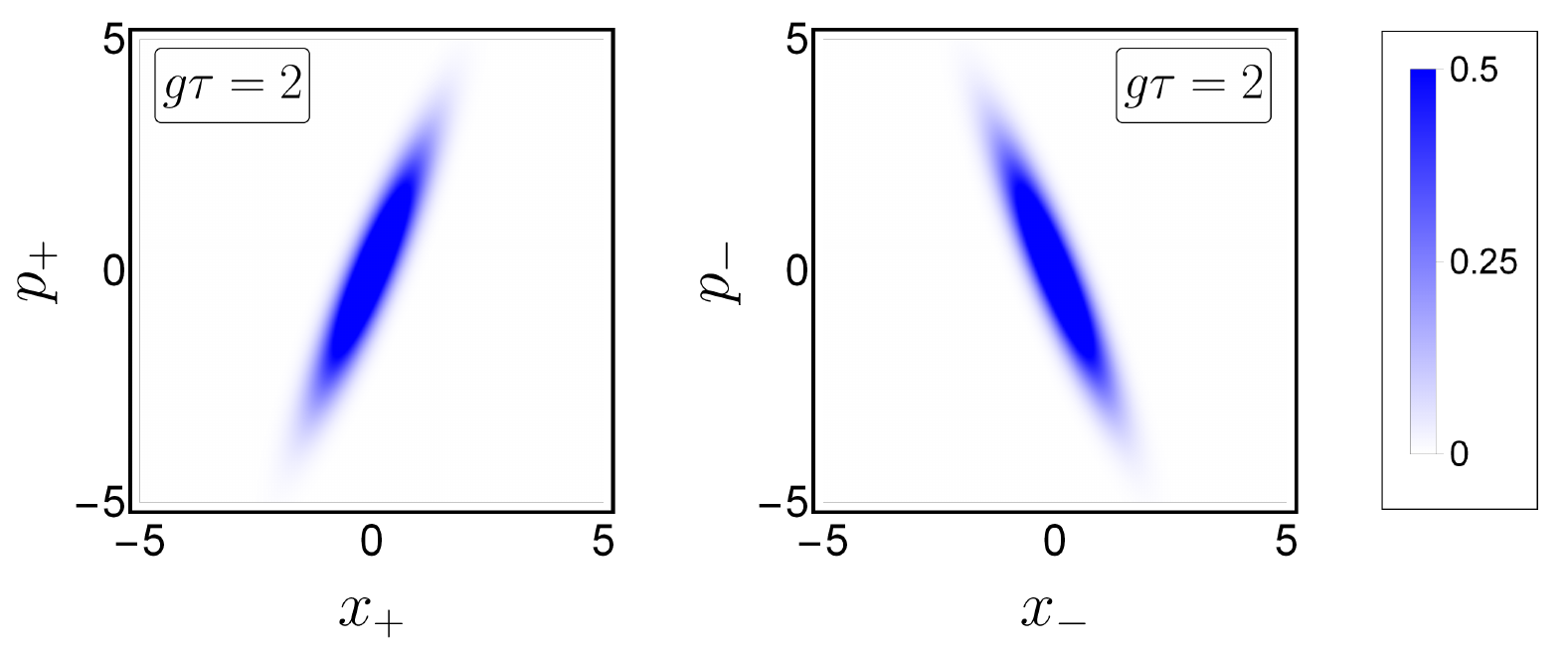}
    \caption{We show the Wigner function in the hybrid phase spaces $(\hat{x}_+,\hat{p}_+)$ (left) and $(\hat{x}_-,\hat{p}_-)$ (right) for the state $\ket{\psi_\text{in}}$, after the application of the first ODF pulse. In these variables the state is clearly squeezed, but the relation to the original spin and boson variables is not straightforward.}
    \label{fig:MethodsWignerFunctions}
\end{figure}

  Let's analyze the state after the first step [Eq.~(\ref{eqn:GenericPsiIn})], which is approximately given by
\begin{equation}\label{eqn:MethodsPsiInLargeN}
    \ket{\psi_\text{in}}=e^{-ig\tau \hat{x}_s\hat{x}_b}\ket{0}_b\otimes\ket{0}_s.
\end{equation}
While this evolution certainly mixes spins and bosons, it has a very simple description in terms of the hybrid quadratures
\begin{equation}
    \hat{x}_+=\frac{\hat{x}_s+\hat{x}_b}{\sqrt{2}}\hspace{2cm}\hat{x}_-=\frac{\hat{x}_s-\hat{x}_b}{\sqrt{2}},
\end{equation}
whereby 
\begin{equation}
    \ket{\psi_\text{in}}=e^{-i\frac{g\tau}{2}(\hat{x}_+^2-\hat{x}_-^2)}\ket{0}_+\otimes\ket{0}_-,
\end{equation}
and $\ket{0}_+\otimes\ket{0}_-=\ket{0}_s\otimes\ket{0}_b$ is the vacuum in either pair of variables, with wavefunction $\psi (0)\propto e^{-(x_+^2+x_-^2)/2}=e^{-(x_s^2+x_b^2)/2}$. Consider the $x_+$ quadrature. Since everything is gaussian, we can compute the Wigner function~\cite{walls_quantum_2008} at time $\tau$:
\begin{equation}
    W(x_+,p_+)=\frac{1}{\pi}\exp\big[-x_+^2-(p_+-g\tau x_+)^2\big]=\frac{1}{\pi}\exp\bigg[-\frac{\bar{x}^2}{\xi^2}-\xi^2\bar{p}^2\bigg],
\end{equation}
where 
\begin{align}\begin{split}
    \bar{x}=x_+\cos\phi+p_+\sin\phi\hspace{2cm}\bar{p}=p_+\cos\phi-x_+\sin\phi,
\end{split}\end{align}
are rotated quadratures about an angle $\phi>0$:
\begin{equation}
    \tan\phi=\frac{\sqrt{4+(g\tau)^2}-g\tau}{2},
\end{equation}
and $\xi^2$ is the squeezing parameter, given by 
\begin{equation}
    \xi^2=\frac{1}{1+g^2\tau^2+g\tau\sqrt{2+g^2\tau^2}}\approx \frac{1}{2g^2\tau^2}\hspace{1cm} g\tau\gg 1.
\end{equation}
This state is represented in Fig.~\ref{fig:MethodsWignerFunctions} (left) and shows squeezing along the $\bar{x}$ direction. We also show the Wigner function of the $(x_-,p_-)$ degree of freedom (right) which is just a mirror reflection of the previous one. 
A displacement of $x_b$, which is what we are trying to measure, manifests as mirror shifts of the Wigner functions in the horizontal direction. Taking advantage of this squeezing would require us to measure the $\bar{x}$ quadrature, which in terms of the original variables is very complicated indeed:
\begin{equation}
    \hat{\bar{x}}\approx \frac{\hat{x}_b\cos\phi+\hat{p}_b\sin\phi}{\sqrt{2}}+\frac{\hat{J}_z\cos\phi-\hat{J}_y\sin\phi}{\sqrt{N}}.
\end{equation}
Not only is this a time dependent observable ($\phi$ is a function of $\tau$), but also  we would  need to measure both bosons and spins, which is not possible with the current technical constraints of this experiment, which can only access the spins. Measuring only the spins immediately after the displacement of the bosons (implemented by $e^{\beta(\hat{a}^\dagger-\hat{a})}$) is useless because $e^{-\beta(\hat{a}^\dagger-\hat{a})}\hat{O}e^{\beta(\hat{a}^\dagger-\hat{a})}=\hat{O}$ for any spin observable $\hat{O}$. Measuring the bosons only is also useless, as we can see by computing their reduced density matrix $\hat{\rho}_B$:
\begin{equation}\label{eqn:MethodsBosonDensityMatrix}
    \braket{x_b|\hat{\rho}_B|x_b'}=\rho_B(x_b,x_b')=\frac{1}{\sqrt{\pi}}e^{-(x_b^2+x_b'^2)/2}e^{-(x_b-x_b')^2g^2\tau^2/4},
\end{equation}
whose Wigner function
\begin{equation}\label{eqn:MethodsReducedBosonWigner}
    W_S(x_b,p_b)=\frac{1}{\pi\sqrt{1+(g\tau)^2}}e^{-x_b^2}e^{-p_b^2/(1+g^2\tau^2)},
\end{equation}
shows only antisqueezing in the $p_b$ quadrature.
\subsection{Entanglement dynamics}
Given the crucial role that entanglement plays in the protocol, let us compute an entanglement measure as a function of the time $t$ in a sequence of duration $2\tau$. The state is given by
 \begin{equation}
    \ket{\psi_t}=\begin{cases}e^{-it\hat{H}_{\text{ODF}}}\ket{0}_{b}\otimes\ket{\left( N/2 \right)_x}&\text{if }0<t<\tau\\[5pt]
    e^{i(t-\tau)\hat{H}_{\text{ODF}}}\,e^{\beta(\hat{a}^\dagger-\hat{a})}\,e^{-i\tau\hat{H}_{\text{ODF}}}\ket{0}_b\otimes\ket{\left( N/2 \right)_x}&\text{if }\tau<t<2\tau\end{cases},
\end{equation}
and we choose the second Renyi entropy
\begin{equation}
    S_2=-\log[\mathrm{Tr}(\hat{\rho}_B^2)]
\end{equation}
as our entanglement proxy. The state during the first half ($t<\tau$) is of the form of $\ket{\psi_{\text{in}}}$ with the replacement $\tau\to t$, so the Renyi entropy can be directly calculated from the reduced density matrix [Eq.~(\ref{eqn:MethodsBosonDensityMatrix})] as
\begin{equation}\label{eqn:MethodsRenyiPsiIn}
    S_2(t)=-\log\bigg[\int \,dx_b\,dx_b'\,\rho_B(x_b,x_b')^2\bigg]=\frac{1}{2}\log(1+g^2t^2).
\end{equation}
For the second half ($\tau<t<2\tau$)
\begin{equation}
   \ket{\psi_t}= e^{i(t-\tau)\hat{H}_{\text{ODF}}}\,e^{\beta(\hat{a}^\dagger-\hat{a})}\,e^{-i\tau\hat{H}_{\text{ODF}}}\ket{0}_b\otimes\ket{\left( N/2 \right)_x}=e^{\beta(\hat{a}^\dagger-\hat{a})}\,e^{\frac{2ig\beta\hat{J}_x}{\sqrt{N}}}\,e^{-i(2\tau-t)\hat{H}_{\text{ODF}}}\ket{0}_b\otimes\ket{\left( N/2 \right)_x},
\end{equation}
where we have used again the fact $e^{-i\tau\hat{H}_{\text{ODF}}}$ is an $\hat{J}_z$ dependent displacement. Since the first two exponentials on the left act on independent degrees of freedom, they do not modify entanglement measures. Thus $S_2$ depends only on the rightmost exponential, whose effect we have already computed. Thus, we have that
\begin{equation}\label{eqn:MethodsRenyiEntropy}
    S_2(t)=\begin{cases}\frac{1}{2}\log\big[1+g^2t^2\big]^{-1/2}&\text{if } 0<t<\tau\\
    \frac{1}{2}\log\big[1+g^2(t-2\tau)^2\big]^{-1/2}&\text{if } \tau<t<2\tau\end{cases}.
\end{equation}
At then end ($t=2\tau$) there is no spin-boson entanglement, in agreement with Eq.~(\ref{eqn:MethodsPsiFinalIdeal}). The initial build up of entanglement ($t<\tau$) is crucial to keep the fluctuations $\text{Var}(\hat{J}_y)_f$ small at the end of the protocol. We explore the consequences of omitting this first step in the next subsection.
\subsection{Read-out protocol}\label{sec:Readout}
The protocol we will study in this subsection is characterized by omitting the first step of the full time reversal protocol described in the main text. Mathematically
\begin{equation}
    \ket{\psi_f'}=\underbrace{e^{i\tau\hat{H}_{\text{ODF}}}}_{\text{ODF interaction}}\hspace{0.2cm}\underbrace{e^{\beta(\hat{a}^\dagger-\hat{a})}}_{\text{Displacement}}\hspace{0.2cm}\underbrace{\ket{0}_b\otimes\ket{\left( N/2 \right)_x}}_{\text{Initial state}}.
\end{equation}
This has been implemented before for off-resonant electric field sensing~\cite{Gilmore2017,Affolter2020}, and the logic justifying this particular sequence of steps is semiclassical. The displacement operation transforms the bosonic vacuum into a coherent state: $\ket{0}_b\to\ket{\beta}_b$. Naively doing the replacement $\hat{a},\hat{a}^\dagger\to\beta$ in $\hat{H}_{\text{ODF}}$ then makes $\hat{H}_{\text{ODF}}\to 2g\beta\hat{J}_z/\sqrt{N}$ which generates a rotation of the spin degree of freedom. Measuring $\hat{J}_y$ should then give information about $\beta$. While this argument captures correctly the result of the measurement signal, it omits the fact that the entanglement created by $\hat{H}_{\text{ODF}}$ also increases the variance of $\hat{J}_y$, which we now calculate. Working again in the large $N$ limit, the evolution of the operator $\hat{J}_y$ in the Heisenberg picture is given by
\begin{align}\label{eqn:MethodsSyTime}
    \begin{split}
   e^{-\beta(\hat{a}^\dagger-\hat{a})} e^{-i\tau\hat{H}_{\text{ODF}}}\hat{J}_ye^{i\tau\hat{H}_{\text{ODF}}}e^{\beta(\hat{a}^\dagger-\hat{a})}&\longrightarrow -\sqrt{\frac{N}{2}} \Big(e^{i\beta\sqrt{2}\hat{p}_b}\Big)\Big(e^{-ig\tau\hat{x}_s\hat{x}_b}\Big)\,\hat{p}_s\Big(e^{ig\tau\hat{x}_s\hat{x}_b}\Big)\Big(e^{-i\beta\sqrt{2}\hat{p}_b}\Big)\\
   &=-\sqrt{\frac{N}{2}}e^{i\beta\sqrt{2}\hat{p}_b}(\hat{p}_s+g\tau\hat{x}_b)e^{-i\beta\sqrt{2}\hat{p}_b}\\
   &=-\sqrt{\frac{N}{2}}(\hat{p}_s+g\tau\hat{x}_b+g\tau\beta\sqrt{2}).
\end{split}
\end{align}
Since the initial state in this approximation is $\ket{0}_b\otimes\ket{0}_s$, it follows that
\begin{align}
    \begin{split}
        \braket{\hat{J}_y}_f&\approx -\sqrt{N}\beta g\tau\\
        \text{var}(\hat{J}_y)_f&\approx\frac{N}{4}(1+g^2\tau^2).
    \end{split}
\end{align}
We see that while the slope of the signal increases with $g\tau$, the variance of $\hat{J}_y$ also increases with $g\tau$ in a way that completely nullifies the effect of the increased signal. In fact, the sensitivity is
\begin{equation}
    (\Delta\beta)^2=\frac{1}{4g^2\tau^2}+\frac{1}{4}>\frac{1}{4},
\end{equation}
and it never gets below the SQL. As can be seen from the last line of Eq.~(\ref{eqn:MethodsSyTime}), this increase in variance is a result of the vacuum noise from the bosons being fed into the spins, which in turn is a consequence of the build up of entanglement.

\section{Generalized treatment of sensing with technical noise}
To incorporate the effects of various sources of technical noise and adequately describe both the displacement and electric field sensing protocols of the main text, we adopt a model based on the more general time-dependent Hamiltonian: 
\begin{equation}
    \hat{H}_{\eta}(t)=-\delta\hat{a}^\dagger\hat{a}+\frac{g(t)}{\sqrt{N}}(\hat{a}^\dagger+\hat{a})\hat{J}_z+i\eta(t)(\hat{a}^\dagger-\hat{a}). \label{eqn:HSensGeneral}
\end{equation}
Here, we have: i) included a time-dependent ODF interaction strength [$g(t)$] which can include, e.g., the sign change necessary for time-reversal, and ii) a time-dependent coherent drive of strength $\eta(t)$ to account for the concurrent application of the displacement and ODF drive in the electric field sensing protocol. Moreover, we assume that $\eta(t)$ is sufficiently small so that it can be accounted for within perturbation theory. Within Eq.~(\ref{eqn:HSensGeneral}), different sensing protocols simply correspond to different time profiles for $\eta(t)$ and $g(t)$. Thus, we will first work with a generic time dependence before specializing the obtained results to particular scenarios. 

The inclusion of the static detuning $\delta=\mu-\omega_z$, where $\mu$ is the ODF drive frequency and $\omega_z$ is the COM mode frequency, accounts for the observed deviations of this quantity away from the ideal $\delta = 0$ in the experiment. Specifically, we assume that $\delta$ fluctuates randomly between experimental trials but is fixed over the duration of a single experimental trial. We model $\delta$ as a Gaussian random variable with mean zero and variance $\sigma^2$. 

Experimentally, we also find the ions are not perfectly initialized in their motional ground-state. Limited by Doppler cooling, the COM mode is initially well described as a thermal bosonic state with an occupation $\bar{n}$. Combined with the spin-state after the initial microwave pulse, the initial state of the spin-phonon system is thus the density matrix $\hat{\rho}(0)=\rho_{\text{th}}(\bar{n})\otimes \ketbra{\left( N/2 \right)_x}{\left( N/2 \right)_x}$.
The final state after application of $\hat{H}_{\eta}$ for a time $t$ is then
\begin{equation}
   \hat{ \rho}_{\eta}(t)=\hat{U}_{\eta}(t)\hat{\rho}(0)U_{\eta}^\dagger(t),
\end{equation}
where $\hat{U}_{\eta}(t)$ is the time evolution operator associated to $\hat{H}_{\eta}(t)$. The achievable sensitivity $\Delta\eta$ to the applied drive is, for the specific observable $\hat{J}_y$ and at time $t$, constructed from two quantities: 
\begin{enumerate}
    \item $\text{Var}\big[\hat{J}_y(t)\big]$, the variance of the observable at the end of the evolution when $\eta(t)=0$ (equivalent to $\beta=0$ in the previous section).
    \item $\delta_{\eta}\braket{\hat{J}_y(t)}=\braket{\hat{J}_y(t)}_{\eta}-\braket{\hat{J}_y(t)}_{0}$, the change in the observable at the end of the protocol with respect to the unperturbed evolution to first order in $\eta(t)$ (equivalent to $\beta\partial_\beta\braket{\hat{J}_y}$ in the previous section).
\end{enumerate}
However, for calculational purposes it is more convenient to work with $\braket{\hat{J}_+(t)}$, $\braket{\hat{J}_+^2(t)}$ and $\delta_{\eta}\braket{\hat{J}_+(t)}$.

\subsection{Derivation}
The general results for $\text{Var}(\hat{J}_y(t))$ and $\delta\braket{\hat{J}_y(t)}$ are better expressed in terms of the Heisenberg picture operators
\begin{equation}
    \hat{J}_+(t)\equiv \hat{U}_0^\dagger(t)\hat{J}_+\hat{U}_0(t)\hspace{2cm}\hat{a}(t)\equiv \hat{U}_0^\dagger(t)\hat{a}\,\hat{U}_0(t).
\end{equation}
Note that this Heisenberg picture is defined in the absence of the perturbation ($\eta=0$). The relevant quantities can then be calculated using time-dependent perturbation theory with $\eta$ as a perturbation parameter:
\begin{align}\label{eqn:Signal&Variance}
    \begin{split}
        \braket{\hat{J}_+^n(t)}&=\mathrm{Tr}\Big\{\hat{J}_+^n(t)\hat{\rho}(0)\Big\}\\
        \delta_\eta\braket{\hat{J}_+(t)}&=-i\,\mathrm{Tr}\bigg\{\hat{\rho}(0)\int_0^t i\eta(s)\,ds\,\Big[\hat{J}_+(t),\hat{a}(s)^\dagger-\hat{a}(s)\Big]\bigg\},
    \end{split}
\end{align}
with $n=1,2$.
The evolution of the operators $\hat{J}_+(t)$ and $\hat{a}(t)$ is given by the following Heisenberg equations of motion:
\begin{align}
    i\dot{\hat{a}}(t)&=i\delta\hat{a}(t)-\frac{ig(t)}{\sqrt{N}}\hat{J}_z\\
    i\dot{\hat{J}}_+(t)&=-\frac{g(t)}{\sqrt{N}}\big[\hat{a}(t)^\dagger+\hat{a}(t)\big]\hat{J}_+(t).
\end{align}
As $\hat{J}_z$ is a conserved quantity the equation for $\hat{a}(t)$ is closed. After formal solution of $\hat{a}(t)$ we subsequently solve for $\hat{J}_+(t)$. The final result is
\begin{align}\begin{split}\label{eqn:HESolutions}
    \hat{a}(t)&=\hat{a}e^{i\delta t}-\frac{ih(t) e^{i\delta t}}{\sqrt{N}}\hat{J}_z\\
    \hat{J}_+(t)&=\exp\bigg[\frac{i}{\sqrt{N}}(h(t)\hat{a}^\dagger+\bar{h}(t)\hat{a})\bigg]\exp\bigg[-\frac{2ip(t)}{\sqrt{N}}\bigg(\hat{J}_z+\frac{1}{2}\bigg)\bigg]\hat{J}_+,
\end{split}\end{align}
where we have introduced the functions
\begin{align}
    \begin{split}
        h(t)&=\int_0^tg(s)e^{-i\delta s}\,ds ,\hspace{2cm}
        p(t)=\frac{i}{2}\int_0^t(\dot{\bar{h}}(s)h(s)-\bar{h}(s)\dot{h}(s))\,ds.
    \end{split}
\end{align}
Expectation values are taken with respect to the initial state, yielding,
\begin{align}\begin{split}\label{eqn:ExpectationValues}
    \braket{\hat{J}_+(t)}&=\frac{N}{2}\exp\bigg[-\frac{|h(t)|^2}{N}\bigg(\bar{n}+\frac{1}{2}\bigg)\bigg]\cos\bigg(\frac{p(t)}{N}\bigg)^{N-1} , \\[10pt]
    \braket{\hat{J}_+^2(t)}&=\frac{N(N-1)}{4}\exp\bigg[-\frac{4|h(t)|^2}{N}\bigg(\bar{n}+\frac{1}{2}\bigg)\bigg]\cos\bigg(\frac{2p(t)}{N}\bigg)^{N-2}.
\end{split}\end{align}
The last thing we need is $\delta\braket{\hat{J}_+(t)}$, which can be calculated using Eq.~(\ref{eqn:Signal&Variance}), the solutions in Eq.~(\ref{eqn:HESolutions}) and systematic application of commutation relations:
\begin{equation}
    \delta_{\eta}\braket{\hat{J}_+(t)}=i\sqrt{N}q(t)\exp\bigg[-\frac{|h(t)|^2}{N}\bigg(\bar{n}+\frac{1}{2}\bigg)\bigg]\cos\bigg(\frac{p(t)}{N}\bigg)^{N-1} ,
\end{equation}
where 
\begin{equation}
    q(t)=\int_0^t\int_0^s g(s)\eta(u)\cos[\delta(u-s)]\,du\,ds.
\end{equation}
The previous formulas include all the Hamiltonian effects we will take into account. However, the system is also subject to light scattering whenever the ODF interaction is on, and is modelled by the following dissipative Lindbladian:
\begin{equation}
    \mathcal{L}\hat{\rho}=\frac{\Gamma}{4}\sum_{i=1}^N\Big(\hat{\sigma}_z^i\hat{\rho}\hat{\sigma}_z^i-\hat{\rho}\Big).
\end{equation}
Since the Hamiltonian evolution commutes with $\mathcal{L}$ (in the superoperator sense), the consequences of this dissipative process can be calculated at the end and independently from the effects of $\hat{H}$ \cite{Rey_2008}. Thus, we need to replace $\braket{\hat{J}_+^n(t)}\to \braket{\hat{J}_+^n(t)}e^{-n\Gamma t_{\text{odf}}/2}$, where $t_{\text{odf}}$ is the amount of time the ODF is on. In addition, the observable $\braket{\hat{J}_+\hat{J}_-+\hat{J}_-\hat{J}_+}/2=N(N+1)/4$, which was conserved throughout the protocol in the absence of $\Gamma$, becomes
\begin{equation}
    \frac{1}{2}\braket{\hat{J}_+\hat{J}_-+\hat{J}_-\hat{J}_+}\to\frac{N}{2}+\frac{N(N-1)}{4}e^{-\Gamma t_{\text{odf}}}
\end{equation}
\subsection{Summary of results}
Using the definition of $\hat{J}_y$ in terms of $\hat{J}_+$ and $\hat{J}_-$ we thus find that
\begin{align}\begin{split}\label{eqn:MethodsSummaryofResults}
    \braket{\hat{J}_y(t)}&=0\\
    \braket{\hat{J}_y^2(t)}&=\frac{N}{4}+\frac{N(N-1)e^{-\Gamma t_{\text{odf}}}}{8}\Bigg\{1-\exp\bigg[-\frac{4|h(t)|^2}{N}\bigg(\bar{n}+\frac{1}{2}\bigg)\bigg]\cos\bigg(\frac{2p(t)}{N}\bigg)^{N-2}\Bigg\}\\
    \delta_{\eta}\braket{\hat{J}_y(t)}&=q(t)\sqrt{N}e^{-\Gamma t_{\text{odf}}/2}\exp\bigg[-\frac{|h(t)|^2}{N}\bigg(\bar{n}+\frac{1}{2}\bigg)\bigg]\cos\bigg(\frac{p(t)}{N}\bigg)^{N-1}\\
    \braket{\hat{J}_x(t)}&=\frac{Ne^{-\Gamma t_{\text{odf}}/2}}{2}\exp\bigg[-\frac{|h(t)|^2}{N}\bigg(\bar{n}+\frac{1}{2}\bigg)\bigg]\cos\bigg(\frac{p(t)}{N}\bigg)^{N-1}
\end{split}\end{align}
We see that $h(t)$ is associated to thermal and zero-point bosonic motion, $p(t)$ is a squeezing contribution that increases the variance and $q(t)$ is some kind of detuning averaged total displacement. We also show $\braket{\hat{J}_x(t)}$ because it is necessary for the experimental calibration of $\sigma$, the COM frequency variance.
\subsection{Displacement sensing}
\begin{figure}
     \centering
    \includegraphics[width=0.45\textwidth]{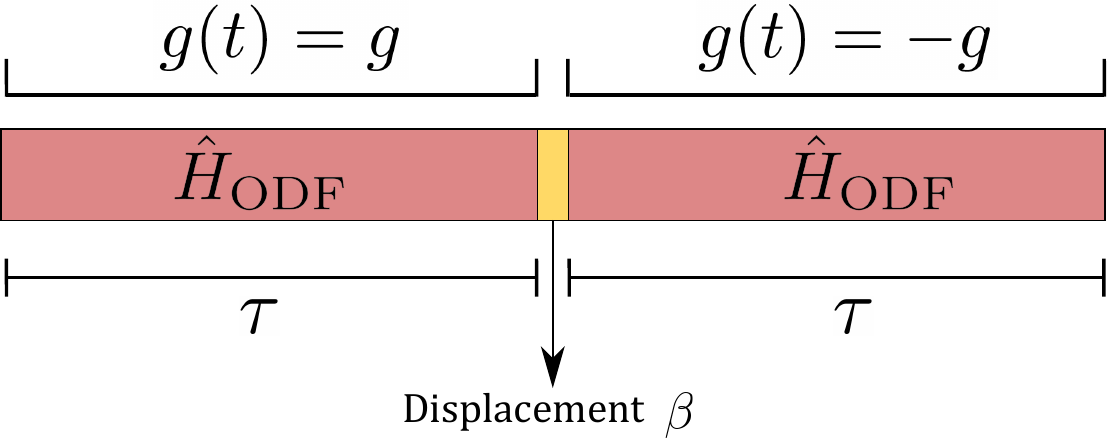}\hspace{1cm}
     \includegraphics[width=0.45\textwidth]{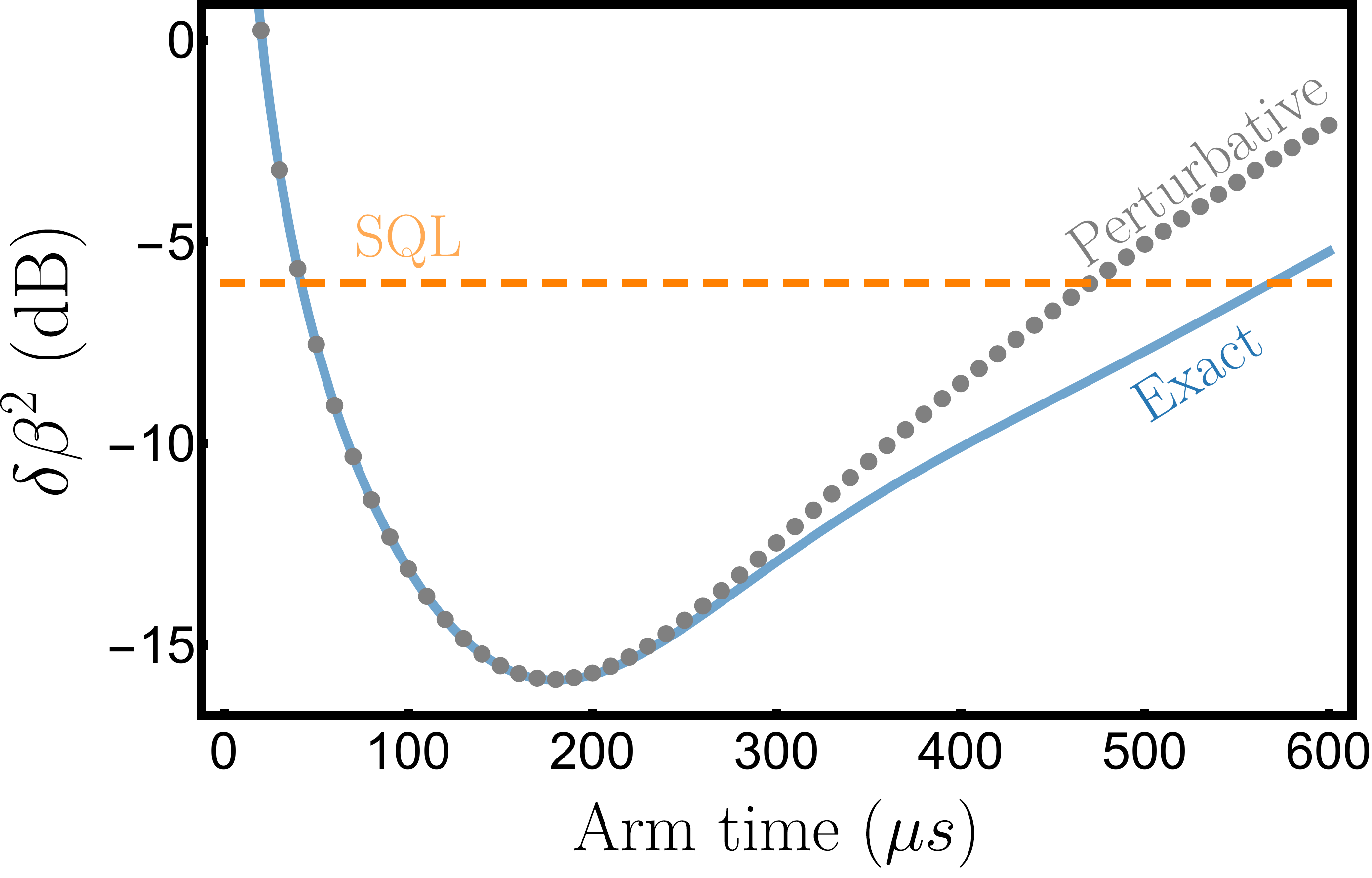}
    \caption{(Left) Schematic of the time-reversal protocol for displacement sensing. The ODF is turned on for a time $\tau$, after which a displacement $\beta$ is quickly applied. Finally, the second ODF pulse is applied for a time $\tau$. (Right) Sensitivity $(\delta\beta)^2$ for experimentally realistic parameters $N=150$, $g=2\pi\times 3.91$ kHz, $\bar{n}=5$, $\Gamma_{\text{el}}=610\text{ s}^{-1}$ and $\sigma=2\pi\times 40$ Hz. We compare the exact result obtained from numerics (light blue) versus the perturbative expression Eq.~(\ref{eqn:MethodsDisplacementSensitivity}) (gray) and see that the optimal sensitivity is captured correctly by the perturbative result.}
    \label{fig:MethodsDisplacementSchematic}
\end{figure}

The scheme is shown in Fig. \ref{fig:MethodsDisplacementSchematic}, with total protocol duration $2\tau$. The time profiles are $g(t)=g\,\text{sign}(t-\tau)$, where the sign function accounts for the time reversal step in the latter half of the protocol ($\tau<t<2\tau$), and $\eta(t)=\beta\delta(t-\tau)$ indicates that the drive instantaneously generates a displacement $\beta$ just before the time reversal step ($t=\tau$). Plugging this into the definitions of $h(t)$, $p(t)$ and $q(t)$ we obtain
\begin{align}\begin{split}\label{eqn:MethodsDisplacementParameters}
    |h{(2\tau)}|^2=\frac{16g^2}{\delta^2}\sin\bigg(\frac{\delta \tau}{2}\bigg)^4\hspace{1.5cm} p{(2\tau)}=\frac{g^2}{\delta^2}\big[4\sin(\delta \tau)-\sin(2\delta \tau)-2\delta \tau\big]\hspace{1.5cm} q{(2\tau)}=-\frac{\beta g}{\delta}\sin(\delta \tau).
\end{split}\end{align}
We then numerically average $\braket{\hat{J}_y(2\tau)}$ and $\delta\braket{\hat{J}_y(2\tau)}$ over $\delta$ and calculate the sensitivity according to 
\begin{equation}
    (\Delta\beta)^2=\frac{\braket{\hat{J}_y^2(2\tau)}^{\text{av}}}{\big(\frac{\partial\braket{\hat{J}_y(2\tau)}^{\text{av}}}{\partial\beta}\big)^2},
\end{equation}
where $\braket{...}^{\text{av}}$ indicates the average over detunings. For short protocol times, $2\sigma \tau\ll 1$, we can calculate perturbatively the effect of COM fluctuations. Keeping terms up to second order in $\delta$ and lowest order in $1/N$, we obtain after averaging
\begin{align}\begin{split}
    \braket{\hat{J}_y^2(2\tau)}^{\text{av}}&\approx\frac{N}{4}+Ng^2\sigma^2\tau^4e^{-2\Gamma \tau}\bigg(\frac{\bar{n}+1/2}{2}+\frac{g^2\tau^2}{9}\bigg)\\
    \frac{\delta_\eta\braket{\hat{J}_y(2\tau)}^{\text{av}}}{\beta}&\approx -\sqrt{N}g\tau e^{-\Gamma \tau}\bigg(1-\frac{\sigma^2\tau^2}{6}\bigg).
\end{split}\end{align}
Hence the sensitivity is
\begin{equation}\label{eqn:MethodsDisplacementSensitivity}
    (\Delta\beta)^2\approx\underbrace{\frac{e^{2\Gamma \tau}}{4g^2 \tau^2}}_{\text{Ideal result}}+\underbrace{\frac{\sigma^2e^{2\Gamma \tau}}{12g^2}}_{\text{Signal reduction}}+\underbrace{\frac{\sigma^2\tau^2(\bar{n}+1/2)}{2}}_{\text{Spin-phonon correlations}}+\underbrace{\frac{g^2\sigma^2\tau^4}{9}}_{\text{Spin-spin correlations}}.
\end{equation}

We have indicated the nature of each of the terms involved in the sensitivity. The first term corresponds (when setting $\Gamma=0$) to the ideal result obtained in Eq.~(\ref{eqn:MethodsIdealSensitivity}). The next term arises from a reduction in the signal due to the presence of $\sigma$. Nevertheless, it usually is the smallest contribution. More relevant is the third term, which is a manifestation of the remaining spin-phonon correlations at the end of the protocol as a consequence of the imperfect echo. It has the typical $\bar{n}+1/2$ factor associated to the size of the boson wavefunction. Finally, the imperfect echo also results in spin squeezing, which creates spin-spin correlations and an increase in the variance of $\hat{J}_y$. This is the content of the last term. We compare Eq.~(\ref{eqn:MethodsDisplacementSensitivity}) against exact numerics for the sensitivity in Fig.~\ref{fig:MethodsDisplacementSchematic} (right panel). There is good agreement at short times and, importantly, the optimal sensitivity is captured correctly. Discrepancies appear only at longer times. 

From Eq.~(\ref{eqn:MethodsDisplacementSensitivity}) we obtain Eq.~(5) in the main text by neglecting the signal reduction term, which we find does not affect any of the results presented. The application of the perturbative formulas is justified for Fig. 3(a) in the main text since the signal-to-noise ratio is computed at the time of optimal sensitivity.
\subsection{Electric Field Sensing}
Here, we are interested in sensing the drive strength itself [$\eta(t)$ in Eq.~(\ref{eqn:HSensGeneral})] since it is proportional to the applied electric field. We want the drive to act for as long as possible, so we keep $\eta(t)=\eta$ constant throughout the full evolution and add ODF evolutions for short spans of time at the end and/or the beginning of the sensing protocol. We will consider two schemes here, which we dub as ``Classical Protocol" and ``Quantum Protocol". In the former, the ODF interaction is only used as a readout mechanism at the end of the sequence, closely following the discussion in Sec.~\ref{sec:Readout}, thus building spin-phonon entanglment even under ideal conditions and limiting the achievable sensitivity. We improve this situation in the ``Quantum Protocol" by adding an extra ODF pulse at the beginning that creates an initially entangled state, and allows for an enhancement in sensitivity.
\subsubsection{Classical Protocol}
\begin{figure}
    \centering
    \includegraphics[width=0.42\textwidth]{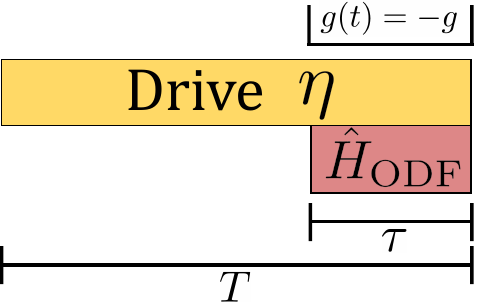}\hspace{2cm}
    \includegraphics[width=0.42\textwidth]{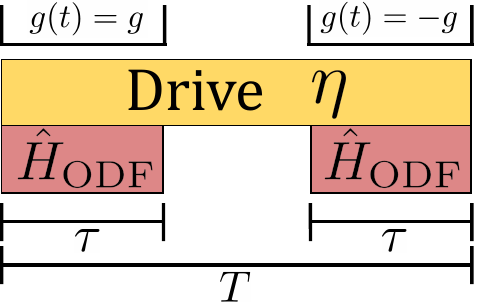}
    \caption{Theoretical schematics of the electric field sensing protocols. The classical protocol is shown in the left, with a drive $\eta$ applied for a time $T$ and the ODF interaction used as a read-out and applied for a time $\tau$ in the end. The quantum protocol is shown in the right and includes an extra application of the ODF Hamiltonian for a time $\tau$ at the beginning.}
    \label{fig:MethodsElectricFieldSensingProtocol}
\end{figure}

The scheme is shown in Fig. \ref{fig:MethodsElectricFieldSensingProtocol} (left) with total protocol duration $T$ and the ODF interaction is only applied at the end of the protocol for a time $\tau$. Again, we calculate $h(T)$, $p(T)$ and $q(T)$:
\begin{equation}
    |h(T)|^2=\frac{4g^2}{\delta^2}\sin\bigg(\frac{\delta\tau}{2}\bigg)^2\hspace{1.2cm}p(T)=\frac{g^2}{\delta^2}[\sin(\delta\tau)-\delta\tau]\hspace{1.2cm}    q(T)=\frac{\eta g}{\delta^2}\big\{\cos(\delta T)-\cos\big[\delta(T-\tau)\big]\big\}
\end{equation}
As in the previous case, we numerically average the expectation values over $\delta$ and calculate the sensitivity according to
\begin{equation}
    (\Delta\eta)^2=\frac{\braket{\hat{J}_y^2(T)}^{\text{av}}}{\big(\partial\braket{\hat{J}_y(T)}^{\text{av}}/\partial\eta\big)^2}.
\end{equation}
To second order in $\sigma$, we find after averaging that
\begin{align}\begin{split}
    \braket{\hat{J}_y^2(T)}^{\text{av}}&\approx\frac{N}{4}+\frac{Ng^2\tau^2e^{-\Gamma \tau}}{8}\bigg[4(\bar{n}+1/2)-\frac{\sigma^2\tau^2}{3}(\bar{n}+1/2)+\frac{g^2\sigma^2\tau^4}{18}\bigg]\\
    \frac{\delta_\eta\braket{\hat{J}_y(T)}^{\text{av}}}{\eta}&\approx \frac{\sqrt{N}e^{-\Gamma \tau/2}g\tau(\tau-2T)}{2}\bigg[1-\frac{\sigma^2}{12}(2T^2-2\tau T+\tau^2)\bigg].
\end{split}\end{align}
Hence the sensitivity is
\begin{align}\begin{split}\label{eqn:MethodsSensitivityClassicalProtocol}
    (\delta\eta)^2_{\text{classical}}&\approx\underbrace{\frac{e^{\Gamma \tau}}{g^2 \tau^2(2T-\tau)^2}}_{\text{Ideal result}}\hspace{1cm}+\underbrace{\frac{2\bar{n}+1}{(2T-\tau)^2}}_{\text{Ideal spin-phonon correlations}}\\[10pt]
    &+\underbrace{\frac{\sigma^2e^{\Gamma \tau}(2T^2-2\tau T+\tau^2)}{6g^2\tau^2(2T-\tau)^2}}_{\text{Signal reduction}}\hspace{0.5cm}+\underbrace{\frac{\sigma^2(\bar{n}+1/2)}{6}}_{\text{Spin-phonon correlations}}+\underbrace{\frac{g^2\sigma^2\tau^4}{36(2T-\tau)^2}}_{\text{Spin-spin correlations}}.
\end{split}\end{align}
As in the analysis of the displacement protocol, we have separated the different contributions to the sensitivity according to their origin. The first line includes effects present in the absence of COM frequency fluctuations ($\sigma=0$): the first term is the increase in sensitivity due to the acquisition of a bigger signal. However, as discussed in section~\ref{sec:Readout}, this is accompanied by an increase in variance at the end of the protocol due to spin-phonon entanglement, and this creates the second term in the first line. The second line includes the effects due to COM frequency fluctuations and are of a similar nature as those present in the displacement protocol. Note that
\begin{equation}\label{eqn:MethodsEfieldSQL}
    (\delta\eta)^2_{\text{classical}}>\frac{1}{4T^2}=(\delta\eta)^2_{\text{SQL}},
\end{equation}
and so we never beat the SQL with this scheme. The achievable sensitivity is even much worse:
\begin{equation}\label{eqn:MethodsEfieldThermalSQL}
    (\delta\eta)^2_{\text{classical}}>\frac{2\bar{n}+1}{4T^2}=(\delta\eta)^2_{\text{th}},
\end{equation}
which includes thermal effects. We thus see that thermal occupation of the bosons very negatively impacts the ``Classical Protocol".

\begin{figure}
    \centering
    \includegraphics[width=1.0\textwidth]{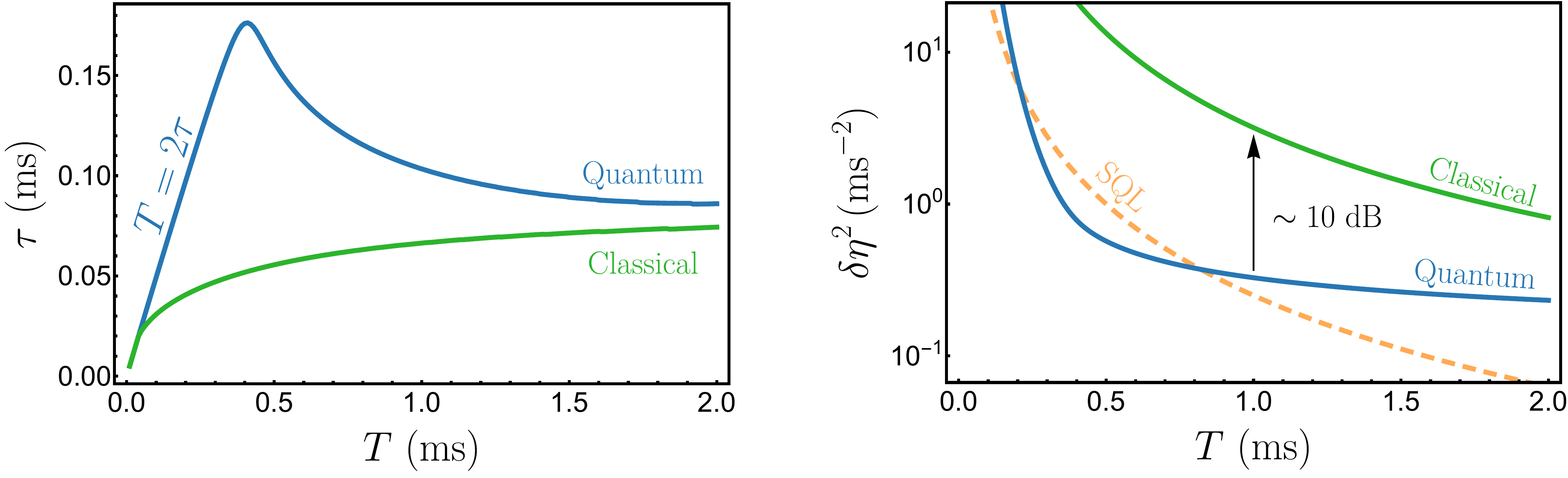}
    \caption{We show the optimal ODF interaction time $\tau$ (left) and the resulting optimal sensitivity (right) as a function of the protocol time $T$ for $g=2\pi\times 3.88$ kHz, $N=150$, $\bar{n}=5$, $\Gamma=520\text{ s}^{-1}$, and $\sigma=2\pi\times 40$ Hz. For short $T$, $\tau$ is exactly half of the protocol time, especially in the quantum protocol. For large $T$, $\tau$ saturates. The right panel shows that the classical protocol is generically about 10 dB worse than the quantum one. We also include the SQL for comparison, defined as $\delta\eta^2_{\text{SQL}}=1/(4T^2)$, and note that the quantum protocol does get below it.}
    \label{fig:MethodsQCLSensitivity}
\end{figure}

\subsubsection{Quantum Protocol}
The scheme is shown in Fig.~\ref{fig:MethodsElectricFieldSensingProtocol} (right), with total protocol duration $T$ and the ODF interaction is applied both at the beginning and at the end of the protocol for a time $\tau$ in each section. As in the previous case, $\eta(t)=\eta$, a constant throughout the evolution. Again, we calculate $h(T)$, $p(T)$ and $q(T)$:
\begin{align}\begin{split}
    |h(T)|^2&=\frac{16g^2}{\delta^2}\sin\bigg(\frac{\delta \tau}{2}\bigg)^2\sin\bigg[\frac{\delta(T-\tau)}{2}\bigg]^2\\[8pt]
    p(T)&=-\frac{2g^2}{\delta^2}\bigg\{\delta \tau -\sin(\delta\tau)-2\sin\bigg(\frac{\delta \tau}{2}\bigg)^2\sin[\delta(T-\tau)]\bigg\}\\[8pt]
    q(T)&=-\frac{4g\eta}{\delta^2}\sin\bigg(\frac{\delta\tau}{2}\bigg)\sin\bigg[\frac{\delta(T-\tau)}{2}\bigg]\cos\bigg(\frac{\delta T}{2}\bigg).
\end{split}\end{align}
As in the previous case, we numerically average the expectation values over $\delta$ and calculate the sensitivity according to
\begin{equation}
    (\delta\eta)^2=\frac{\braket{\hat{J}_y^2(T)}^{\text{av}}}{\big(\partial\braket{\hat{J}_y(T)}^{\text{av}}/\partial\eta\big)^2}.
\end{equation}
To second order in $\sigma$, we find after averaging that
\begin{align}\begin{split}
    \braket{\hat{J}_y^2(T)}^{\text{av}}&\approx\frac{N}{4}+\frac{Ng^2\sigma^2\tau^2e^{-2\Gamma \tau}}{8}\bigg[4(\bar{n}+1/2)(T-\tau)^2+2g^2\tau^2\bigg(T-\frac{4\tau}{3}\bigg)^2\bigg]\\
    \frac{\delta_\eta\braket{\hat{J}_y(T)}^{\text{av}}}{\eta}&\approx -\sqrt{N}e^{-\Gamma \tau}g\tau(T-\tau)\bigg[1-\frac{\sigma^2}{12}(2T^2-\tau T+\tau^2)\bigg].
\end{split}\end{align}
Hence the sensitivity is
\begin{align}\begin{split}
    (\delta\eta)^2_{\text{quantum}}&\approx\underbrace{\frac{e^{2\Gamma \tau}}{4g^2 \tau^2(T-\tau)^2}}_{\text{Ideal result}}\\[10pt]
    &+\underbrace{\frac{\sigma^2e^{2\Gamma \tau}(2T^2-\tau T+\tau^2)}{24g^2\tau^2(T-\tau)^2}}_{\text{Signal reduction}}+\underbrace{\frac{\sigma^2(2\bar{n}+1)}{4}}_{\text{Spin-phonon correlations}}+\underbrace{\frac{g^2\sigma^2\tau^2(3T-4\tau)^2}{36(T-\tau)^2}}_{\text{Spin-spin correlations}}.
\end{split}\end{align}
We have again included the effects of COM frequency fluctuations only in the second line. Note that, in contrast to the ``Classical Protocol", the ideal result is not bounded (when $\Gamma=0$) and thus can surpass the SQL [Eq.~(\ref{eqn:MethodsEfieldSQL})] and the thermal bound [Eq.~(\ref{eqn:MethodsEfieldThermalSQL})], irrespective of the thermal occupation of the bosons. The COM frequency dependent contributions are of a similar nature as those in the displacement sensing protocol.

In both the classical and quantum electric field sensing protocols we obtain the sensitivity as a function of two variables, $T$ and $\tau$. For each $T$ we numerically optimize over $\tau$ for realistic values of $g$, $N$ and $\sigma$. We show the optimal value of $\tau$ as a function of $T$ in Fig.~\ref{fig:MethodsQCLSensitivity} (left) and the optimal sensitivity in Fig.~\ref{fig:MethodsQCLSensitivity} (right). For the parameters used in the main text, the perturbative expressions for $(\delta\eta)_{\text{classical}}$ and $(\delta\eta)_{\text{quantum}}$ are accurate only for $T\leq 0.6$ ms so the construction of the theory curves in Fig.~4(b) of the main text required us to use the full numerical treatment.

\section{Experimental details}

Here we outline some details of the Penning trap system used for the measurements presented in the main text \cite{Biercuk2009, Affolter2020}. Subsequent sections in this supplementary material will address experimental calibrations and data analysis. 

\begin{figure}
    \centering
    \includegraphics[scale = 0.5]{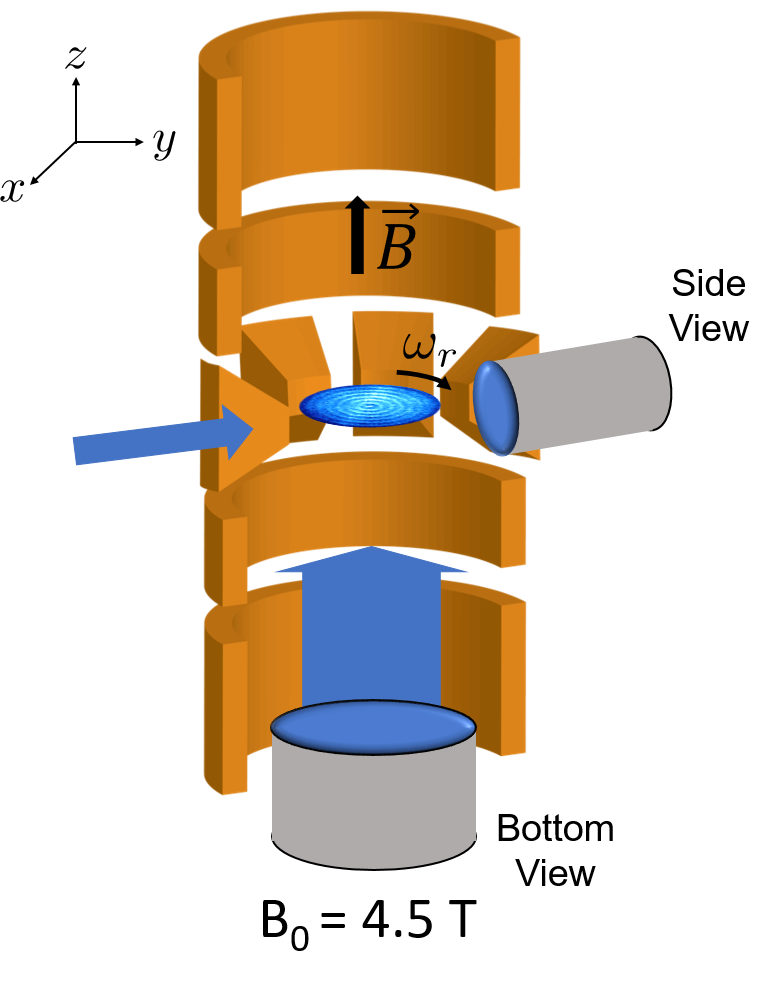}
  \caption[Trap]{Illustration of the Penning trap electrodes and imaging. The view is cross-sectional and the electrodes are cylindrically symmetric. Doppler cooling is performed along the trap axis and perpendicular to it. Photons are imaged on to the `side view' PMT for global state-dependent readout of the spins.}\label{trap_imaging}
\end{figure}

The Penning trap consists of a stack of cylindrical electrodes (see cross-sectional schematic in Fig. \ref{trap_imaging}) which are used to generate an axially confining electrostatic potential. The trap is inserted into the bore of a $B= 4.5$ T superconducting magnet oriented vertically such that the magnetic field is aligned with the $\hat{z}-$axis defined by the trap. Radial confinement of the ions results from the $E \times B$ induced rotation of the ions at a frequency $\omega_r$. In a frame rotating at $\omega_r$, the confining potential is

\begin{equation}
    q \Phi_{trap} = \frac{1}{2} M \omega_z^2 (Z^2 + \beta_r r^2),
\end{equation}
where $Z $ is the axial distance and $r$ the cylindrical radius from the trap center, $w_z/(2\pi) = 1.59$ MHz is the COM mode frequency, which characterizes the strength of the confining potential, and $\beta_r = \frac{\omega_r(\Omega_c - \omega_r)}{\omega_z^2} - \frac{1}{2}$ is the relative strength of the radial confinement for rotation frequency $\omega_r$ and cyclotron frequency $\Omega_c/(2\pi) = 7.6$ MHz. The rotation frequency of the crystal is precisely  with a rotating dipole electric field \cite{Huang1998b} and, for the $N \approx 150$ ions typical of the work described in the main text, set to $\omega_r/(2\pi) \approx 180$ kHz. At this rotation frequency, the crystal is a 2-D single-plane array of ions.

\begin{figure}
    \centering
    \includegraphics[scale = 0.5]{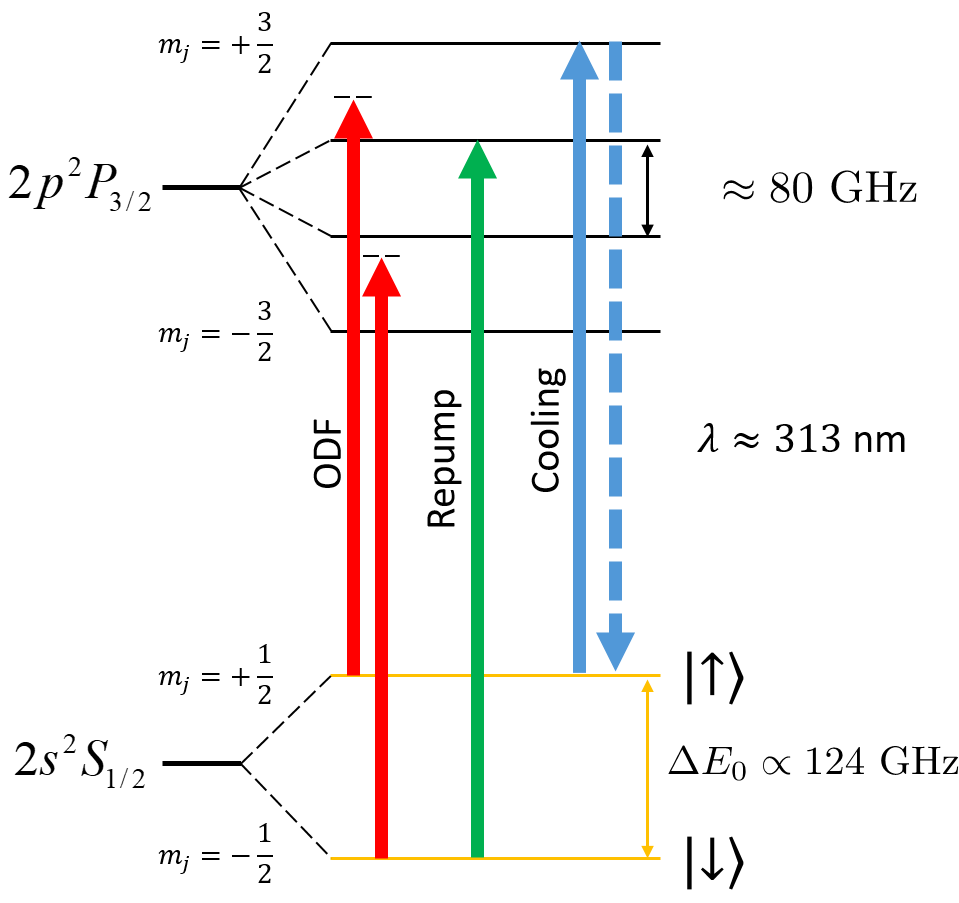}
  \caption[Levels]{Relevant energy levels for $^{9}$Be$^{+}$ at $B = 4.5 $ T. The splitting between $\ket{\uparrow}$ and $\ket{\downarrow}$ is 124 GHz. All lasers are in the UV with $\lambda \approx 313$ nm. Doppler cooling and state-dependent projective measurements are performed on the cycling transition $\ket{\uparrow} \rightarrow {}^2P_{3/2}(m_J=+3/2)$. A repump laser pumps spin population out of the $\ket{\downarrow}$ state and prepares the state $\ket{\uparrow}$ with high fidelity. A pair of off-resonant lasers couple the qubit states to states in the ${}^2P_{3/2}$ manifold, producing a spin-dependent optical-dipole force.}\label{be_levels}
\end{figure}

Figure \ref{be_levels} shows the relevant energy levels for $^{9}$Be$^{+}$ at high magnetic field ($4.5 ~T$). Though $^{9}$Be$^{+}$ has nuclear spin $I=3/2$, we optically pump the nuclear spin to the $m_I = +3/2$ level, where it remains throughout the duration of an experiment and which allows for simplifying the level structure as shown in Fig. \ref{be_levels}. We define the qubit states as two levels in the $S_{1/2}$ manifold:  $\ket{\uparrow} \equiv \ket{m_J = +1/2}$ and $\ket{\downarrow} \equiv \ket{m_J = -1/2}$. The qubit is, essentially, the ${}^9\mathrm{Be}^+$ valence electron spin in the ground state. For a magnetic field of $B = 4.46$ T, the splitting between the qubit states $\ket{\uparrow}$ and $\ket{\downarrow}$ is approximately 124 GHz. A low phase noise microwave source at 124 GHz allows for global rotations of the spins. The duration of a $\pi-$pulse (the time to flip $\ket{\downarrow}$ to $\ket{\uparrow}$) is $45~\mu$s. 

Doppler cooling and state-selective readout is performed on the cycling transition $\ket{\uparrow} \rightarrow {}^2P_{3/2}(m_J=+3/2)$. State initialization is performed with optical pumping into the $\ket{\uparrow}$ state with the Doppler cooling and repump lasers (see Fig. \ref{be_levels}). For the work presented here, all readout is made via projective measurements of the ion spin states through state-dependent global resonance fluorescence. Photons are collected on a photomultiplier tube (PMT) from the global fluorescence of ions in the $\ket{\uparrow}$ state in a direction perpendicular to the magnetic field (the `side view' in Fig. \ref{trap_imaging}). The PMT counts are integrated over $2.5$ ms to reduce the impact of photon shot noise.

\section{Experimental calibrations}

This section will address the various experimental calibrations that provide input to the theoretical expressions.

\subsection{Applied electric field calibration}

\begin{figure}
    \centering
    \includegraphics[scale = 0.5]{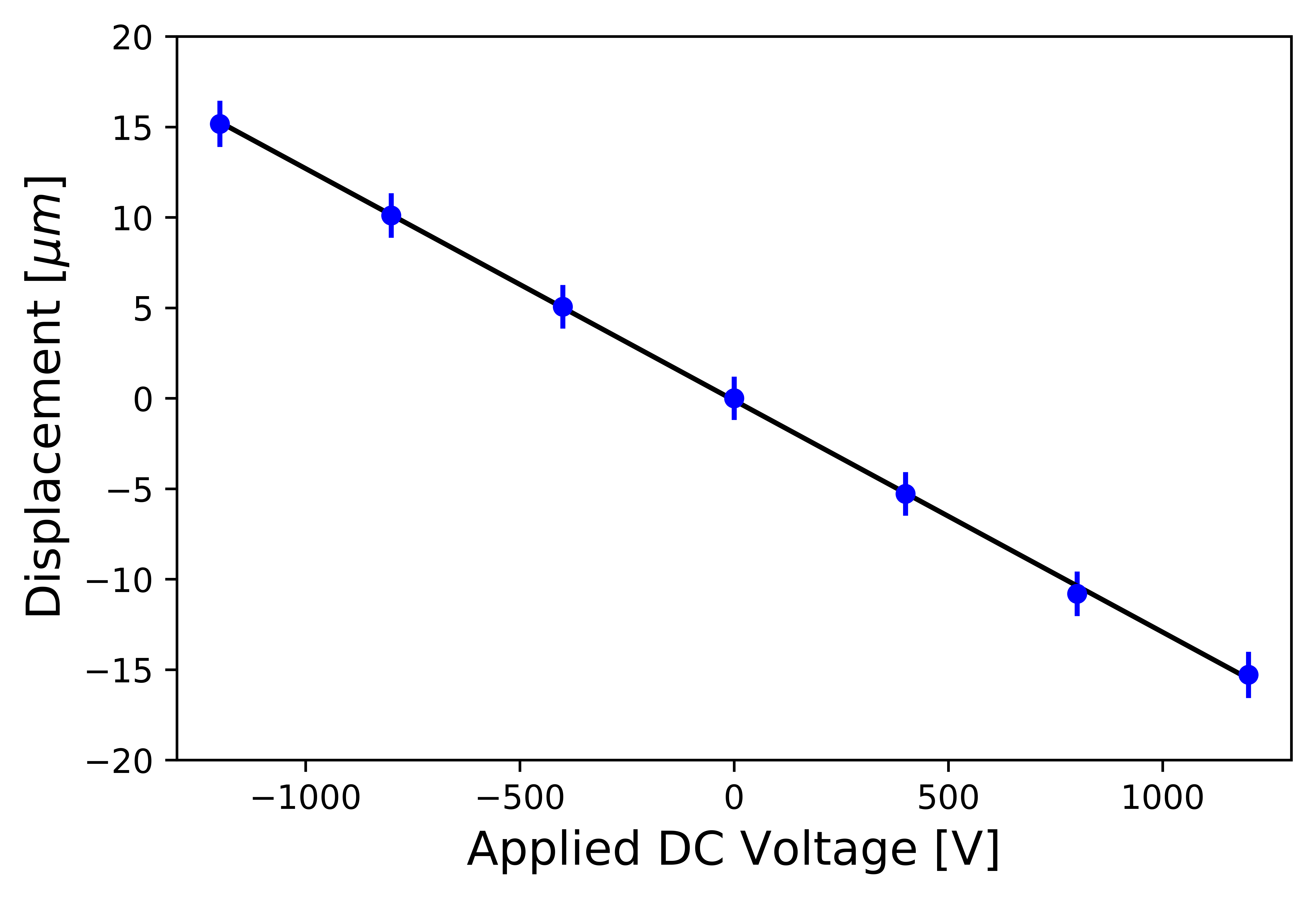}
  \caption[Disp]{Voltage-to-displacement calibration of a nearby trap electrode.  A linear fit to the measured axial displacement of the crystal versus applied DC voltage gives a calibration of $12.9 \pm 0.46$~nm/V. }\label{DispCal}
\end{figure}

The COM mode is driven by applying a resonant rf voltage to a nearby trap electrode.  To calibrate the strength of this driving force at the ions, we measure the axial displacement of the ion crystal when a large DC voltage is applied.  Applying hundreds of volts displaces the crystal by several microns, which we measure on a sideview imaging system with a magnification of $11.6 \pm 0.2$.  Figure~\ref{DispCal} shows these measured displacements over a range of voltages, and the linear fit used to extract the voltage-to-displacement calibration of $12.9 \pm 0.46$~nm/V.

Since the ions are confined axially in a quadratic potential, this displacement is linearly related to the force applied to the ions
\begin{equation}
F_{D}=-m_{\rm Be}\omega_{z}^2z.
\label{DriveForce}
\end{equation}
Through Eq.~\ref{DriveForce} and the voltage-to-displacement calibration, we are able to calculate the driving force applied to the ions for a given rf voltage and therefore the amplitude of the COM mode that is excited for a given drive duration. The uncertainty in the voltage-to-displacement calibration is the largest systematic uncertainty in these experiments.

\subsection{Spin depolarization due to photon scattering}

\begin{figure}
    \centering
    \includegraphics[scale = 0.5]{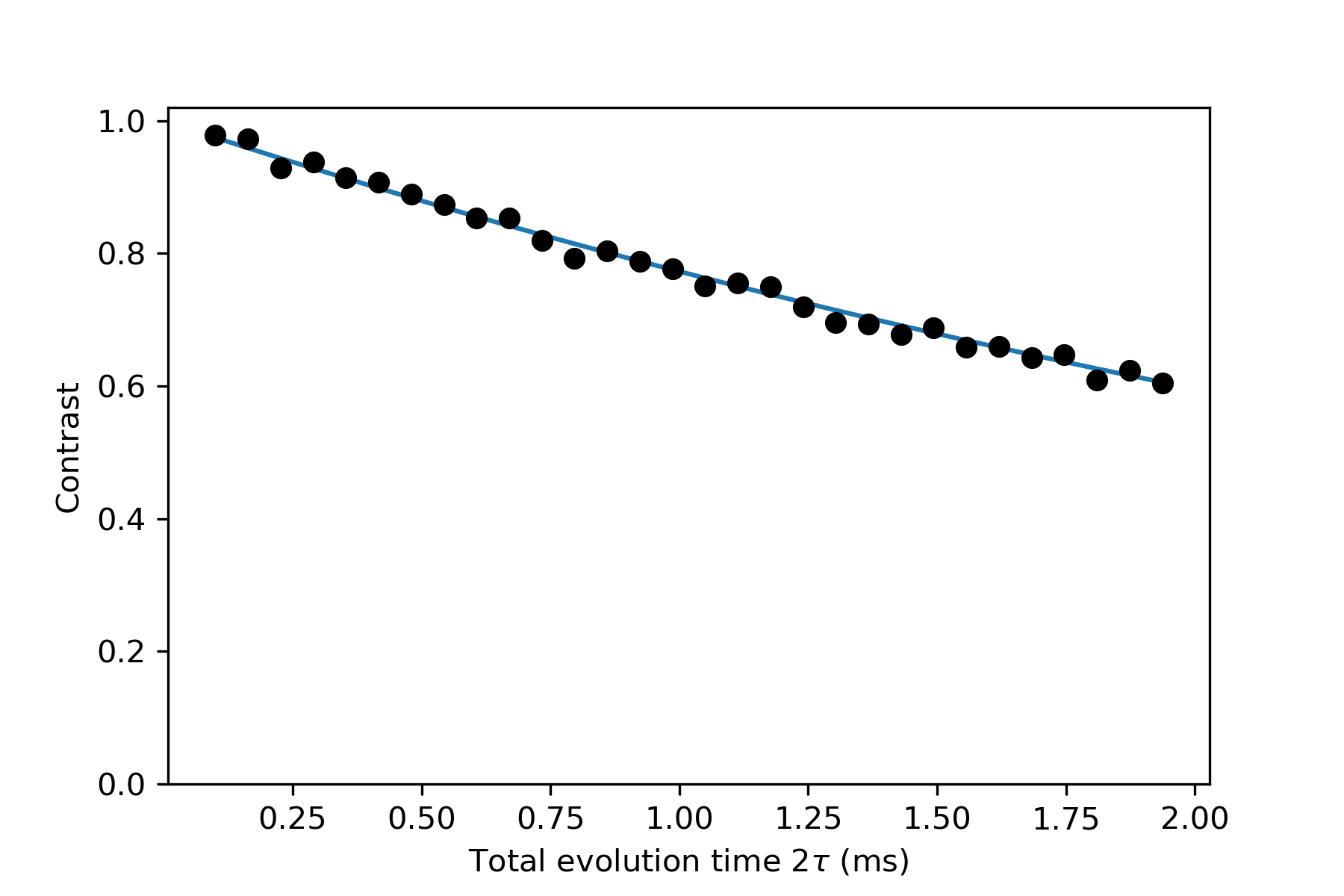}
  \caption[Measurement of depolarization from photon scattering]{Depolarization due to photon scattering from the off-resonance laser beams that produce the ODF. The black points are data for the decoherence rate due to ODF beams. The blue lines are fits to the data. In this case, $\Gamma_{tot} = 250 \pm 3 ~s^{-1}$, respectively.}\label{depolarization}
\end{figure}

An important source of noise that must be calibrated is the spin depolarization due to photon scattering from the off-resonance ODF laser beams. For the conditions of the NIST Penning trap set-up, Rayleigh scattering is the dominant source of decoherence over Raman scattering \cite{Uys2010}. Typically, $\Gamma_{el} \approx 4 \Gamma_{ram}$. As a result, the dominant effect of photon scattering here is spin depolarization rather than spin-flip errors. To empirically measure this depolarization rate, a spin-echo Ramsey sequence is used with the ODF turned on during the two free evolution periods each of time $\tau$ (for a total interaction time of $2 \tau$). The ODF detuning $\delta$ is set $50$ kHz above the COM mode. In Fig. \ref{depolarization} the contrast is the measured length of the composite Bloch vector of the spins. The contrast is given by $\braket{\hat{\sigma}_x} = \exp(- 2 \Gamma_{\mathrm{tot}}  \tau)$, where $\Gamma_{\mathrm{tot}} = \frac{1}{2}(\Gamma_{ram}+\Gamma_{el})$. In the main text and elsewhere in this supplementary material, the simplifying convention $\Gamma \approx \Gamma_{el}$ is used. Figure \ref{depolarization} is an example of such a dataset. As a result of the ACSS being nulled as well as the amplitudes stabilized for each beam, decoherence due to laser power fluctuations is negligible.

\subsection{Spin-motion coupling strength}

To couple the spin and motional degrees of freedom, a spin-dependent optical dipole force (ODF) is generated by a pair of off-resonant laser beams. The two beams intersect at the ions and create a one-dimensional traveling wave potential at a frequency $\mu$ equal to the resulting beatnote frequency. The relevant Hamiltonian for this interaction is:

\begin{equation}
H_{\text{SDF}} = F_0 \cos(\mu t) \sum_i \hat{z}_i \hat{\sigma}_i^z
\label{H_sdf},
\end{equation}
where $\hat{\sigma}_i^{z}$ is the $z$-component of the Pauli spin matrix for ion $i$, and $\hat{z}_i$ is the axial position of ion $i$. Details of the generation of this spin-dependent force are given in the supplementary information of \cite{Britton2012, Bohnet2016}. The axial motion of the crystal can be decomposed into $N$ modes with frequencies $\omega_m$ (with $\omega_z$, the COM frequency, being the highest) and normalized eigenfunctions $\vec{b}_m$ (with components $b_m^i$) that can be calculated as detailed in the supplementary material of~\cite{Britton2012}. 

We measure the strength of $F_0$ by tuning the spin-dependent force to a frequency close to the COM frequency $\omega_z $ such that the detuning $\delta=\mu-\omega_z=2\pi/\tau$  (where $\tau$ is
the interaction time) is chosen to decouple the spin and COM motion degrees of freedom at the end of the sequence \cite{Leibfried2003}.  This stroboscopically realizes a pure Ising spin interaction.  The effect of the remaining modes (other than the COM) is also captured by an Ising interaction by virtue of their large detuning with respect to the spin-dependent force frequency.  The net effect of the ODF on the spins is then well described by the Ising Hamiltonian
\begin{equation}\label{eqn:MethodsIsing}
    \hat{H}_{\text{Ising}}=\frac{\hbar}{N} \sum_{i<j} J_{i,j} \hat{\sigma}_i^z \hat{\sigma}_i^z
\end{equation}
where the amplitudes $J_{i,j}$ are related to the mode parameters through
\begin{equation}
    J_{i,j}=\frac{ F_0^2}{4\hbar\,m_{Be}}\sum_m\frac{Nb_m^ib_m^j}{\omega_m(\mu-\omega_m)}.
\end{equation}
The only unknown quantity is $F_0$, the strength of the spin-dependent force. We calibrate it through a measurement of the so-called mean-field spin precession, the spin precession induced by the spin-motion coupling in a regime where the spins decouple from the motion and the induced spin-spin interaction is weak, resulting in minimal spin correlations. Under this approximation, each spin precesses about an effective magnetic field created by the rest of the spins: $\hat{H}_{\text{mf}}=\sum_i B_i\hat{\sigma}_z^i/2$, with $B_i=2\hbar\sum_{j\neq i}J_{i,j}\braket{\hat{\sigma}_z^j}/N$. By applying the Hamiltonian in Eq.~(\ref{eqn:MethodsIsing})
 and measuring the mean-field precession of the collective Bloch vector $\braket{\hat{J}}$ we can then extract $F_0$~\cite{Britton2012,Bohnet2016}.
 
In the main text we concentrate only on the COM mode and so the spin-motion Hamiltonian reduces to
 \begin{equation}
     \hat{H}=\hbar\,\delta \hat{a}^\dagger\hat{a}+\frac{\hbar g}{\sqrt{N}}(\hat{a}^\dagger+\hat{a})\hat{J}_z.
 \end{equation}
 In this notation, the COM mode coupling strength is $g=F_0z_0/\hbar$ and $z_0=\sqrt{\hbar/(2m_{\rm Be}\omega_z)}$. For this work, typical values range from $g / (2\pi) = 3.7 $ to $4.1$ kHz. Specifically, for the amplitude sensing experiment in the main text $g/(2\pi)=3.91$ kHz, while $g/(2\pi)=3.88$ kHz for electric field sensing.

\subsection{Center-of-mass frequency fluctuations}

\begin{figure}
    \centering
    \includegraphics[width=0.8\columnwidth]{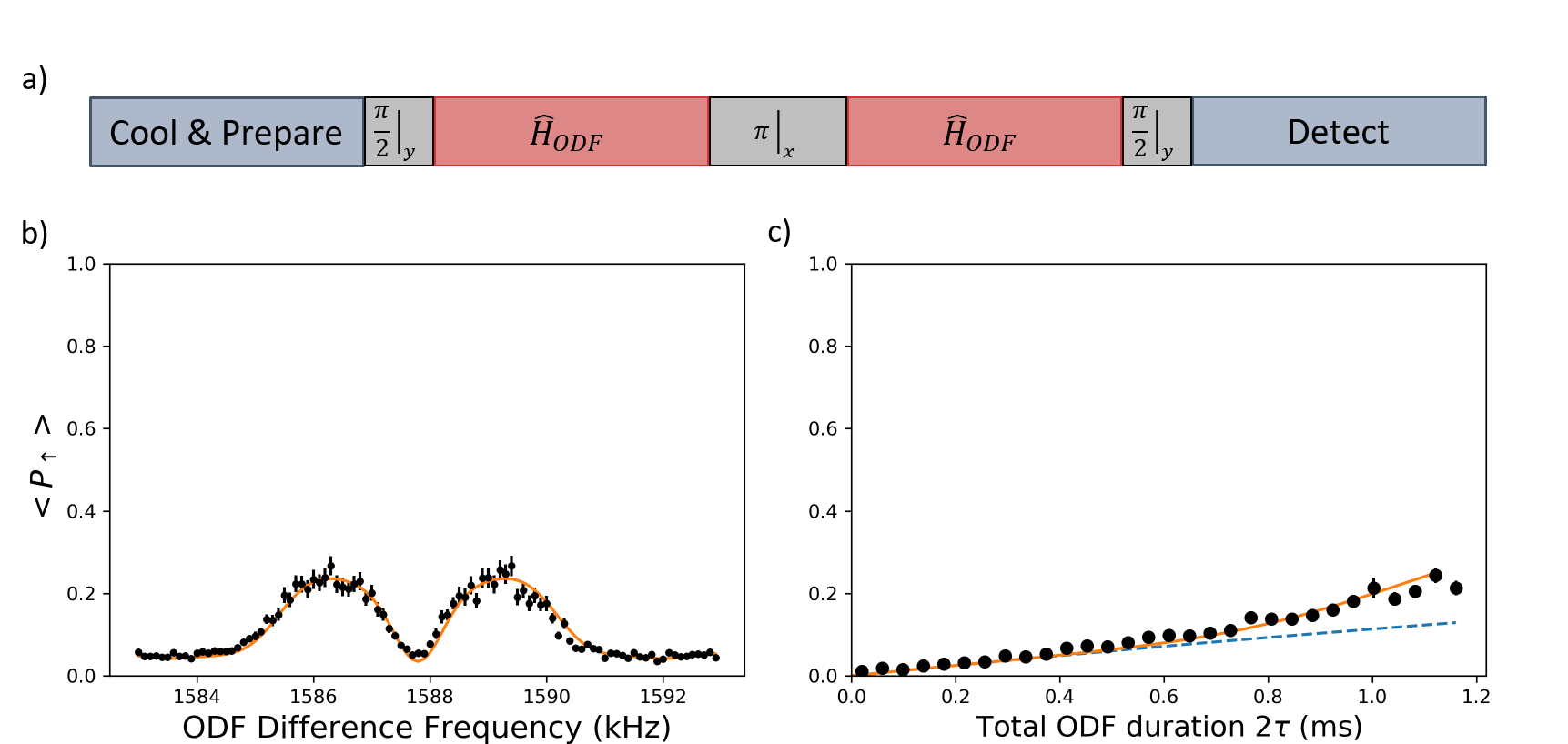}
  \caption[COM mode and fluctuations]{a) Sequence used to measure the COM mode phonon occupation and frequency fluctuations. b) Scan over COM mode lineshape, used to determine the phonon occupation of the mode. By fitting to the data, a value of $\bar{n} = 4.4$ is extracted. c) Measurement of COM frequency fluctuations by nominally setting $\delta = 0$ and measuring the increase of $<P_{\uparrow}>$ above the expectation from spin depolarization as a function of the protocol duration. The orange line is a fit to the data (black points), yielding rms frequency fluctuations of $\sigma/(2 \pi) = 30 $ Hz. The blue dashed line is theory including only spin depolarization due to photon scattering.}\label{com_sig}
\end{figure}

The frequency fluctuations of the center-of-mass mode were determined by measuring the inability of a time reversal to undo the spin-motion entanglement in the sequence shown in Fig. \ref{com_sig}(a). If no spin-dependent force is applied the spins are rotated to the down state at the end of the sequence, resulting in zero bright fraction. Figure \ref{com_sig}(b) shows the signal line shape as the frequency $\mu$ of the spin-dependent force is swept across the center-of-mass mode frequency $\omega_z$. On resonance ($\delta = \mu-\omega_z$) the spin-motion entanglement generated in the first pulse is taken out by the second pulse, resulting in a sharp dip in the line shape.

However, if the COM mode frequency changes from one experimental trial to the next, this cancellation will be imperfect. We characterize this effect by measuring $\langle P_{\uparrow}\rangle$ with $\delta $ nominally set to zero while varying the duration of the ODF (Fig. \ref{com_sig}(c)). Frequency fluctuations of the COM mode will result in the bright fraction $\langle P_{\uparrow}\rangle$ increasing above the expected background due to photon scattering. We obtain the amplitude of frequency fluctuations by fitting this increase to $\braket{P_{\uparrow}}=\frac{1}{2}[1-\braket{\hat{J}_x(2\tau)}]$, where $\braket{\hat{J}_x(2\tau)}$ is obtained from Eqs.~(\ref{eqn:MethodsSummaryofResults}) . Since the sequence of steps used in the calibration is essentially the same as the displacement sensing protocol without the displacement (see Fig.~\ref{fig:MethodsDisplacementSchematic}), $|h(2\tau)|^2$ and $p(2\tau)$ are given by Eq.~(\ref{eqn:MethodsDisplacementParameters}), giving us the formula
\begin{equation}
    \braket{P_{\uparrow}}=\frac{1}{2}-\frac{1}{2}\frac{e^{-2\Gamma_{\mathrm{tot}}\tau}}{\sqrt{1+\frac{g^2\sigma^2\tau^4(2\bar{n}+1)}{N}+\frac{4g^4\sigma^2\tau^2}{9N}}},
\end{equation}
which we find agreement with at both short and long times. The size of the measured frequency fluctuations varied slowly over a time period of an hour and from day to day. Typically, the rms amplitude of frequency fluctuations obtained through this fitting is $\sigma / (2 \pi) \approx 40 \pm 20$ Hz. 

\subsection{Center-of-mass mode damping \& ring-down}

\begin{figure}
    \centering
    \includegraphics[scale=0.4]{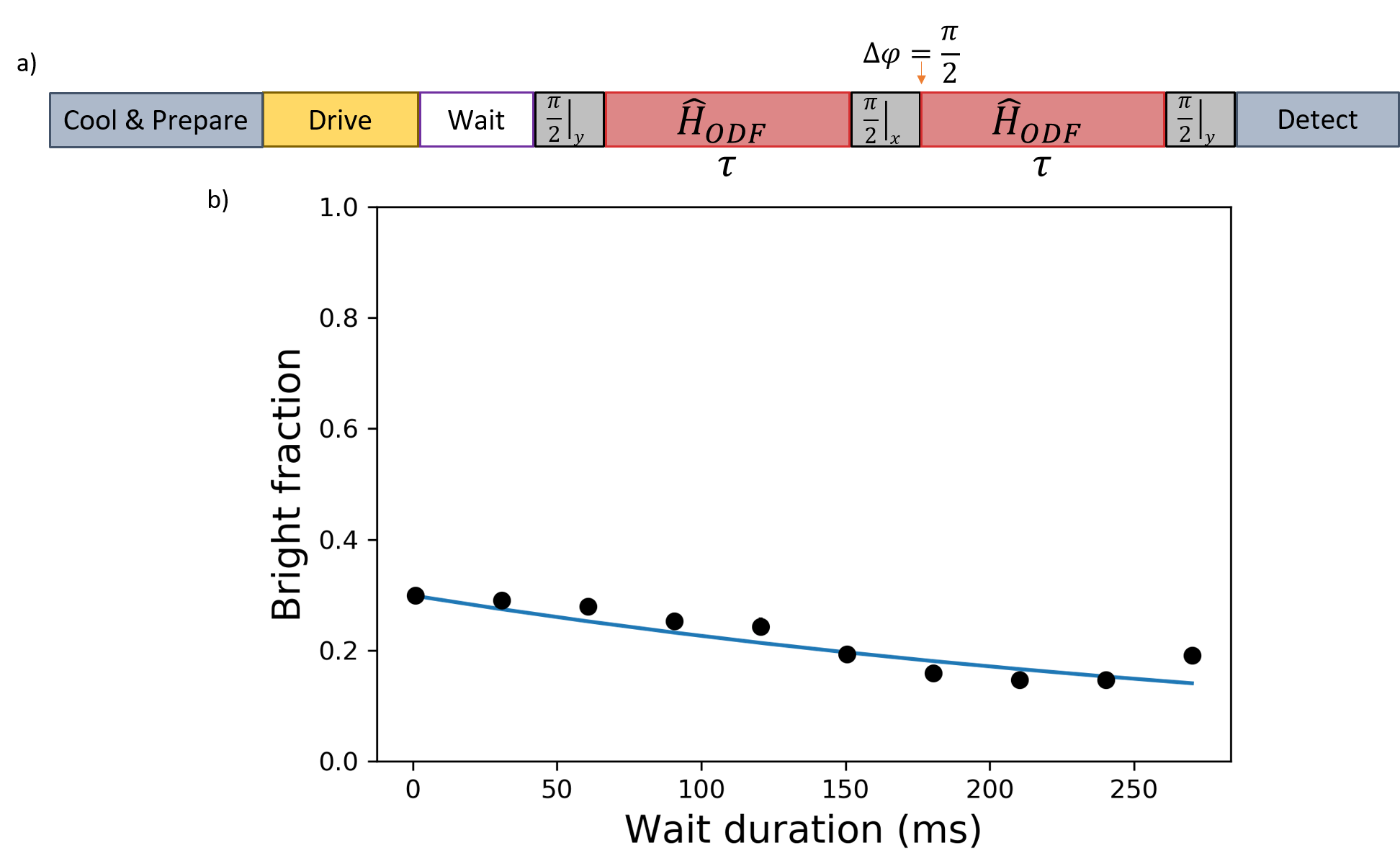}
  \caption[Ring down experimental data]{Experimental data for the ring down sequence described in the text. An exponential decay is fit to the data and yields a decay rate of $\kappa = 1 / ( 300 \: \mathrm{ms})$.}\label{ring_down}
\end{figure}

A potential source of noise that could impact the amplitude and electric field measurements is damping of the COM mode. This might occur by energy transfer out of the COM mode and to the other axial or in-plane modes. To assess whether this is an issue for this experiment, we perform a ring down experiment wherein a large coherent motion is excited by a drive applied to an endcap electrode and the excited COM mode amplitude is subsequently measured following a variable wait duration. The amplitude of the motion is calibrated to be $Z_c \approx 50$ nm, about an order of magnitude larger than the thermal axial extent of the COM mode $z_{\mathrm{cm}} = \sqrt{\frac{\hbar}{2 N m_{\mathrm{Be}} \omega_z} (2 \bar{n} + 1)} = 5.5$ nm, with $N = 120$, $\omega_z / (2 \pi) = 1.59$ MHz, and $\bar{n} = 4.6$. The motion is read out by mapping it onto the spin state with a spin-dependent ODF and performing a projective measurement of the spins. The phase of the driven motion relative to the ODF is left uncontrolled and varies randomly from one trial of the experiment to the next \cite{Gilmore2017}. The protocol used in \cite{Gilmore2017} measures a single quadrature of the motion, but here we measure the rms sum of both quadratures with the sequence shown in Fig. \ref{ring_down} so that the sensitivity to the relative phase $\phi$ is removed and the signal-to-noise ratio is not limited by fluctuations in $\phi$. Note that the signal is sensitive to the rms sum of the spin precession in each arm of the sequence. By applying a $\pi/2$ rotation about the $x$-axis in the middle of the sequence followed by a $\pi/2$ phase shift of the ODF, both quadratures are mapped into rotations of the spins: $\theta_1 = \theta_{\mathrm{max}} \cos(\phi)$ and $\theta_2 = \theta_{\mathrm{max}} \sin(\phi)$, where $\theta_{\mathrm{max}} = F_0 Z_c \tau \hbar$ with $\tau$ the duration of each ODF arm. The signal, then, is $\left< P_{\uparrow} \right> = \frac{1}{2}\left[1-e^{-\Gamma_{\mathrm{tot}} 2 \tau} \left< \cos(\theta_1) \cos(\theta_2) \right>\right] \approx \frac{1}{2}\left[1-e^{-2\Gamma_{\mathrm{tot}}  \tau} \cos(\theta_{\mathrm{max}})\right]$, where this approximation is valid for small angles. Figure \ref{ring_down}(b) shows the results of an experiment where the previously described sequence is applied and the wait duration following excitation of the COM mode is varied. Typically, no reduction in the signal is observed for over $100$ ms. By fitting to this data with an exponential decay, a value of $1/\kappa = 300$ ms is extracted, where $\kappa$ is the exponential decay rate. Calculations show that this weak damping is sufficiently small that it does not impact the displacement and electric field sensitivity studied in this manuscript.

\subsection{Limits on stochastic electric fields}

Ion traps, in particular small rf ion traps, are known to exhibit stochastic electric fields that heat the motional degrees of freedom of a trapped ion crystal. These stochastic electric fields could be due to noisy external circuits connected to the trap electrodes. In general, this source of noise can be filtered, but there remains an anomalous source of stochastic electric fields due to fluctuating potentials on the surfaces of the trap electrodes \cite{Brownnutt2015}. Measurements of the ion crystal heating rate can be used to place limits on the size of a stochastic electric field sensed by the trapped-ion crystal.  

The heating rate of the COM mode was determined by applying the sequence discussed is Sec. S5.D to measure the COM mode phonon occupation, but with a variable wait time after the state preparation. The mode will heat during this wait, and by measuring the mode occupation for various wait times the heating rate can be extracted. Figure 13 shows an example of this heating rate measurement following EIT cooling to near the motional ground state. A fit to this data yields a heating rate of $58 \pm 5$ quanta/s. The measured heating rate would vary from day to day, with an upper bound of $100$ quanta/s. Note that the COM mode heating rate due to electric field noise increases linearly with the number of ions. Therefore, the upper bound of 100 quanta/s performed with $N \approx 100$ ions provides an upper bound on a single-ion heating rate due to stochastic electric fields of $1$ quanta/s.

The $1$ quanta/s heating rate, if entirely due to stochastic electric fields, corresponds to a power spectral density of electric field noise of $S_E(2\pi\times 1.6 \mathrm{MHz}) \approx 0.25\times 10^{-14}~ (\mathrm{V/m})^2~ \mathrm{s}$ \cite{Brownnutt2015} or a $50$~nV m$^{-1}/\sqrt{\mathrm{Hz}}$ stochastic electric field spectral density. We emphasize that this is an upper limit, as the measured heating rate could be due to sources other than stochastic electric fields. Our demonstrated $240$ ~nV m$^{-1}/\sqrt{\mathrm{Hz}}$ sensitivity for measuring a coherent electric field is large compared to this stochastic electric field upper bound. 

\begin{figure}
    \centering
    \includegraphics[scale = .6]{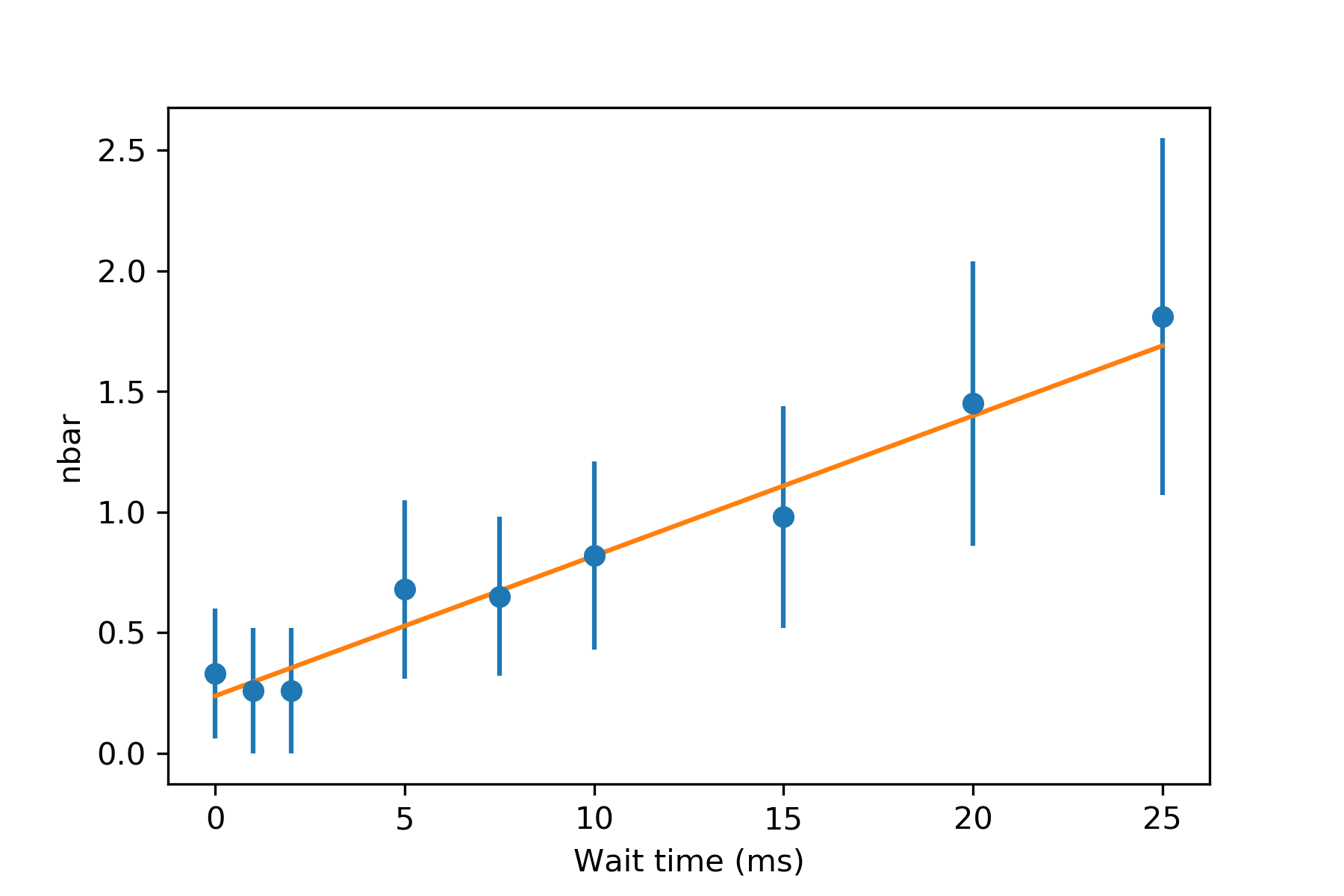}
  \caption[COM mode heating rate]{By measuring the COM mode occupation after a variable wait, the heating rate can be extracted. Blue points are data following a fit to the COM mode lineshape. The orange line is a fit to the data and gives a heating rate of $58 \pm 5$ quanta/s.}\label{heating_rate}
\end{figure}

\subsection{Excess noise}

To measure any excess experimental noise beyond projection noise, we perform the displacement sensing experiment detuned $50$~kHz from the COM mode where the effects of thermal and frequency fluctuations of this mode are negligible.  In this off-resonant experiment, an $18\%$ increase in the noise above projection noise is measured.  This excess noise is reduced as the strength of the ODF is decreased, which points towards effects due to departure from the Lamb-Dicke regime as the likely cause.  For short ODF arm times, this noise dominates over COM mode frequency fluctuations.  When comparing theory and experimental data, we have reduced the signal-to-noise of the theory by including this excess noise in the theory curves.

\section{Experimental analysis}

To compare with theory, we relate the unitless parameter $\beta$ to a displacement $Z_c$ through $\beta = Z_c\sqrt{N}/(2z_0)$, where $z_0 = \sqrt{\hbar/(2mw_z)}$ is the ground state wavefunction size for a single ion and $Z_c = F_D T / (2 m_{Be} w_z)$ is the zero-to-peak amplitude of the COM motion after a drive interval $T$ with a force per ion $F_D$. The force $F_D$ or equivalently the applied electric field at the ions is calibrated as discussed in Sec. S5. A. The signal-to-noise ratio for a single experiment $Z_c / \Delta Z_c \equiv \beta / \Delta \beta $ is extracted by repeatedly measuring $\hat{J}_y$, which is related to the spin population in the $\ket{\uparrow}$ state by $\braket{P_{\uparrow}}=\frac{1}{2}[1-\braket{\hat{J}_y}]$. We take the difference between pairs of subsequent measurements with a $\pi$ phase shift in the amplitude $\beta$ such that $\hat{J}_y^{(1)} = -\hat{J}_y^{(2)}$. The signal-to-noise data (Fig. 3(a) of the main text) is produced by taking the average of the measurement pair differences ${\braket{\hat{J}_y^{(1)}} - \braket{\hat{J}_y^{(2)}}}$, dividing by the standard deviation of the set of difference measurements, and further dividing by $\sqrt{2}$ to produce the single-measurement signal-to-noise ratio. The experimental single-measurement signal-to-noise is defined as

\begin{equation}
    \frac{Z_c}{\delta Z_c}|_{\mathrm{single, exp}} \equiv \frac{\beta}{\Delta \beta}  = \frac{\braket{J_y^{(1)}} - \braket{J_y^{(2)}}}{\sqrt{2} ~ \sigma(J_y^{(1)} - J_y^{(2)})}.
\end{equation}

The variance $(\Delta \beta)^2$ of a single measurement of $\beta$, which is plotted in Fig. 3(b) of the main text,  is calculated from the experimentally determined signal-to-noise ratio for a single measurement of an amplitude $Z_c$ through the expression

\begin{equation}
(\Delta \beta)^2 = \frac{1}{4} \left( \frac{\sqrt{N}Z_c}{z_0 \frac{Z_c}{\delta Z_c}|_{\mathrm{single}}} \right)^2
\label{sens_exp}
\end{equation}
where $N$ is the number of ions and $z_0 = \sqrt{\hbar/2 m_{\rm Be} \omega_z}$.

Similarly, $(\Delta \eta)^2$ (see Fig. 4(b) of the main text) is calculated through

\begin{equation}
(\Delta \eta)^2 = \frac{1}{4} \left( \frac{\sqrt{N}Z_c}{z_0 T \frac{Z_c}{\delta Z_c}|_{\mathrm{single}}} \right)^2,
\label{sens_exp_eta}
\end{equation}
where $T$ is the duration of the spin-independent drive. The standard deviation of a single measurement of the electric field, then, is

\begin{equation}
\Delta \epsilon = \frac{2 Z_c m_{\rm Be} \omega_{z}}{q T \frac{Z_c}{\delta Z_c}|_{\mathrm{single}}}.
\label{efield_exp}
\end{equation}

\bibliography{library}